\documentclass[twoside, a4paper, notoc]{JHEP3}
\usepackage{amsfonts}
\usepackage{amssymb}
\usepackage{comment}
\usepackage{epsfig}
\usepackage{graphicx}
\usepackage{amsmath}

\newcommand{\be}{\begin{equation}}
\newcommand{\ee}{\end{equation}}
\newcommand{\bea}{\begin{eqnarray}}
\newcommand{\eea}{\end{eqnarray}}
\newcommand{\mbb}{\mathbb}

\newcommand{\gsim}{\gtrsim}
\newcommand{\lsim}{\lesssim}

\newcommand{\beqa}{\begin{eqnarray}}
\newcommand{\eeqa}{\end{eqnarray}}
\newcommand{\V}{{\cal{V}}}

\newcommand{\OmegA}{{\cal X}}
\newcommand{\exd}{{\rm d}}
\newcommand{\efold}{$e$\,-fold}
\def\pref#1{(\ref{#1})}

\title{Fibre Inflation:
 Observable Gravity  Waves from \\IIB String Compactifications}

\author{M. Cicoli,$^{\small{1}}$ C.P. Burgess$^{\small{2,3,4}}$
and F. Quevedo$^{\small{1}}$ \\
$^1$ DAMTP, Centre for Mathematical Sciences, \\
\phantom{$^1$} \; Wilberforce Road, Cambridge, CB3 0WA, UK. \\
$^2$ Perimeter Institute for Theoretical Physics,
Waterloo ON, N2L 2Y5, Canada.\\
$^3$ Physics $\&$ Astronomy, McMaster University, Hamilton ON, L8S
4M1, Canada.
\\$^4$ Theory Division, CERN, CH-1211 Geneva 23, Switzerland.

Email: \email{M.Cicoli@damtp.cam.ac.uk,
cburgess@perimeterinstitute.ca, F.Quevedo@damtp.cam.ac.uk}}

\abstract{ We introduce a simple string model of inflation, in
which the inflaton field can take trans-Planckian values while
driving a period of slow-roll inflation. This leads naturally to a
realisation of large field inflation, inasmuch as the inflationary
epoch is well described by the single-field scalar potential $V =
V_0 \left( 3-4 e^{-\hat\varphi/\sqrt{3}}\right)$. Remarkably, for
a broad class of vacua all adjustable parameters enter only
through the overall coefficient $V_0$, and in particular do not
enter into the slow-roll parameters. Consequently these are
determined purely by the number of \efold ings, $N_e$, and so are
not independent: $\varepsilon \simeq \frac32 \eta^2$. This implies
similar relations among observables like the primordial
scalar-to-tensor amplitude, $r$, and the scalar spectral tilt,
$n_s$: $r \simeq 6(n_s - 1)^2$. $N_e$ is itself more
model-dependent since it depends partly on the post-inflationary
reheat history. In a simple reheating scenario a reheating
temperature of $T_{rh}\simeq 10^{9}$ GeV gives $N_e\simeq 58$,
corresponding to $n_s\simeq 0.970$ and $r\simeq 0.005$, within
reach of future observations. The model is an example of a class
that arises naturally in the context of type IIB string
compactifications with large-volume moduli stabilisation, and
takes advantage of the generic existence there of K\"ahler moduli
whose dominant appearance in the scalar potential arises from
string loop corrections to the K\"{a}hler potential. The inflaton
field is a combination of K\"{a}hler moduli of a K3-fibered
Calabi-Yau manifold. We believe there are likely to be a great
number of models in this class -- `high-fibre models' -- in which
the inflaton starts off far enough up the fibre to produce
observably large primordial gravity waves.}

\preprint{DAMTP-2008-59}


\begin{document}

\tableofcontents

\bigskip

\section{Introduction}

Much progress has been made over the past few years towards the
goal of finding cosmological inflation amongst the controlled
solutions of string theory \cite{GloriousHistory}. Part of the
motivation for so doing has been the hope that observable
predictions might emerge that are robust to all (or many)
realizations of inflation in string theory, but not generic to
inflationary models as a whole. The amplitude of primordial
gravity waves has recently emerged as a possible observable of
this kind \cite{BMcA,SmallGW}, with unobservably small predictions
being a feature of most of the known string-inflation proposals.

The prediction arises because the tensor amplitude is related (see
below) to the distance traversed in field space by the inflaton
during inflation, and this turns out to have upper limits in
extant models, despite there being a wide variation in the nature
of the candidate inflaton fields considered: including brane
separation \cite{BBbarInfl}; the real and imaginary parts of
K\"{a}hler moduli \cite{KModInfl,kahler}; Wilson lines
\cite{WLInfl}; the volume \cite{volume,cklq} and so on.
Furthermore, the same prediction appears also to be shared by some
of the leading proposed alternatives to inflation, such as the
cyclic/ekpyrotic models.

Since the observational constraints on primordial tensor
fluctuations are about to improve considerably --- with
sensitivity reaching down to $r \simeq 0.001$ (for $r = T/S$ the
ratio of amplitudes of primordial tensor and scalar fluctuations)
\cite{Verde,rBounds} --- it is important to identify precisely how
fatal to string theory would be the observation of primordial
gravity waves at this level. This has launched a search amongst
theorists either to prove a no-go theorem for observable $r$ from
string theory, or to derive explicit string-inflationary scenarios
that can produce observably large values of $r$. Silverstein and
Westphal \cite{SW} have taken the first steps along these lines,
proposing the use of monodromies in a particular class of IIA
string compactifications. In such models the inflaton field
corresponds to the position of a wrapped D4 brane that can move
over a potentially infinite range, thereby giving rise to
observably large tensor perturbations.

In this article we provide a concrete example of large field
inflation in the context of moduli stabilisation within the well
studied IIB string compactifications. Working within such a
framework allows us to use the well-understood properties of
low-energy 4D supergravity, with the additional control this
implies over the domain of validity of the inflationary
calculations.

More generally, we believe the inflationary model we propose to be
the simplest member of a broad new family of inflationary
constructions within the rich class of IIB stabilisations known as
the LARGE\footnote{The capitalisation of LARGE is a reminder that
the volume is exponentially large, and not simply large enough to
trust the supergravity limit.} Volume Scenario (LVS) \cite{LVS}.
Within this framework complex structure moduli are fixed
semiclassically through the presence of branes and fluxes, while
K\"ahler moduli are stabilised by an interplay between
non-perturbative corrections to the low-energy superpotential,
$W$, and perturbative $\alpha'$ and string-loop corrections to the
K\"{a}hler potential, $K$, of the effective low-energy 4D
supergravity. In particular, the LARGE volume that defines these
scenarios naturally arises as an exponentially large function of
the small parameters that control the calculation.

Most useful for inflationary purposes is the recent classification
\cite{ccq2} within the LVS framework of the order in the $\alpha'$
and string-loop expansions that governs the stabilisation of the
various K\"ahler moduli for general IIB Calabi-Yau
compactifications. In particular it was found that for K3-fibred
Calabi-Yaus, LVS moduli stabilisation only fixes the overall
volume and blow-up modes if string loop corrections to $K$ are
ignored. The fibre modulus --- call it $\OmegA$, say -- then
remains with a flat potential that is only lifted once string loop
corrections are also included. Consequently $\OmegA$ has a flatter
potential than does the overall volume modulus, making it
systematically lighter, and so also an attractive candidate for an
inflaton. Our proposal here is the first example of the family of
inflationary models which exploits this flatness mechanism, and
which we call {\it Fibre Inflation}.

This class of models is also attractive from the point of view of
obtaining large primordial tensor fluctuations. This is because
the relatively flat potential for $\OmegA$ allows it to traverse a
relatively large distance in field space compared with other
K\"ahler moduli. In this note we use these LVS results to
explicitly derive the inflaton potential in this scenario, where
the range of field values is large enough to easily give rise to
60 \efold ings of slow-roll inflation.

Unlike most string-inflation models (but similar to K\"ahler
modulus inflation (KMI) \cite{kahler}) slow roll is ensured by
large field values rather than tuning amongst parameters in the
potential. Most interestingly, within the inflationary regime all
unknown potential parameters appear only in the normalization of
the inflaton potential and not in its shape. Consequently,
predictions for the slow-roll parameters (and for observables
determined by them) are completely determined by the number of
\efold ings, $N_e$, between horizon exit and inflation's end.
Elimination of $N_e$ then implies the slow-roll parameters are
related by $\varepsilon \simeq \frac32 \, \eta^2$, implying a
similar relation between $r$ and the scalar spectral tilt: $r
\simeq 6(n_s - 1)^2$. [By contrast, the corresponding predictions
for KMI are $\varepsilon \simeq 0$ and so $r \simeq 0$, leaving
$n_s \simeq 1 + 2\eta$.]

Since the value of $N_e$ depends somewhat on the post-inflationary
reheat history, the precise values of $r$ and $n_s$ are more model
dependent, with larger $N_e$ implying smaller $r$. In a simple
reheat model (described in more detail below) $N_e$ is correlated
with the reheat temperature, $T_{rh}$, and the inflationary scale,
$M_{inf}$, through the relationship:
\begin{equation}
 N_e \simeq 62 + \ln\left(\frac{M_{inf}}{10^{16}
 GeV}\right) - \frac{\left(1-3w\right)}{3\left(1+w\right)}
 \ln\left(\frac{M_{inf}}{T_{rh}}\right) \,,
\end{equation}
where $w = p/\rho$ parameterizes the equation of state during
reheating. Numerically, choosing $M_{inf}\simeq 10^{16} GeV$ and
$T_{rh}\simeq 10^{9}$ GeV (respectively chosen to provide
observably large primordial scalar fluctuations, and to solve the
gravitino problem), we find that $N_e \simeq 58$, and so $n_s
\simeq 0.970$ and $r \simeq 0.005$. Tensor perturbations this
large would be difficult to see, but would be within reach of
future cosmological observations like EPIC, BPol or CMBPol
\cite{Verde,rBounds}.

Our preliminary investigations reveal several features likely to
be common to the broader class of Fibre Inflation models. On one
hand, as already mentioned, slow roll is ultimately controlled by
the large values of the moduli rather than on the detailed tuning
of parameters in the scalar potential. On the other hand, large
volumes imply low string scales, $M_s$, and this drives down the
inflationary scale $M_{inf}$. This is interesting because it may
lead to inflation even at low string scales but could be  a a
problem inasmuch as it makes it more difficult to obtain large
enough scalar fluctuations to account for the primordial
fluctuations seen in the CMB. (It also underlies the well-known
tension between TeV scale supersymmetry and the scale of inflation
\cite{cklq,kl}.) This suggests studying alternative methods to
generate density fluctuations within these models,\footnote{We
thank Toni Riotto for numerous discussions of this point.}, to
allow lower inflationary scales to co-exist with observably large
primordial fluctuations. Although fluctuations generated in this
way would not produce large tensor modes, they might be testable
through their predictions for non-gaussianities.

The biggest concern for Fibre Inflation and K\"{a}hler Modulus
Inflation is whether higher-loop contributions to the potential
might destabilize slow roll. In KMI this problem arises already at
one loop, and leads to the requirement that no branes wrap the
inflationary cycle (from which the dangerous contributions arise).
Fibre Inflation models do not have the same problems, and this is
likely to simplify greatly the ultimate reheating picture in these
models. They may yet have similar troubles once contributions from
blow-up modes or higher loops can be estimated,\footnote{We thank
Markus Berg for conversations about this.} but we find that
current best estimates for these corrections are not a
problem.

Finally, it is relatively simple in these models to obtain large
hierarchies amongst the size of the moduli, in a way that leads to
some dimensions becoming larger than others (rather than making
the extra dimensions into a frothy Swiss cheese). This potentially
opens up the possibility of `sculpting' the extra dimensions, by
having some grow relatively slowly compared to others as the
observed four dimensions become exponentially large.

After a short digression, next, summarising why large $r$ has
proven difficult to obtain in past constructions, the remainder of
the paper is devoted to explaining Fibre Inflation, and why it is
possible to obtain in it $r \simeq 0.005$. The description starts,
in \S2, with a review of various LVS tools which are used in
subsequent sections. Aficionados of the LVS can skip this
discussion, jumping right to \S3 which describes both the special
case of K3 fibrations, and the inflationary potentials to which
they give rise. Our conclusions are briefly summarised in \S4.

\subsection{The Lyth bound}

What is so hard about obtaining observably large primordial tensor
fluctuations in string constructions? In 1996 David Lyth
\cite{Lyth} derived a general correlation between the ratio $r$
and the range of values through which the (canonically normalized)
inflaton field, $\varphi$, rolls in single-field slow-roll models:
\be
    r = 16 \, \varepsilon = \frac{8}{N_{\rm eff}^2}
    \left(\frac{\Delta\varphi}{M_p}\right)^2 \,,
\ee
where $\varepsilon = \frac12 (V'/V)^2$ is the standard first
slow-roll parameter, and
\be
    N_{\rm eff} = \int_{t_{\rm he}}^{t_{\rm end}}
    \left( \frac{\xi}{r} \right)^{1/2}
    H \exd t \,.
\ee
Here $\xi(t) = 8(\dot\varphi/H M_p)^2$ is the quantity whose value
at horizon exit gives the observed tensor/scalar ratio, $r =
\xi(t=t_{\rm he})$, $H(t) = \dot a/a$ is (as usual) the Hubble
parameter, and the integral runs over the $N_e \gsim 50$ \efold
ings between horizon exit and inflation's end. Notice in
particular that $N_{\rm eff} = N_e$ if $\xi$ is a constant. Lyth's
observation was that the validity of the slow roll and
measurements of the scalar spectral index, $n_s-1$, constrain
$N_{\rm eff} \gsim 50$, and so $r \gsim 0.01$ requires the
inflaton to roll through a transplanckian range, $\Delta \varphi
\gsim M_p$.

This observation has proven useful because the inflaton usually
has some sort of a geometrical interpretation when inflationary
models are embedded into string theory, and this allows the
calculation of its maximum range of variation. For instance,
suppose inflation occurs due to the motion of the position, $x$,
of a brane within 6 extra dimensions, each of which has length
$L$. Then expressing the geometric upper limit $\Delta x < L$ in
terms of the canonically normalized inflaton field, $\varphi =
M_s^2 x$, gives $\Delta \varphi/M_p < M_s^2 L/M_p$, where $M_s$ is
of order the string scale. But $L$ is not itself independent of
$M_s$ and the 4D Planck constant, $M_p$. For instance, in the
absence of warping one often has $M_p^2 \simeq M_s^8 L^6$, which
allows one to write $\Delta \varphi/M_p < (M_s/M_p)^{2/3}$.
Finally, consistency of calculations performed in terms of a
(higher-dimensional) field theory generally require the hierarchy,
$1/L \ll M_s$ which implies $M_s/M_p \simeq (M_s L)^{-3} \ll 1$,
showing that $\Delta \varphi/M_p \ll 1$.

More careful estimates of brane motion within an extra-dimensional
throat, with the condition that it geometrically cannot move
further than the throat itself is long, lead to similar
constraints \cite{BMcA}. It is considerations such as these that
show (on a case-by-case basis) for each of the extant
string-inflation constructions that the distance moved by the
inflaton is too small to allow $r \gsim 0.01$. However, in the
absence of a no-go theorem, there is strong motivation to find
stringy examples which evade these kinds of constraints, and allow
the inflaton to undergo large excursions.

\section{The Large Volume Framework}
\label{Section3}

In this section we provide a brief boilerplate review of LVS
moduli stabilisation in Type IIB string compactifications, with a
view to setting the stage for the K3 fibrations considered as
examples in the next section. We start with a reminder of the
basics of Type IIB flux compactifications themselves.

\subsection{Type IIB flux compactifications}
\label{2}

The Type IIB flux compactifications of interest are those
preserving $\mathcal{N} = 1$ supersymmetry in 4D, leading to
compactifications on Calabi-Yau three-folds $X$ \cite{gkp}. The
low-energy 4D theory obtained at low energies is described by an
$\mathcal{N}=1$ supergravity, and so is described by a K\"ahler
potential, $K$, superpotential, $W$ and gauge kinetic function,
$f_{ab}$.

\subsubsection{Lowest-order expressions}

To leading order in the string-loop and $\alpha'$ expansions, the
resulting low-energy K\"{a}hler potential has the form:
\begin{equation}
  K_{tree}=-2\ln \mathcal{V} -\ln
  \left(S+\bar{S}\right) -\ln \left( -i\int\limits_{X}\Omega \wedge
  \bar{\Omega}\right). \label{eqtree}
\end{equation}
In (\ref{eq}) $S$ is the axio-dilaton, $S=e^{-\phi }+i C_{0}$,
$\Omega $ is the Calabi-Yau's holomorphic (3,0)-form, and
$\mathcal{V}$ is its volume, measured with an Einstein frame
metric $g^{\scriptscriptstyle E}_{\mu \nu} = e^{-\phi/2} \,
g^s_{\mu \nu}$, and expressed in units of the string length,
$l_{s} = 2\pi \sqrt{\alpha'}$. In general, the complex fields of
the 4D theory include $S$ and the complex moduli of the Calabi Yau
geometry, including its complex structure moduli, $U_{\alpha}$,
$\alpha=1,...,h_{2,1}(X)$, and K\"ahler moduli, $T_i$,
$i=1,...,h_{1,1}(X)$. In eq.~\pref{eqtree} $\Omega$ is to be read
as implicitly depending on the $U_{\alpha}$'s, and $\mathcal{V}$
as depending implicitly on the $T_i$'s.

The values of $S$ and the complex structure moduli, $U_\alpha$,
can become fixed once background fluxes, $G_{3} = F_{3} +iS
H_{3}$, are turned on, where $F_{3}$ and $H_{3}$ are respectively
Type IIB supergravity's RR and NSNS 3-form fluxes (for recent
reviews on flux compactifications see: \cite{fluxes}). The
potential energy for this is captured by the resulting low-energy
superpotential, which is given by
\begin{equation}
  W_{tree}=\int\limits_{X}G_{3}\wedge \Omega \,.
  \label{yujtree}
\end{equation}
These fluxes may, but need not, break the remaining 4D
$\mathcal{N}=1$ supersymmetry, corresponding to whether or not the
resulting scalar potential is minimized where $D_\alpha W =
\partial_\alpha W + W \partial_\alpha K$ vanishes at the minimum.

The K\"ahler moduli $T_i$ do not appear in $W_{tree}$ and so
remain precisely massless at leading semiclassical order. The
supergravity describing this massless sector is obtained after
eliminating the heavier fields $S$ and $U_\alpha$ at the classical
level. Provided this can be done in a supersymmetric way
\cite{IntOut}, by solving $D_\alpha W = 0$, the supergravity
description of the remaining K\"ahler moduli is specified by a
constant superpotential, $W = W_0 = \langle W_{tree} \rangle$, and
the K\"ahler potential $K=K_{cs} - 2\ln\left({2}/{g_{s}}\right) +
K_0$, with
\be
  K_0 = -2\ln \mathcal{V}
  \qquad \hbox{and} \qquad
  e^{-K_{cs}} = \left\langle -i\int\limits_{X}\Omega \wedge
  \bar{\Omega}\right\rangle\,. \label{K0expr}
\ee

To express $K_0$ explicitly in terms of the fields $T_i$ write the
volume in terms of the K\"{a}hler form, $J$, expanded in a basis
$\{ \hat{D}_{i} \}$ of $H^{1,1}(X,\mathbb{Z})$ as $J =
\sum_{i=1}^{h_{1,1}} t^{i} \hat{D}_{i}$ (we focus on orientifold
projections such that $h_{1,1}^{-} = 0 \Rightarrow h_{1,1}^{+} =
h_{1,1}$). This gives
\begin{equation}
  \mathcal{V} = \frac{1}{6}\int\limits_{X}J\wedge J\wedge
  J = \frac{1}{6} k_{ijk}t^{i}t^{j}t^{k}. \label{IlVolume}
\end{equation}
Here $k_{ijk}$ are related to the triple intersection numbers of
$X$ and the $t^{i}$ are 2-cycle volumes. The quantities $t^i$
appearing here are related to the components of the chiral
multiplets $T_{i}$ as follows. Writing $T_i =\tau _{i} + ib_{i}$,
$\tau _{i}$ turns out to be the Einstein-frame volume (in units of
$l_s$) of the divisor $D_{i}\in H_{4}(X,\mathbb{Z})$, which is the
Poincar\'{e} dual to $\hat{D}_{i}$. Its axionic partner $b_{i}$ is
the component of the RR 4-form $C_{4}$ along this cycle:
$\int_{D_{i}} C_{4} = b_{i}$. The 4-cycle volumes $\tau _{i}$ are
related to the 2-cycle volumes $t^{i}$ by:
\begin{equation}
 \tau _{i}=\frac{\partial \mathcal{V}}{\partial
 t^{i}}=\frac{1}{2}\int\limits_{X}\hat{D}_{i}\wedge J\wedge
 J=\frac{1}{2} k_{ijk}t^{j}t^{k}. \label{defOfTau}
\end{equation}
$K_0$ is now given as a function of $T_i$ by solving these
equations for the $t^i$ as functions of the $\tau_i = \frac12(T_i
+\overline{T}_i)$, and substituting the result into
eq.~\pref{K0expr} using eq.~\pref{IlVolume} to evaluate
$\mathcal{V}$.

The $\mathcal{N}=1$ F-term supergravity scalar potential for $T_i$
which results is given in terms of $K$ and $W$ (in 4D Planck
units) by:
\begin{equation}
  V=e^{K}\left\{K^{i\bar\jmath} D_{i}W D_{\bar\jmath}\overline{W}
 -3\left\vert
 W\right\vert ^{2}\right\} ,  \label{b}
\end{equation}
where $K^{i\bar\jmath}$ is the inverse of the K\"ahler metric
$K_{\bar\imath j} = \partial_{\bar\imath} \partial_j K$ and, as
before, $D_{i}W = \partial_{i}W + W \partial_{i} K$. Notice that
the above procedure ensures that $\mathcal{V}$ is a homogeneous
function of degree $\frac32$ in the $\tau_i$'s, and so also
ensures $K_0$ satisfies $K_0(\lambda \tau_i) \equiv K_0(\tau_i) -
3\ln\lambda$ as an identity for all $\lambda$ and $\tau_i$. It
follows from this that $K_0$ satisfies the no-scale identity:
$K_0^{i\bar\jmath} \partial_i K_0 \partial_{\bar\jmath} K_0 \equiv
3$ for all $\tau_i$. This in turn guarantees the potential,
\pref{b}, constructed using $K_0$ is completely flat, $V \equiv
0$, as is required for agreement with the microscopic
compactification since the fluxes did not stabilize the K\"ahler
moduli to leading order.

\subsection{Corrections to the leading approximation} \label{22}

Because the leading contributions to the potential for the $T_i$'s
identically vanish, we must work to sub-leading order in $\alpha'$
and $g_s$ (string loops) in order to determine its shape. (The
same is not required for $S$ and $U_\alpha$, whose potential is
dominated by the leading order contribution.) For the present
purposes there are three important corrections to track.

\subsubsection{Superpotential corrections}

Since the superpotential receives no contributions at any finite
order in $\alpha'$ and $g_s$, its first corrections arise
non-perturbatively. These can be generated either by Euclidean D3
branes (ED3) wrapping 4 cycles, or by gaugino condensation by the
supersymmetric gauge theories located on D7 branes also wrapping
4-cycles in the extra dimensions. The resulting superpotential is
\begin{equation}
  W=W_{0}+\sum\limits_{i}A_{i}e^{-a_{i}T_{i}},
 \label{yuj}
\end{equation}
where the sum is over the 4-cycles generating nonperturbative
contributions to $W$, and as before $W_{0}$ is independent of
$T_i$. The $A_{i}$ correspond to threshold effects and can depend
on the complex structure moduli and positions of D3-branes. The
constants $a_i$ in the exponential are given by $a_{i}=2\pi $ for
ED3 branes, or $a_{i}=2\pi /N$ for gaugino condensation with gauge
group $SU(N)$. There may additionally be higher instanton effects
in (\ref{yuj}), but these can be neglected so long as each
$\tau_{i}$ is stabilised such that $a_{i}\tau_{i}\gg 1$.

The presence of such a superpotential generates a scalar potential
for $T_i$, of the form
\begin{eqnarray}
 \delta V_{(sp)} &=& e^{K_0}  K_{0}^{j\bar\imath} \Bigl[
 a_{j}A_{j} \, a_{i} \bar{A}_{i} e^{-\left(
 a_{j}T_{j} + a_{i}\overline{T}_{i}\right) } \nonumber\\
 && \qquad\qquad  -\left(
 a_{j}A_{j}e^{-a_{j}T_{j}}\overline{W}
 \partial_{\bar\imath} K_{0}+a_{i}\bar{A}_{i} e^{-a_{i}\overline{T}_{i}} W
 \partial_j K_{0}\right) \Bigr].  \label{scalarWnp}
\end{eqnarray}

\subsubsection{Leading $\alpha'$ corrections}

Unlike the superpotential, the K\"ahler potential receives
corrections order-by-order in both the $\alpha'$ and string-loop
expansions. The leading $\alpha'$ corrections for the Type IIB
flux compactifications of interest lead to a K\"{a}hler potential
for the K\"ahler moduli of the form
\begin{equation}
  K= -2\ln \left( \mathcal{V}+\frac{\xi }{2g_{s}^{3/2}}\right) \,,
 \label{eq}
\end{equation}
up to an irrelevant $T_i$-independent constant. Here $\xi$ is
given by $ \xi =-\frac{\chi (X)\zeta (3)}{2(2\pi )^{3}}$, $\chi$
is the Euler number of the Calabi-Yau $X$, and the relevant value
for the Riemann zeta function is $\zeta(3) \approx 1.2$. The
leading contribution to the scalar potential that follows from
this correction is given by (defining $\hat{\xi}\equiv
\xi/g_{s}^{3/2}$):
\begin{equation}
 \delta V_{(\alpha')} = 3\, \hat{\xi}e^{K_0} \frac{\left( \hat{\xi}^{2}+7\hat{\xi}
 \mathcal{V}+\mathcal{V}^{2}\right) }{\left( \mathcal{V}-\hat{\xi}
 \right) \left( 2\mathcal{V}+\hat{\xi} \right)^{2}} \left\vert
 W_0\right\vert ^{2}
 \approx \frac{3 \,\hat\xi}{4 \mathcal{V}^3} \, |W_0|^2 \,,
 \label{scalaralpha'}
\end{equation}
where the validity of the $\alpha'$ and loop expansions require
$\mathcal{V} \gg \hat\xi \gg 1$.

\subsubsection{String-loop corrections} \label{5}

$K$ also receives corrections from string loops, and although
there is at present no explicit computation of string scattering
amplitudes on a generic Calabi-Yau background it has proven
possible to argue what the $T_i$-dependence is likely to be for
the leading string loop corrections \cite{07040737, cicq}. This
section briefly reviews these arguments.

The only explicit string computation of loop corrections to $K$ is
available for $\mathcal{N}=1$ compactifications on $T^{6}/(
\mathbb{Z}_{2}\times \mathbb{Z}_{2})$ \cite{bhk} and gives:
\begin{equation}
 \delta K_{(g_{s})}=\delta K_{(g_{s})}^{KK}+\delta K_{(g_{s})}^{W},
\end{equation}
where $\delta K_{(g_{s})}^{KK}$ comes from the exchange between D7
and D3-branes of closed strings which carry Kaluza-Klein momentum,
and reads (for vanishing open string scalars)
\begin{equation}
 \delta K_{(g_{s})}^{KK}= -\frac{1}{128\pi^{4}}
 \sum\limits_{i=1}^{3}
 \frac{\mathcal{E}_{i}^{KK}(U,\bar{U})}{\hbox{Re}\left( S\right)
 \tau _{i}}. \label{KK}
\end{equation}
In the previous expression we assumed that all the three 4-cycles
of the torus are wrapped by D7-branes and $\tau_{i}$ denotes the
volume of the 4-cycle wrapped by the $i$-th D7-brane. The other
correction $\delta K_{(g_{s})}^{W}$ is interpreted in the closed
string channel as due to exchange of winding strings between
intersecting stacks of D7-branes. It takes the form
\begin{equation}
 \delta K_{(g_{s})}^{W}=-\frac{1}{128\pi^{4}}
 \sum\limits_{i=1}^{3}\left.
 \frac{\mathcal{E}_{i}^{W}(U,\bar{U})}{\tau _{j}\tau_{k}}
 \right\vert _{j\neq k\neq i},  \label{W}
\end{equation}
where $\tau_{i}$ and $\tau_{j}$ denote the volume of the 4-cycles
wrapped by the $i$-th and the $j$-th intersecting D7-branes. Note
that in both cases there is a very complicated dependence of the
corrections on the $U$ moduli, encoded in the functions
$\mathcal{E}_{i}(U,\bar{U})$, but a very simple dependence on the
$T$ moduli.

These observations were used by \cite{07040737} to conjecture a
generalisation of these formulae to 1-loop corrections on general
Calabi-Yau three-folds. Given that these corrections can be
interpreted as the tree-level propagation of a closed KK string,
and that a Weyl rescaling is always necessary to convert the
string computation to Einstein frame, they proposed
\begin{equation}
 \delta K_{(g_{s})}^{KK}\sim \sum\limits_{i=1}^{h_{1,1}}
 \frac{\mathcal{C}_{i}^{KK}(U,\bar{U})m_{KK}^{-2} }{\hbox{Re}
 \left( S\right) \mathcal{V}} \sim \sum\limits_{i=1}^{h_{1,1}}
 \frac{\mathcal{C}_{i}^{KK}(U,\bar{U})\left( a_{il}t^{l}\right)
 }{\hbox{Re} \left( S\right) \mathcal{V}},  \label{UUU}
\end{equation}
where $a_{il}t^{l}$ is a linear combination of the basis 2-cycle
volumes $t_{l}$ that is transverse to the 4-cycle wrapped by the
$i$-th D7-brane. A similar line of argument for the winding
corrections gives
\begin{equation}
 \delta K_{(g_{s})}^{W}\sim \sum\limits_{i}
 \frac{\mathcal{C}_{i}^{W}(U,\bar{U})m_{W}^{-2}}{\mathcal{V}}\sim
 \sum\limits_{i}\frac{\mathcal{C} _{i}^{W}(U,\bar{U})}{\left(
 a_{il}t^{l}\right) \mathcal{V}}, \label{UUUU}
\end{equation}
with $a_{il}t^{l}$ now being the 2-cycle where the two D7-branes
intersect. $\mathcal{C}^{KK}_i $ and $\mathcal{C}^W_i$ are unknown
functions of the complex structure moduli, which may be simply
regarded as unknown constants for the present purposes because the
complex structure moduli are already flux-stabilised by the
leading-order dynamics. What is important is the leading order
dependence on K\"{a}hler moduli, which the conjecture explicitly
displays.

Ref. \cite{cicq} used these expressions to work out the
implications of eqs. (\ref{UUU})-(\ref{UUUU}) for the effective
scalar potential. The result is a relatively simple formula that
is expressible in terms of the tree-level K\"{a}hler metric
$K_{0}=-2\ln \mathcal{V}$ and the winding correction to the
K\"{a}hler potential, as follows:
\begin{equation}
 \delta V_{\left( g_{s}\right) }^{1-loop}=\left(
 \frac{ \left( \mathcal{C} _{i}^{KK} \right)^2}
 {\hbox{Re}(S)^{2}} \, a_{ik} a_{ij}
 K_{k\bar\jmath}^{0} - 2 \sum_{i} \delta
 K_{(g_{s}),t_{i}}^{W}\right) \frac{W_{0}^{2}}{\mathcal{V}^{2}}.
 \label{V at 1-loop}
\end{equation}
For branes wrapped only around the basis 4-cycles (such as we
consider below) the combination appearing in the first term
degenerates to $a_{ik} a_{ij} K^0_{k\bar\jmath} =
K^0_{i\bar\imath}$. As emphasized in ref.~\cite{07040737}, this
contribution is subdominant in powers of $1/\mathcal{V}$ relative
to the leading $\alpha'$ correction, eq.~\pref{scalaralpha'}, in
the limit of current interest, where $\mathcal{V}$ is large.

\subsection{Exponentially large volumes}

The observation of the LVS \cite{LVS} is that these corrections
can generate a potential for the volume modulus, $\mathcal{V}$,
with a minimum at exponentially large values. To see this it is
necessary to include corrections of more than one type, and given
the subdominance of the string-loop contribution, eq.~\pref{V at
1-loop}, relative to the $\alpha'$ correction,
eq.~\pref{scalaralpha'}, it is useful to consider initially only
the contributions $\delta V_{(sp)}$ and $\delta V_{(\alpha')}$. To
this end, combining (\ref{scalarWnp}) in (\ref{scalaralpha'})
gives the total potential for $\mathcal{V}$ of the form
\begin{eqnarray}
 V &=& e^{K_0} \left\{
 K_{0}^{j\bar\imath} \Bigl[
 a_{j}A_{j} \, a_{i} \bar{A}_{i} e^{-\left(
 a_{j}T_{j} + a_{i}\overline{T}_{i}\right) }
  \phantom{\frac12}\right.\nonumber\\
 && \qquad\qquad  - \left. \left(
 a_{j}A_{j}e^{-a_{j}T_{j}}\overline{W}
 \partial_{\bar\imath} K_{0}+a_{i}\bar{A}_{i} e^{-a_{i}\overline{T}_{i}} W
 \partial_j K_{0}\right) \Bigr]
 + \frac{3 \,\hat\xi}{4 \mathcal{V}} \, |W_0|^2 \right\}.
 \label{scalar}
\end{eqnarray}

Since the parameters (like $\xi$) appearing in this potential are
related to the topology of the underlying Calabi-Yau space, the
choices required for the existence of a minimum at large volume
imply conditions on this underlying topology. These conditions are
analysed in ref.~\cite{ccq2} and can be summarised
as follows:
\begin{enumerate}
\item{} The Euler number of the Calabi Yau manifold must be
negative, or more precisely: $h_{12} > h_{11} > 1$. This ensures
the coefficient $\hat\xi$ is positive, which in turn guarantees
that $V$ goes to zero from below (in a particular direction) as
$\mathcal{V}$ goes to infinity \cite{LVS}. This is a both
sufficient and necessary condition for the existence of a minimum.
\item{} To remain within the domain of approximation, it is also
necessary to have a second modulus whose exponential contributions
to $W$ can balance the inverse powers of $\mathcal{V}$ arising
from $\alpha'$ corrections. This requires the Calabi-Yau manifold
to have at least one blow-up mode corresponding to a 4-cycle
modulus that resolves a point-like singularity \cite{ccq2}. This
4-cycle must  have positive first Chern class, and so positive
curvature, as a del Pezzo surface. This implies it can be shrunk
to zero size, re-obtaining the singularity resolved by the
blow-up.
\end{enumerate}

Although these conditions ensure the stabilization of
$\mathcal{V}$, in general the above potential is insufficient to
stabilize all of the K\"ahler moduli. In particular, if there are
$N_{small}$ blow-up modes and $L = (h_{11} - N_{small} - 1)$ modes
which do not resolve point-like singularities or correspond to the
overall volume modulus, then the above potential can stabilise all
of the $N_{small}$ blow-up moduli (at values large in string
units) and the overall volume (at values that are exponentially
large in the blow-up moduli). But the other $L$ K\"{a}hler moduli
are {\it not} fixed.

The lifting of these remaining flat directions occurs with the
inclusion of string loop corrections, which for these modes are
always dominant compared to non-perturbative effects. Since the
overall volume is stabilised, the internal moduli space is
compact, implying a finite range for these remaining moduli.
Consequently we expect that the loop-generated potential does not
simply generate a runaway for these remaining fields, but must
instead generically induce a minimum. The K3 fibration example
used below to derive inflation is an explicit illustration of this
picture, with the inflaton being one of the moduli whose potential
is loop-generated.

For instance, the original example of an exponentially large
volume minimum was realised explicitly in \cite{LVS} for the
Calabi Yau $\mathbb{C}P^{4}_{[1,1,1,6,9]}[18]$, whose volume is
given by
\begin{equation}
 \mathcal{V} = \frac{1}{9\sqrt{2}}\left(\tau_{b}^{3/2}
 -\tau_{s}^{3/2}\right).
\end{equation}
In the absence of fine tuning the tree-level superpotential is of
order $W_{0}\sim\mathcal{O}(1)$, and so the $\alpha'$ and
non-perturbative corrections compete naturally to give an
exponentially large volume (AdS) minimum that breaks SUSY, located
at
\begin{equation}
 \mathcal{V} \sim W_{0} \, e^{a_{s}\tau_{s}} \gg
 \tau_{s} \sim \hat{\xi}^{2/3} \gg 1 \,.
\end{equation}
Inclusion of the string loop corrections to $V$ do not appreciably
alter this minimum since due to the subleading dependence on
$\mathcal{V}$ remarked on above \cite{07040737}.

For phenomenological applications it is usually necessary to
up-lift this minimum from AdS to allow Minkowski (or slightly de
Sitter) 4D geometries. This can be done using one of the various
methods proposed in the literature (inclusion of $\overline{D3}$
branes \cite{kklt}, D-terms from magnetised D7 branes \cite{bkq},
F-terms from a hidden
sector \cite{ss}, etc.).

An immediate generalisation of the
$\mathbb{C}P^{4}_{[1,1,1,6,9]}[18]$ model that is useful for
inflationary applications is the so called `Swiss-cheese'
Calabi-Yaus, whose volume is given by
\begin{equation}
 \mathcal{V}=\alpha\left(\tau_{b}^{3/2}
 -\sum_{i=1}^{N_{small}}\lambda_{i}\tau_{i}^{3/2}\right),\text{ \ }
 \alpha > 0,\text{ \ }\lambda_{i} > 0\text{ \ }\forall
 i=1,...,N_{small}.
\end{equation}
Examples having this form with $h_{1,1}=3$ are the Fano three-fold
$\mathcal{F}_{11}$, the degree 15 hypersurface embedded in
$\mathbb{C}P^{4}_{[1,3,3,3,5]}$ and the degree 30 hypersurface in
$\mathbb{C}P^{4}_{[1,1,3,10,15]}$. In this case the potential
stabilises the various 4-cycles, $\tau_{i}$, (that control the
size of the `holes' of the Swiss-cheese) at comparatively small
values (though still much larger than the string scale), $\tau_{i}
\sim \mathcal{O}(10)$, $\forall i=1,...,N_{small}$. By contrast,
$\tau_{b}$ (which controls the overall size of the Calabi-Yau) is
stabilised at the exponentially large value, $\mathcal{V} \sim
e^{a_{i} \tau_{i}}$.

\subsection{K\"{a}hler modulus inflation}

Finally, we briefly review the mechanism of K\"{a}hler moduli
inflation \cite{kahler}, since many of the features of the model
presented here draw on this example. The starting point for this
model is a Swiss cheese Calabi-Yau manifold, which must have at
least {\it two} blow-up modes ($N_{small} \ge 2$ and so $h_{1,1}
\ge 3$), such as is true, for instance, for the $\mathbb{C}
P_{[1,3,3,3,5]}$ model \cite{blumenhagen, ccq2}.

Assuming the minimal three K\"ahler moduli of this form, our
interest is in that part of moduli space for which these satisfy
$\tau_b \gg \tau \gg \tau_s$, where $\tau$ and $\tau_s$ are the
blow-up modes while $\tau_b$ controls the overall volume. As a
first approximation neglect string loop corrections as well as
exponentials of the large moduli $\tau_b$ and $\tau$ in $V$. Then
one finds that $\tau_b$ and $\tau_s$ can both be stabilised with
$\mathcal{V} \sim e^{a_s \tau_s}$ and $\tau_s \gg 1$.

Fixing these to their stabilised values, but now considering the
subdominant dependence on $\tau$, the potential for the remaining
modulus takes the form:
\be
 V= A \frac{\sqrt{\tau}\,e^{-2a\tau}}{\V}-B\frac{\tau
 e^{-a\tau}}{\V^2}+C\frac{\hat\xi}{\V^3} \,,
\ee
where the volume $\V$ should be regarded as being fixed. Varying
$\tau$ with $\V$ fixed (this is the reason why $h_{11}\geq 3$ is
needed), the potential for large $\tau$ is dominated for the last
two terms, which is naturally very flat due to its exponential
form.

The above potential gives rise to slow-roll inflation, {\it
without} the need for fine-tuning parameters in the potential. For
the canonically normalised inflaton,
\be
\varphi = \sqrt{4 \lambda /
(3\V)} \; \tau^{3/4},
\ee
 the above potential becomes
\be
 V\simeq V_0 -\beta\left(\frac{\varphi}{\V}\right)^{4/3}\,
 e^{-a'\V^{2/3}\varphi^{4/3}} \,,
\ee
with $V_0 \sim \hat\xi/\V^3$. This is very similar to textbook
models of large-field inflation \cite{liddle}, although with the
search for observably large tensor modes in mind, one must also
keep in mind an important difference. This is because although in
both cases slow roll requires a large argument for the exponential
in the potential, this is accomplished differently in the two
cases. In the textbook examples the argument of the exponential is
typically given by $\varphi/M_p$, and so slow roll requires
$\varphi \gsim M_p$. In the present case, however, slow roll is
typically accomplished for small $\varphi$, due to the large
factor of $\V^{2/3}$ in the exponent. In this crucial way, what we
have is actually a small-field model of inflation.

\subsubsection{Naturalness}

It is remarkable that this is one of the only string-inflation
models that does not suffer from the $\eta$ problem, inasmuch as
slow roll does not require a delicate adjustment amongst the
parameters in the scalar potential.
However, one worries that the extreme flatness of the potential
might be affected by sub-leading corrections not yet included in
the scalar potential, such as string-loop corrections to the
K\"{a}hler potential.

Although a definitive analysis requires performing a string loop
calculation, some conclusion may be drawn using the conjectured
modulus dependence \cite{07040737,cicq,ccq2} discussed above. In
fact, examination of the previous formulae shows that dangerous
contributions can arise if D7 branes wrap the inflationary cycle,
since in this case string-loop corrections take the form
\be
 \delta V_{1-loop}\sim \frac{1}{\sqrt{\tau} \; \V^3}
 \sim \frac{1}{\varphi^{2/3} \V^{10/3}} \,.
\ee
This is dangerous because it gives a contribution to the slow-roll
parameter, $\eta = M_p^2 V''/V$, of the form $\delta \eta \sim
M_p^2 \delta V''/V_0 \sim \varphi^{-8/3} \V^{-1/3} \hat\xi^{-1}$,
which for the typical values of interest, $\varphi \sim \V^{-1/2}
\ll 1$, may be large.

One way out of this particular problem is simply not to wrap D7
branes about the inflationary cycle. In this case the remaining
loop corrections discussed above do not destroy the slow roll.
(Although it is not yet possible to quantitatively characterise
the contributions of higher loops, see Appendix \ref{Appendix A}
for a related discussion of some of the issues.) Of course, if
ordinary Standard Model degrees of freedom reside on a D7, not
wrapping D7s on the inflationary cycle is likely to complicate the
eventual reheating mechanism because it acts to decouple the
inflaton from the observable sector. However we do not regard this
particular objection as being too worrisome, since a proper study
of reheating in these (and most other models) of string inflation
remains a long way off \cite{warpedreheat}.

\section{Fibre Inflation}

We now return to our main line of argument, and describe the
simplest K3-Fibration inflationary model. We regard this model as
being a representative of a larger class of constructions (Fibre
Inflation), which rely on choosing the inflaton to be one of those
K\"ahler moduli whose potential is first generated at the
string-loop level.\footnote{Even though these moduli are also
K\"ahler moduli, their behaviour is very different from the volume
and in particular the blowing-up modes that drive K\"ahler moduli
inflation. In this sense the previous scenario might be more
properly called `blow-up inflation' to differentiate it from the
later `volume' inflation \cite{volume} and `fibre' inflation
developed here.}

\subsection{K3 fibration Calabi-Yaus} \label{examples}

To describe the model we first require an explicit example of a
Calabi-Yau compactification which has a modulus that is not
stabilized by nonperturbative corrections to $W$ together with the
leading $\alpha'$ corrections to $K$. The simplest such examples
are given by Calabi Yaus which have a K3 fibration structure.

For our present purposes, a K3 fibered Calabi-Yau can be regarded
as one whose volume is linear in one of the 2-cycle sizes, $t_j$
\cite{oguiso}. That is, when there is a $j$ such that the only
non-vanishing coefficients are $k_{jlm}$ and $k_{klm}$ for $k,l,m
\neq j$, then the Calabi-Yau manifold is a K3 fibration having a
$\mbb{C}P^1$ base of size $t_j$, and a K3 fibre of size $\tau_j$.
The simplest such K3 fibration has two K\"ahler moduli, with $\V=
\tilde t_1 \tilde t_2^{\,2} + \frac{2}{3}\tilde t_2^{\,3}$. This
becomes $\V = \frac12 \sqrt{\tilde\tau_1} \left( \tilde\tau_2 -
\frac23 \tilde\tau_1 \right)$ when written in terms of the 4-cycle
volumes $\tilde\tau_1 = \tilde t_2^{\,2}$ and $\tilde\tau_2 =
2(\tilde t_1 + \tilde t_2) \tilde t_2$, corresponding to the
geometry $\mbb{C}P^4_{[1,1,2,2,6]}[12]$ \cite{Candelas}. For later
convenience we prefer to follow a slightly different basis of
cycles in this geometry,
\begin{equation}
 \tau_1 = \tilde\tau_1, \textit{ \ \ \ \ \
 } \tau_2 = \tilde\tau_2 - \frac{2}{3}\tilde \tau_1,
\end{equation}
with a similar change in the 2-cycle basis, $\{ \tilde t_i \} \to
\{ t_i \}$. In terms of these the overall volume becomes
\begin{equation}
 \V = t_1 t_2^2
 = \frac{1}{2}\sqrt{\tau_1} \; \tau_2\textit{ \ \ \
 }\Leftrightarrow\textit{ \ \ \ } \V = t_1\tau_1,
\end{equation}
where $t_1$ is the base and $\tau_1$ the K3 fiber.

For inflationary purposes we also require a third K\"ahler
modulus, which we can achieve by simply adding an extra blow-up
mode, as is required in any case to guarantee the existence of
controlled large volume solutions. We therefore begin by assuming
a compactification whose volume is given in terms of its three
K\"ahler moduli in the following way:
\begin{equation}
 \mathcal{V} = \lambda_{1}t_{1}
 t_{2}^{2} + \lambda_{3}t_{3}^{3}
 = \alpha \left( \sqrt{\tau _{1}}\tau _{2} - \gamma \tau _{3}^{3/2}\right)
 = t_1\tau_1-\alpha\gamma\tau_3^{3/2},  \label{hhh}
\end{equation}
where the constants $\alpha $ and $\gamma $ are given in terms of
the model-dependent numbers, $\lambda_i$, by $\alpha = \frac12 \,
\lambda_1^{-1/2}$ and $\gamma = (4\lambda_1/27 \lambda_3)^{1/2}$,
related to the two independent intersection numbers, $d_{122}$ and
$d_{333}$, by $\lambda_1 = \frac12 \, d_{122}$ and $\lambda_3 =
\frac16 \, d_{333}$. (Clearly, including more blow-up modes than
we have done here is straightforward.) Given that eq. (\ref{hhh})
simply expresses the addition of the blow-up mode $\tau_3$, to the
geometry $\mathbb{C}P^{4}_{[1,1,2,2,6]}[12]$ \cite{ccq2}, we do
not expect there to be any obstruction to the existence of a
Calabi-Yau manifold with these features.

We further assume that $h_{2,1}(X) > h_{1,1}(X) = 3$, thus
satisfying the other general LVS condition. Since we seek
stabilisation with $\V$ large and positive, we work in the
parameter regime
\begin{equation} \label{V0hier}
 \V_0 := \alpha\sqrt{\tau_1}
 \; \tau_2 \gg \alpha\gamma\tau_3^{3/2} \gg 1 \,,
\end{equation}
with the constant $\gamma$ taken to be positive and order unity.
This limit keeps the volume of the Calabi-Yau large, while the
blow-up cycle remains comparatively small. Regarding the relative
size of $\tau_1$ and $\tau_2$, we consider two situations in what
follows: $\tau_2 \gsim \tau_1 \gg \tau_3$ and $\tau_2 \gg \tau_1
\gg \tau_3$. (We notice in passing that the second case
corresponds to $t_1 \sim \tau_2/\sqrt{\tau_1} \gg t_2 \sim
\sqrt{\tau_1} \gg t_3 \sim \sqrt{\tau_3}$, corresponding to
interesting geometries having the two dimensions of the base,
spanned by the cycle $t_1$, hierarchically larger than the other
four of the K3 fibre, spanned by $\tau_1$.) The similarity of
eq.~\pref{V0hier} with the `Swiss cheese' Calabi-Yaus of previous
sections,
\begin{equation}
\mathcal{V}=\alpha \underset{\tau
_{big}^{3/2}}{(\underbrace{\sqrt{\tau _{1}}\tau _{2}}}-\gamma \tau
_{3}^{3/2}) \,,
\end{equation}
leads us to expect (and our calculations below confirm) that the
scalar potential has an AdS minimum at exponentially large volume,
together with $( h_{1,1} - N_{small} - 1) = 1$ flat directions.

\subsubsection{The potential without string loops}
\label{3modK3noLoopCalc}

We start by considering the scalar potential computed using the
leading $\alpha'$ corrections to the K\"{a}hler potential, as well
as including nonperturbative corrections to the superpotential.
\begin{equation} \label{3}
 K = K_{0} + \delta K_{(\alpha')} = -2 \ln \left(
 \mathcal{V} + \frac{\hat{\xi}}{2} \right) \qquad \hbox{and}
 \qquad
 W = W_{0} + \sum_{k=1}^3 A_{k} e^{-a_{k}T_{k}} \,.
\end{equation}
Because our interest is in large volume $\V_0 \gg
\alpha\gamma\tau_3^{3/2}$, we may to first approximation neglect
the dependence of $T_{1,2}$ in $W$ and use instead
\begin{equation} \label{sp}
 W \simeq W_{0} + A_{3} e^{-a_{3} T_{3}} \,.
\end{equation}
In the large volume limit the K\"{a}hler metric and its inverse
become
\begin{equation}
 K_{i\bar\jmath}^{0}=\frac{1}{4\tau_2^2}\left(
 \begin{array}{ccccc}
 \frac{\tau_2^2}{\tau _{1}^{2}} && \gamma\left(
 \frac{\tau_{3}}{\tau_{1}}\right)^{3/2} && -\frac{3 \gamma
 }{2}\frac{\sqrt{\tau _{3}}}{
 \tau _{1}^{3/2}}\tau_{2} \\
 \gamma\left(\frac{\tau_{3}}{\tau_{1}}\right)^{3/2} && 2 &&
 -3\gamma \frac{\sqrt{\tau_{3}}}{\sqrt{\tau_{1}}} \\
 -\frac{3\gamma }{2}\frac{\sqrt{\tau_{3}}}{\tau_{1}^{3/2}}\tau_{2} && -3\gamma
 \frac{\sqrt{\tau_{3}}}{\sqrt{\tau_{1}}} && \frac{3\alpha \gamma
 }{2}\frac{\tau_2^2}{\mathcal{V}\sqrt{\tau_{3}}}
 \end{array}
 \right), \label{LaDiretta}
\end{equation}
and
\begin{equation}
 K_{0}^{\bar\imath j}=4\left(
 \begin{array}{ccccc}
 \tau _{1}^{2} && \gamma\sqrt{\tau_{1}}\tau_{3}^{3/2} &&
 \tau _{1}\tau _{3} \\
 \gamma\sqrt{\tau_{1}}\tau_{3}^{3/2} && \frac12
 \, \tau_2^2 &&
  \tau_2 \tau _{3} \\
 \tau _{1}\tau _{3} && \tau_2
 \tau_{3} && \frac{2}{3\alpha \gamma }\mathcal{V}\sqrt{ \tau _{3}}
 \end{array}
 \right) \,,  \label{Kinverse}
\end{equation}
where we systematically drop all terms that are suppressed
relative to those shown by factors of order
$\sqrt{\tau_3/\tau_2}$. In particular, here (and below), $\V$ now
denotes $\V_0 = \alpha \sqrt{\tau_1}\tau_2$ rather than the full
volume, $\V_0 - \alpha \gamma \tau_3^{3/2}$.

We now use these expressions in eq.~\pref{scalarWnp}, adding the
linearisation of $\delta V_{(\alpha')}$ in $\hat\xi$,
eq.~\pref{scalaralpha'}. The following identity (to the accuracy
of eqs.~\pref{LaDiretta} and \pref{Kinverse}) proves very useful
when doing so:
\be \label{oddKid}
    K_0^{3\bar 1} K^0_{\bar 1} + K_0^{3\bar 2} K^0_{\bar 2} +
    \hbox{c.c.} = -3\tau_3 \,.
\ee
The result may be explicitly minimised with respect to the $T_3$
axion direction, $b_3 = \hbox{Im} \, T_3$, with a minimum at $b_3
= 0$ if $W_0 < 0$ or at $b_3 = \pi/a_3$ if $W_0 > 0$. Once this is
done, the resulting scalar potential simplifies to
\begin{equation}
 V = \frac{8 \, a_{3}^{2}A_3^2}{3\alpha\gamma}
 \left( \frac{\sqrt{\tau _{3}}}{\mathcal{V}}
 \right) e^{-2a_{3}\tau_3}
 -4 W_{0}a_{3} A_3 \left( \frac{\tau _{3}}{\mathcal{V}^{2}}
 \right) \, e^{-a_{3}\tau_3}
 +\frac{3 \, \hat\xi W_0^2}{4 \mathcal{V}^{3}} \,,
 \label{ygfdo}
\end{equation}
where we take $W_0$ to be positive and neglect terms that are
subdominant relative to the ones displayed by inverse powers of
$\V$ without compensating powers of $e^{a_3\tau_3}$.

\begin{figure}[ht]
\begin{center}
\epsfig{file=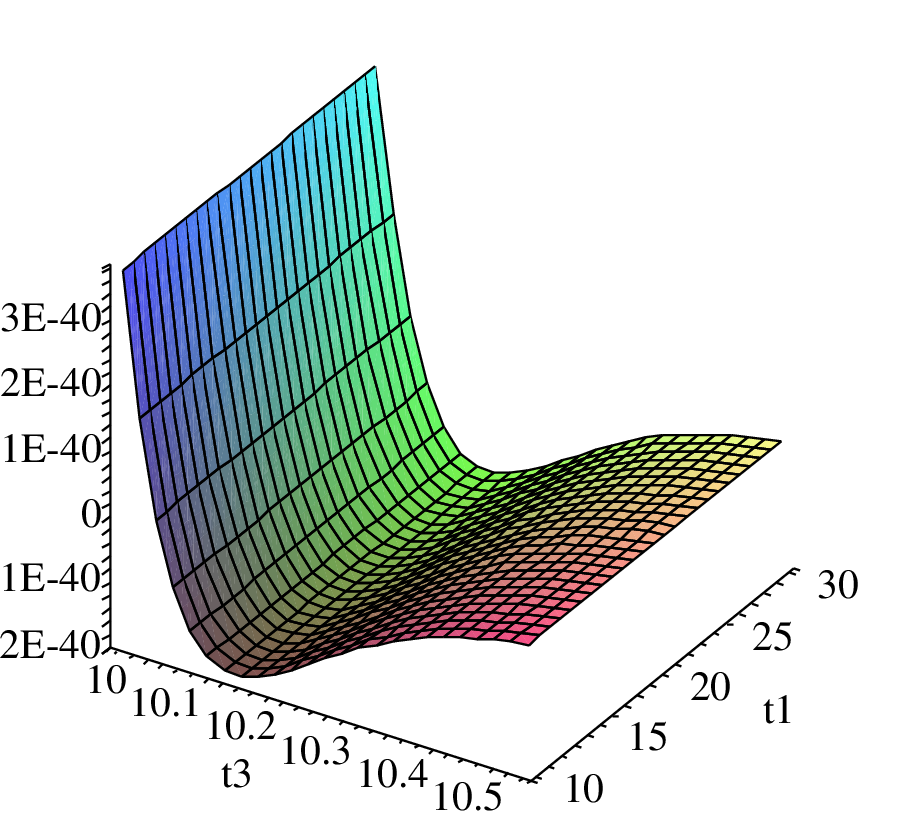, height=70mm,width=100mm} \caption{$V$
(arbitrary units) versus $\tau_1$ and $\tau_3$ for one of the
parameter sets discussed in the text, with $\mathcal{V}$ evaluated
at its minimum.} \label{Fig:sofa}
\end{center}
\end{figure}

Now comes the main point. Notice that by virtue of
eq.~\pref{oddKid} $V$ depends only on two of the three moduli on
which it could have depended: $V=V(\V,$\ $\tau_{3})$. This occurs
because we take $a_1 \tau_{1}$ to be large enough to switch off
its non-perturbative dependence in $W$. This observation has two
consequences: First, it implies that within these approximations
there is one modulus --- any combination (call it $\OmegA$, say)
of $\tau_1$ and $\tau_2$ independent of $\V$ --- which describes a
direction along which $V$ is (so far) completely flat. This plays
the r\^{o}le of our inflaton in subsequent sections.

Second, the potential (\ref{ygfdo}) completely stabilises the
combinations $\tau_3$ and $\V$ (and, in fact, has precisely the
same form as the scalar potential of the original $\mathbb{C}P_{[
1,1,1,6,9]}^{4}[18]$ LVS example of \cite{LVS}). In particular,
the only minimum satisfying $a_3\tau_3 \gg 1$ is given explicitly
by $\V = \langle \V \rangle$ and $\tau_3 = \langle \tau_3 \rangle$
with
\begin{equation}
 \langle \tau _{3}\rangle = \left(
 \frac{\hat\xi}{2\, \alpha \gamma } \right) ^{2/3}
 \qquad \hbox{and} \qquad
 \langle
 \mathcal{V}\rangle = \left( \frac{ 3 \,\alpha \gamma }{4a_{3}A_3}
 \right) W_0 \, \sqrt{\langle \tau _{3}\rangle }
 \; e^{a_{3} \langle \tau_3
 \rangle }\text{\ .}  \label{x}
\end{equation}
This is the minimum corresponding to exponentially large
volume\footnote{The two relations (\ref{x}) do not take into
account the shift in the volume minimum due to the up-lifting
term, which are worked out explicitly in Appendix \ref{Appendix A}
(and incorporated in our numerics).}.

The flat direction of the potential eq.~\pref{ygfdo} is manifest
in Figure \ref{Fig:sofa}, which plots this scalar potential with
$\V$ fixed (using the LV parameter set discussed below), as a
function of $\tau_3$ (on the \textit{x}-axis) and $\OmegA$
--- which represents any third field coordinate independent of
$\tau_3$ and $\V$ (such as $\tau_1$, for instance) --- (on the
\textit{y}-axis). In order properly to understand the potential
for $\OmegA$, we must go beyond the approximations that underly
eq.~\pref{ygfdo}, in order to lift this flat direction, such as by
including the leading string-loop contributions to the potential.

\subsubsection*{Sample parameter sets}

In what follows it is useful to follow some concrete numerical
choices for the various underlying parameters. To this end we
track several sets of choices throughout the paper, listed in
Table 1. One of these sets (call it `LV') gives very large
volumes, $\V \simeq 10^{13}$ (and so a string scale of order $M_s
\propto \V^{-1/2} \sim 10^{12}$ GeV), and is representative of
what the LARGE volume scenario likes to give for simple choices of
parameters. The others (`SV1' and `SV2', say) instead have $\V
\sim 10^3$ much smaller (and so with $M_s \sim 10^{16}$ GeV).
While all naturally provide an inflationary regime, the LV choice
has the disadvantage that the value of the classical inflationary
potential turns out too small to provide observable primordial
density fluctuations. The other choices are chosen to remedy this
problem, and to provide illustrations of different inflationary
parameter regimes. We regard all of these choices as being merely
illustrative, and have not attempted to perform a systematic
search through the allowed parameter space.

\begin{figure}[ht]
\begin{center}
\begin{tabular}{c||c|c|c}
  & LV & SV1 & SV2 \\
  \hline\hline
  $\lambda_1$ & 1 & 15 & 21/2 \\
  $\lambda_3$ & 1 & 1/6 & 1/6 \\
  $g_s$ & 0.1 & 0.3 & 0.3 \\
  $\xi$ & 0.409 & 0.934 & 0.755 \\
  $W_0$ & 1 & 100 & 100 \\
  $a_3$ & $\pi$ & $\pi/5$ & $\pi/4$ \\
  $A_3$ & 1 & 1 & 1 \\
  \hline
  $\alpha$ & 0.5 & 0.1291 & 0.1543 \\
  $\gamma$ & 0.385 & 3.651 & 3.055 \\
  $\langle\tau_3\rangle$ & 10.46 & 4.28 & 3.73 \\
  $\langle\V\rangle$ & $2.75 \cdot 10^{13}$ & 1709.55 & 1626.12
\end{tabular}\\
\vspace{0.3cm}{{\bf Table {1}:} Some model parameters (the
up-lifting to a Minkowski minimum has been taken into account).}
\end{center}
\end{figure}

\subsubsection{Inclusion of string loops} \label{bene}

We now specialise the string-loop corrections to the K3 fibration
of interest, using expression (\ref{V at 1-loop}) and working in
the regime $W_{0}\gsim\mathcal{O}(1)$ where the perturbative
corrections are important.

Consider first the contribution coming from stacks of D7 branes
wrapping the blow-up cycle, $\tau_{3}$. The Kaluza-Klein loop
correction to $V$ coming from this wrapping takes the form
\begin{equation}
 \delta V^{KK}_{(g_{s}),\tau _{3}}=
 \frac{g_{s}^{2}(\mathcal{C}_{3}^{KK})^{2}}
 {\sqrt{\tau_{3}} \; \mathcal{V}^{3}}, \label{eq3}
\end{equation}
which does not depend on $\OmegA$, and is subdominant to the
$\alpha'$ correction. These features imply such a term may modify
the exact locus of the potential's minimum, but not the main
features of the model, such as the existence of the flat direction
in $\OmegA$ and the minimization of $\V$ at exponentially large
values.

Similarly, we have seen that the winding-mode contributions to
string-loop corrections arise from the exchange of closed winding
strings at the intersection of stacks of D7 branes. But the form
of the volume (\ref{hhh}) shows that the blow-up mode, $\tau_{3}$,
only has its triple self-intersection number non-vanishing, and so
does not intersect with any other cycle. This is a typical feature
of a blow-up mode which resolves a point-like singularity: due to
the fact that this exceptional divisor is a \textit{local} effect,
it is always possible to find a suitable basis where it does not
intersect with any other divisor. Hence the topological absence of
the required cycle intersections implies an absence of the
corresponding winding-string corrections. In the end, only three
types of loop corrections turn out to arise:
\begin{equation}
 \delta V_{(g_{s})}=\delta V^{KK}_{(g_{s}), \tau_{1}}+\delta
 V^{KK}_{(g_{s}), \tau_{2}}+\delta V^{W}_{(g_{s}),
 \tau_{1}\tau_{2}},
\end{equation}
which have the form
\begin{eqnarray}
 \delta V_{(g_{s}),\tau _{1}}^{KK} &=&g_{s}^{2} \frac{\left(
 C_{1}^{KK}\right) ^{2}}{\tau _{1}^{2}}
  \frac{W_{0}^{2}}{\mathcal{V}^{2}}, \notag \\
 \delta V_{(g_{s}),\tau _{2}}^{KK} &=& 2g_{s}^{2}\frac{\left( C_{2}^{KK}\right)^{2}}{\tau_2^2}
 \frac{W_{0}^{2}}{\mathcal{V}^{2}}, \label{LOOP} \\
 \delta V_{(g_{s}),\tau _{1}\tau _{2}}^{W}
 &=&- \left( \frac{2 \, C_{12}^{W}}{t_{\ast }} \right)
 \frac{W_{0}^{2}}{\mathcal{V}^{3}} \,.
 \notag
\end{eqnarray}
Here the 2-cycle $t_{*}$ denotes the intersection locus of the two
4-cycles whose volumes are given by $\tau_{1}$ and $\tau_{2}$. In
order to work out the form of $t_{*}$, we need the relations:
\begin{equation}
 \tau _{1} = \frac{\partial \mathcal{V}}{\partial t_{1}}
 = \left( \lambda_{1} t_{2}\right) t_{2}
 \qquad\hbox{and} \qquad
 \tau _{2} = \frac{\partial \mathcal{V}}{\partial t_{2}}
 =2t_{1}(\lambda _{1}t_{2}),
 \label{taus}
\end{equation}
and so $t_{*} =\lambda_1 t_{2} = \sqrt{\lambda_{1}\tau_1}$.
Therefore the $g_{s}$ corrections to the scalar potential
(\ref{LOOP}) take the general form:
\begin{equation}
 \delta V_{(g_{s})}= \left(\frac{A}{\tau_{1}^{2}}
 - \frac{B}{\mathcal{V}\sqrt{\tau_{1}}} +\frac{C\tau_{1}}
 {\mathcal{V}^{2}}\right)\frac{W_{0}^{2}}{\mathcal{V}^{2}},
 \label{74}
\end{equation}
where
\begin{eqnarray}
 A &=& \left(g_{s} C_{1}^{KK}\right) ^{2}>0, \nonumber\\
 B &=& 2 \, C_{12}^{W}\lambda _{1}^{-1/2}
 = 4\alpha C_{12}^{W}, \label{GREAT} \\
 C &=& 2\,\left(\alpha g_{s}C_{2}^{KK}\right)^{2}>0. \nonumber
\end{eqnarray}
Notice that $A$ and $C$ are both positive (and suppressed by
$g_s^2$) but the sign of $B$ is undetermined. The structure of
$\delta V_{(g_s)}$ makes it very convenient to use $\OmegA \equiv
\tau_1$ as our parameter along the flat directions at fixed $\V$
and $\tau_3$.

In this way, it is also easier to have a pictorial view of the
inflationary process since the K3 fiber modulus $\tau_1$ will turn
out to be mostly the inflaton. Inflation will correspond to an
initial situation, with the K3 fibre much larger than the base,
which will dynamically evolve to a final situation with the base
larger than the K3 fibre.

For generic values of $A$, $B$ and $C$ we expect the potential of
eq.~\pref{74} to lift the flat direction and so to stabilize
$\OmegA\equiv\tau_1$ at a minimum. Indeed, minimizing $\delta
V_{(g_s)}$ with respect to $\tau_1$ at fixed $\V$ and $\tau_3$
gives
\be
  \frac{1}{\tau_1^{3/2}} = \left( \frac{B}{8 A \V} \right)
  \left[ 1 + (\hbox{sign} \, B) \sqrt{1 + \frac{32 AC}{B^2}}
  \right] \label{tau1soln1} \,,
\ee
which, when $32 AC \ll B^2$, reduces to
\be
  \tau_1 \simeq \left(-\frac{B \V}{2C} \right)^{2/3}
  \quad \hbox{if $B<0$}
  \qquad \hbox{or} \qquad
  \tau_1 \simeq \left(\frac{4A \V}{B} \right)^{2/3}
  \quad \hbox{if $B>0$} \,. \label{tau1soln2}
\ee

Any meaningful minimum must lie within the K\"{a}hler cone defined
by the conditions that no 2-cycle or 4-cycle shrink to zero and
that the overall volume be positive, and so we must check that
this is true of the above solution. Since we take $\tau_{1}$ and
$\tau_{2}$ both much larger than $\tau_{3}$, we may approximate
$\V$ by $\mathcal{V} \simeq \alpha\sqrt{\tau _{1}}\tau _{2}=
\lambda_{1}t_{1}t_{2}^{2}$ where $\lambda_{1}={1}/{4\alpha^{2}} >
0$, and this together with eq.~(\ref{taus}) shows that positive
$t_{1}$ and $t_{2}$ suffices to ensure $\tau_1$, $\tau_2$ and
$\mathcal{V}$ are all positive. Consequently, the boundaries of
the K\"{a}hler cone arise where one of the 2-cycle moduli,
$t_{1,2}$, degenerates to zero. Since in terms of $\mathcal{V}$
and $\OmegA \equiv \tau_1$ we have
\begin{equation}
 t_1 = \frac{\V}{\tau_1} \,,
 \quad
 t_2 = \left( \frac{\tau_1}{\lambda_1} \right)^{1/2}
 \quad\hbox{and}\quad
 \tau_2 = 2\V \left( \frac{\lambda_1}{\tau_1} \right)^{1/2}\,,
 \label{ts}
\end{equation}
the K\"{a}hler cone is given by $0 < \tau_1 < \infty$. At its
boundaries we have
\begin{equation*}
 \tau_{1}\rightarrow 0\Longleftrightarrow \tau _{2}
 \rightarrow \infty \Longleftrightarrow
 t_{1}\rightarrow \infty \Longleftrightarrow t_{2}\rightarrow 0,
\end{equation*}
while
\begin{equation*}
 \tau _{1}\rightarrow \infty\Longleftrightarrow
 \tau _{2}\rightarrow 0
 \Longleftrightarrow t_{1}\rightarrow 0\Longleftrightarrow
 t_{2}\rightarrow \infty.
\end{equation*}
Comparing the solutions of eqs.~\pref{tau1soln2} with the walls of
the K\"{a}hler cone shows that when $32AC \ll B^2$ we must require
either $C> 0$ (if $B<0$) or $A > 0$ (if $B>0$), a condition that
is always satisfied (see (\ref{GREAT})).

In Table 1 we chose for numerical purposes several representative
parameter choices, and these choices are extended to the
loop-generated potential in Table 2. (The entries for
$\langle\tau_3\rangle$ and $\langle\V\rangle$ in this table are
simply carried over from Table 1 for ease of reference.) The LV
case shows that loop corrections can indeed stabilise the
remaining modulus, $\OmegA \equiv \tau_1$, at hierarchically large
values, $\tau_2 \gg \tau_1 \gg \tau_3$ without requiring the
fine-tuning of parameters in the potential, while the SV examples
illustrate cases where $\tau_2 \gg \tau_1 \gsim \tau_3$ (although
$e^{-a_1 \tau_1} \ll e^{-a_3 \tau_3}$).

\begin{figure}[ht]
\begin{center}
\begin{tabular}{c||c|c|c}
  & LV & SV1 & SV2 \\
  \hline\hline
  $C^{KK}_1$ & 0.1 & 0.15 & 0.18 \\
  $C^{KK}_2$ & 0.1 & 0.08 & 0.1 \\
  $C^W_{12}$ & 5 & 1 & 1.5 \\
  \hline
  $A$ & $10^{-4}$ & $2\cdot 10^{-3}$ & $2.9\cdot 10^{-3}$ \\
  $B$ & 10 & 0.52 & 0.93 \\
  $C$ & $5\cdot 10^{-5}$ & $1.9 \cdot 10^{-5}$ & $4.3 \cdot 10^{-5}$ \\
  $\langle\tau_3\rangle$ & 10.46 & 4.28 & 3.73 \\
  $\langle\tau_1\rangle$ & $1.07 \cdot 10^6$ & 8.96 & 7.5 \\
  $\langle\V\rangle$ & $2.75 \cdot 10^{13}$ & 1709.55 & 1626.12
\end{tabular}\\
\vspace{0.3cm}{{\bf Table {2}:} Loop-potential parameters.}
\end{center}
\end{figure}

\subsubsection{Canonical normalisation}

To discuss dynamics and masses requires the kinetic terms in
addition to the potential, which we now display in terms of the
variables $\V$ and $\OmegA \equiv \tau_1$. Neglecting the small
blow-up cycle, $\tau_{3}$, the non canonical kinetic terms for the
large moduli $\tau_1$ and $\tau_2$ are given at leading order by
\begin{eqnarray}
 -\mathcal{L}_{kin} &=& K^0_{i\bar\jmath} \,
 \Bigl( \partial_{\mu } T_{i} \, \partial^{\mu }
 \overline{T}_{j} \Bigr)
 = \frac14 \, \frac{\partial^2 K_0}{\partial \tau_i \partial \tau_j}
 \; \Bigl( \partial_\mu \tau_i \, \partial^\mu \tau_j +
 \partial_\mu b_i \, \partial^\mu b_j \Bigr) \\
 &=& \frac{\partial_{\mu } \tau_{1} \partial^{\mu }
 \tau_{1}}{4\tau_{1}^{2}}
 + \frac{\partial_{\mu } \tau_{2} \partial^{\mu }
 \tau_{2} }{2 \tau_2^2} \;
 + \cdots \,, \nonumber\label{Lkin}
\end{eqnarray}
where the ellipses denote both higher-order terms in $\sqrt{\tau_3
/ \tau_{1,2}}$, as well as axion kinetic terms. Trading $\tau_2$
for $\V$ with eq.~(\ref{ts}), the previous expression becomes
\begin{equation}
 -\mathcal{L}_{kin} =
 \frac{3}{8\tau_{1}^{2}} \; \partial_{\mu } \tau_{1}
 \partial^{\mu } \tau_{1}
 - \frac{1}{2\tau_1 \V} \; \partial_{\mu } \tau_{1}
 \partial^{\mu } \V + \frac{1}{2 \V^2} \;
 \partial_{\mu } \V \partial^{\mu }
 \V + \cdots \,. \label{Lkin2}
\end{equation}
Notice that the kinetic terms in this sector can be made field
independent by redefining $\vartheta_1 = \ln \tau_1$ and
$\vartheta_v = \ln \V$, showing that this part of the target space
is flat (within the approximations used). The canonically
normalized fields satisfy $-{\cal L}_{kin} = \frac12[(\partial
\varphi_1)^2 + (\partial \varphi_2)^2 ]$, and so may be read off
from the above to be given by
\be
 \left( \begin{array}{c}
  \partial_\mu \tau_1/\tau_1 \\ \partial_\mu \V/\V \\
 \end{array} \right) =
 M \cdot
 \left( \begin{array}{c}
  \partial_\mu \varphi_1 \\
  \partial_\mu \varphi_2 \\
 \end{array} \right) \,,
\ee
where the condition
\be
    M^{\scriptscriptstyle T} \cdot
 \left( \begin{array}{rrr}
  \frac34 && -\frac12 \\
  -\frac12 && 1 \\
 \end{array} \right) \cdot M
 = I\,,
\ee
implies $M^2 = \left( \begin{array}{cc} 2 & 1 \\ 1 & \frac32
\end{array} \right)$, and so if $M = \left( \begin{array}{cc} a & b \\ b &
c \end{array} \right)$ then $a_\pm = \sqrt{2-b_\pm^2}$, $c_\pm =
\sqrt{\frac32 - b_\pm^2}$ and $b_\pm^2 = 2/\left(7\pm 4\sqrt
2\right)$ (so explicitly $a_+ \simeq 1.357$, $b_+ \simeq 0.398$,
$c_+ \simeq 1.158$ and $a_- \simeq 0.715$, $b_- \simeq 1.220$,
$c_- \simeq 0.105$).

Finally, we may use these results to estimate the mass of the
propagation eigenstates, $\varphi_{1,2}$, obtained at the
potential's minimum. Before diagonalizing the kinetic terms, but
writing $\vartheta_v = \ln \V$ and $\vartheta_1 = \ln \tau_1$, we
find that the derivatives of the potential at its minimum scale as
$\partial^2 V/\partial \vartheta_v^2 \sim \hat\xi/\V^3$ --- since
it is dominated by contributions from $\delta V_{(\alpha')}$ and
$\delta V_{(sp)}$ --- while $\partial^2 V/\partial \vartheta_1^2
\sim \partial^2 V/\partial \vartheta_v \partial \vartheta_1 \sim
1/\V^{10/3}$ --- since these are dominated by $\delta V_{(g_s)}$.
These properties remain true for the physical mass eigenvalues
after diagonalising the kinetic terms, since this mixing changes
the form of the eigenvectors but not the leading scaling of the
eigenvalues at large $\V$. This confirms the qualitative
expectation that the $\OmegA \equiv \tau_1$ direction is
systematically lighter than $\V$ in the large-$\V$ limit.

\subsection{Inflationary potential} \label{7}

Having established the existence of a consistent LVS minimum of
the potential for all fields, we now explore the inflationary
possibilities that can arise when some of these fields are
displaced from these minima. Since the potential for
$\OmegA\equiv\tau_1$ is systematically flat in the absence of
string loop corrections, it is primarily this field that we
displace in the hopes of finding it to be a good candidate for a
slow-roll inflaton.

In the approximation that string-loop effects are completely
turned off, we have seen that the leading large-$\V$ potential
stabilising both $\V$ and $\tau_3$ is completely flat in the
$\OmegA\equiv\tau_1$ direction. We therefore perform our initial
inflationary analysis within an approximation where both
$\mathcal{V}$ and $\tau_{3}$ remain fixed at their respective
$\tau_1$-independent minima while $\tau_1$ rolls towards its
minimum from initially larger values. In this approximation the
important evolution involves only the single field $\tau_1$,
making it very simple to calculate. This single-field approach
should be an excellent approximation for large enough $\V$, and we
return below to the issue of whether or not $\V$ can be chosen
large enough to call this approximation into question.

\subsubsection{The single-field inflaton}

As before, we choose $\OmegA\equiv\tau_1$ as the coordinate along
the inflationary direction. When $\tau_3 = \langle \tau_3 \rangle$
and $\V = \langle \V \rangle$ are fixed at their
$\tau_1$-independent minima, so that $\partial_{\mu}
\tau_3=\partial_{\mu}\V=0$, \pref{Lkin2} shows that the relevant
dynamics reduces to
\be
 {\cal L}_{inf} = - \frac38 \left( \frac{\partial_\mu \tau_1
 \partial^\mu \tau_1 }{\tau_1^2} \right)\,
 - V_{inf}(\tau_1) \,, \label{LKIN}
\ee
with scalar potential given by
\begin{equation}
 V_{inf} = V_0 + \left(\frac{A}{\tau_{1}^{2}}
 -\frac{B}{\mathcal{V}\sqrt{\tau_{1}}} +\frac{C\tau_{1}}
 {\mathcal{V}^{2}}\right)\frac{W_{0}^{2}}{\mathcal{V}^{2}} \,.
 \label{inflpot}
\end{equation}
Notice that \pref{LKIN} does not depend on the intersection
numbers, $\lambda_1$ and $\lambda_3$, implying that tuning these
cannot help with the search of a canonical normalisation more
suitable for an inflationary roll. The $\tau_1$-independent
constant, $V_0$, of eq.~\pref{inflpot} consists of
\begin{equation}
 V_0 =\frac{8 \,a_{3}^{2}A_3^2\sqrt{\langle \tau_{3}
 \rangle}}{3\alpha\gamma \langle \mathcal{V} \rangle}
 \; e^{-2a_{3} \langle \tau_{3} \rangle} -
 \frac{4 W_{0}a_{3}A_3 \langle \tau_{3} \rangle}{
 \langle \mathcal{V} \rangle^{2}}
 \; e^{-a_{3} \langle \tau_{3} \rangle}
 +\frac{3 \,\hat\xi W_0^2}{4 \langle \mathcal{V}
 \rangle^{3}}
 + \delta V_{up},
 \label{V0pot}
\end{equation}
where $\delta V_{up}$ is an up-lifting potential, such as might be
produced by the tension of an $\overline{D3}$ brane in a warped
region somewhere in the extra dimensions: $\delta V_{up} \sim
\delta_{up}/\V^{4/3}$. For the present purposes, what is important
about this term is that it does not depend at all on $\tau_1$ once
$\V$ is fixed. We imagine $\delta_{up}$ to be tuned to ensure the
complete vanishing of $V$ (or a tiny positive value) at the
minimum, with $\delta_{up} \sim 1/\langle \V \rangle^{5/3}$
required to cancel the non-perturbative and $\alpha'$-correction
parts of the potential (which together scale like $\langle \V
\rangle^{-3}$ at their minimum). In addition a second adjustment
($\delta_{up}\to\delta_{up}+\mu_{up}$) of order $\mu_{up}/\langle
\V \rangle^{4/3}  = - \delta V_{(g_{s})}(\langle
\mathcal{V}\rangle, \langle \tau_1\rangle)$ is required to cancel
the loop-generated part of $V$, for which $V_0 \sim
\mathcal{O}\left(1/\langle \mathcal{V}\rangle^{10/3}\right)$.

The canonical inflaton is therefore given by
\begin{equation}
 \varphi =\frac{\sqrt{3}}{2} \, \ln \tau_1 \,,
 \qquad \hbox{and so} \qquad
 \tau_1 = e^{ \kappa \varphi }
 \quad\hbox{with} \quad
 \kappa = \frac{2}{\sqrt3} \,. \label{cambio}
\end{equation}
In terms of this field the walls of the K\"{a}hler cone are
located at
\begin{equation}
 0 < \tau_1 < \infty \Longleftrightarrow
 -\infty < \varphi < +\infty,
\end{equation}
implying that any inflationary dynamics can in principle take
place over an \textit{infinite} range in field space. The
potential \pref{inflpot} becomes
\begin{eqnarray}
 V_{inf} &=& V_0 + \frac{W_{0}^{2}}{\mathcal{V}^{2}}
 \left(A \, e^{-2\kappa \varphi}
 -\frac{B}{\mathcal{V}} \, e^{-\kappa\varphi/2}
 +\frac{C} {\mathcal{V}^{2}} \, e^{\kappa \varphi}\right)
 \nonumber\\
 &=& \frac{1}{\left\langle \mathcal{V}
 \right\rangle^{10/3}}\left(\mathcal{C}_0
 \, e^{\kappa \hat{\varphi}}
 - \mathcal{C}_1 \, e^{-\kappa \hat{\varphi}/2}
 + \mathcal{C}_2 \, e^{-2\kappa \hat{\varphi}}
 + \mathcal{C}_{up} \right) \,,
 \label{VVV}
\end{eqnarray}
where we shift $\varphi = \langle \varphi \rangle + \hat\varphi$
by its vacuum value, \pref{tau1soln2}, and adjust $V_0 =
\mathcal{C}_{up}/\langle \V \rangle^{10/3}$ to ensure
$V_{inf}(\langle \varphi \rangle) = 0$. Choosing, for
concreteness' sake, $32AC \ll B^2$\footnote{Notice that this is a
natural choice since for $B>0$, $CA/B^2\sim g_s^4.$ } we have
$\langle \varphi \rangle = \frac{1}{\sqrt3} \ln \left( \zeta \V
\right)$, with $\zeta \simeq -B/2C$ if $B<0$ or $\zeta \simeq
4A/B$ if $B>0$. With these choices the coefficients
$\mathcal{C}_i$ do not depend on $\langle \V \rangle$, being given
by
\begin{equation}
 \mathcal{C}_0 = C W_0^2 \zeta^{2/3}, \quad
 \mathcal{C}_1 = B W_0^2 \zeta^{-1/3},\quad
 \mathcal{C}_2 = A W_0^2 \zeta^{-4/3}
 \quad\hbox{and} \quad
 \mathcal{C}_{up} = \mathcal{C}_1 - \mathcal{C}_0
 - \mathcal{C}_2. \label{Ci2}
\end{equation}
Notice that because $A$ and $C$ are both positive, we know that
$\mathcal{C}_0$ and $\mathcal{C}_2$ must also be. By contrast, not
knowing the sign of $C_{12}^{W}$ precludes having similar control
over the sign of $\mathcal{C}_1$. Table 3 gives the values for
these coefficients as computed using the parameter sets of the
previous tables.

\begin{figure}[ht]
\begin{center}
\begin{tabular}{c||c|c|c}
  & LV & SV1 & SV2 \\
  \hline\hline
  $\mathcal{C}_0$ & $5.8 \cdot 10^{-8}$ & 0.012 & 0.023 \\
  $\mathcal{C}_1$ & 292.4 & 20629.4 & 39786.9 \\
  $\mathcal{C}_2$ & 73.1 & 5157.35 & 9946.73 \\
  $\mathcal{C}_{up}$ & 219.3 & 1200.8 & 29840.2 \\
  $R = \mathcal{C}_0/\mathcal{C}_2$
  & $8\cdot 10^{-10}$ & $2.3\cdot 10^{-6}$
  & $2.3\cdot 10^{-6}$ \\
\end{tabular}\\
\vspace{0.3cm}{{\bf Table {3}:} Coefficients of the inflationary
potential for the various parameter sets discussed in the text.}
\end{center}
\end{figure}

\begin{figure}[ht]
\begin{center}
\epsfig{file=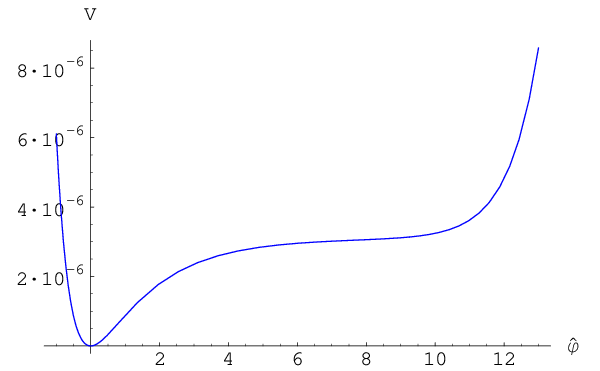, height=70mm,width=80mm}
\caption{$V$ (in
 arbitrary units) versus $\hat{\varphi}$, with $\mathcal{V}$ and
 $\tau_{3}$ fixed at their minima. The plot assumes the parameters
 used in the text (for which $\hat\varphi_{ip} \simeq 0.80$, $\hat\varphi_{end} = 1.0$,
 and $R\equiv\mathcal{C}_0/\mathcal{C}_2\sim 10^{-6}$).} \label{Fig:grafico3}
\end{center}
\end{figure}

Of particular interest is the case where both $A$ and $C$ are
small compared with $|B|$, as might be expected by their explicit
suppression by the factor $g_s^2$. For concreteness we focus in
what follows on the case $B>0$ (and so $\mathcal{C}_1>0$), for
which $\zeta \simeq 4A/B\ll 1$. This leads to two very useful
simplifications. First, it implies that $\mathcal{C}_0 /
\mathcal{C}_1 = \zeta C/B = 4AC/B^2$ and $R := \mathcal{C}_0 /
\mathcal{C}_2 = \zeta^2 C/A = 16 AC/B^2$ and so $\mathcal{C}_0$ is
systematically smaller than either $\mathcal{C}_1$ or
$\mathcal{C}_2$. This observation allows us to neglect completely
the $\mathcal{C}_0 \, e^{\kappa\hat\varphi}$ term of the potential
in the vicinity of the minimum and in most of the inflationary
region, as we shall see in what follows. Second, this limit
implies $\mathcal{C}_1/\mathcal{C}_2 = \zeta B/A = 4$, showing
that $\mathcal{C}_1$ and $\mathcal{C}_2$ are both positive, with a
fixed, order-unity ratio. This observation precludes using the
ratio of these parameters in the next section as a variable for
tuning the inflationary potential. These choices are visible in
Table 3, for which $A,C \ll B$, and so $\mathcal{C}_0$ is small
and $\mathcal{C}_1/\mathcal{C}_2 \simeq 4$. Figure
\ref{Fig:grafico3} plots the resulting scalar potential against
$\varphi$.

\subsection{Inflationary slow roll}
\label{SlowRoll}

We next ask whether the scalar potential (\ref{VVV}) can support a
slow roll, working in the most natural limit identified above,
with $A,C \ll B$ and $B > 0$. As we have seen, this case also
implies $0 < \mathcal{C}_0 \ll \mathcal{C}_1 = 4 \mathcal{C}_2$,
leaving a potential well approximated by
\begin{equation}
 V \simeq \frac{\mathcal{C}_{2}}{\left\langle
 \mathcal{V}\right\rangle ^{10/3}} \left[ (3 - R) - 4
 \left( 1 + \frac16 \, R \right) \,e^{-\kappa
 \hat{\varphi}/2} + \left( 1 + \frac23 \, R \right)
 \, e^{-2\kappa \hat{\varphi}} + R\text{ }
 e^{\kappa \hat{\varphi}}\right] \label{SCALA}
\end{equation}
which uses $\mathcal{C}_{up} \simeq \mathcal{C}_1 - \mathcal{C}_0
- \mathcal{C}_2$ and $\mathcal{C}_1/\mathcal{C}_2 \simeq 4$, and
works to linear order in
\begin{equation}
 R := \frac{\mathcal{C}_{0} }{ \mathcal{C}_{2}}
 = 2 g_s^4 \left(\frac{C_1^{KK} C_{2}^{KK} }{C_{12}^W}
 \right)^2 \ll 1 \,.
\end{equation}
The normalization of the potential may instead be traded for the
mass of the inflaton field at its minimum: $m_\varphi^2 = V''(0) =
4\, \left( 1 + \frac76 \, R \right) {\mathcal{C}_2} / {\left
\langle \mathcal{V} \right \rangle^{10/3}}$.

In practice the powers of $R$ can be neglected in all but the last
term in the potential, where it multiplies a positive exponential
which must eventually become important for sufficiently large
$\hat{\varphi}$. For smaller $\hat\varphi$, $R$ is completely
negligible and the potential is fully determined by its overall
normalisation. Furthermore, the range of $\hat\varphi$ for which
this is true becomes larger and larger the smaller $R$ is, and so
we start by neglecting $R$.

\begin{figure}[ht]
\begin{center}
\epsfig{file=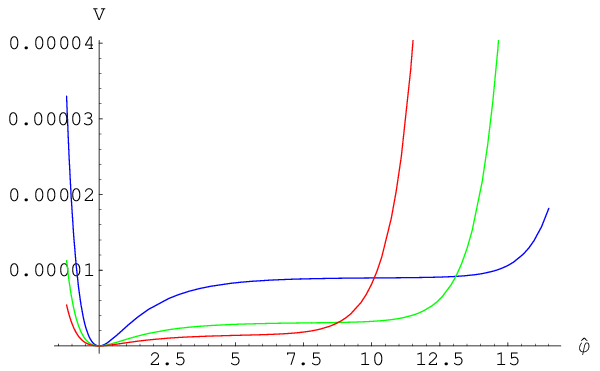, height=50mm,width=47mm}
\epsfig{file=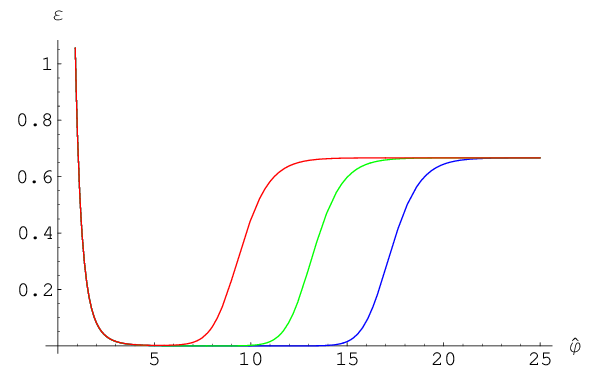, height=50mm,width=47mm}
\epsfig{file=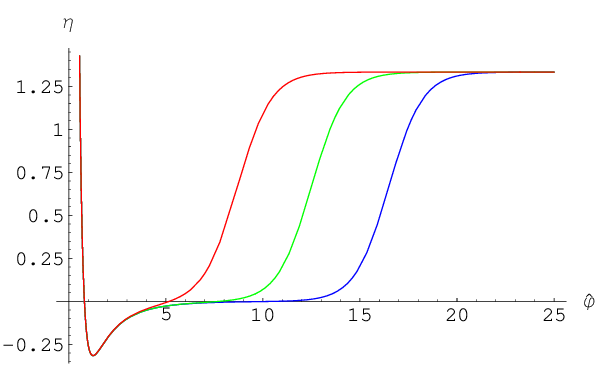, height=50mm,width=47mm} \caption{
 Plots of the potential and the slow-roll parameters $\varepsilon$
  and $\eta$ vs $\hat\varphi$ for $R=10^{-8}$ (blue curve), $R=10^{-6}$ (green
  curve), and $R=10^{-4}$ (red curve).}
   \label{NewFig}
\end{center}
\end{figure}

We seek inflationary rolling focusing on the situation in which
$\hat{\varphi}$ rolls down to its minimum (at $\hat\varphi = 0$)
from positive values. Defining, as usual, the slow-roll
parameters, $\varepsilon$ and $\eta$, by (recalling our use of
Planck units, $M_p = 1$)
\begin{eqnarray}
 \varepsilon  =\frac{1}{2V^{2}}
 \left( \frac{\partial V}{\partial
 \hat{\varphi} }\right)^{2}, \qquad
 \eta  = \frac{1}{V} \left( \frac{\partial^{2}V}{\partial
 \hat{\varphi}^{2}} \right),
\end{eqnarray}
we find (using $\kappa^2 = \frac43$ and keeping $R$ only when it
comes multiplied by $e^{\kappa \hat\varphi}$)
\begin{eqnarray}
 \varepsilon  &\simeq& \frac{8}{3} \left(
 \frac{ e^{-\kappa\hat{\varphi}/2}
 - e^{-2 \kappa\hat{\varphi}}
 + \frac12 \, R \, e^{\kappa \hat\varphi}}
 {3 - 4 \, e^{-\kappa\hat{\varphi}/2} +
 e^{-2\kappa \hat{\varphi}} + R \, e^{\kappa\hat\varphi}}
 \right)^{2}, \label{eps} \\
 \eta  &\simeq& -\frac{4}{3} \left( \frac{
 e^{-\kappa\hat{\varphi}/2}
 - 4 \, e^{-2\kappa\hat{\varphi}} - R \, e^{\kappa \hat\varphi}}
 {3 - 4 \, e^{-\kappa\hat{\varphi}/2}
 + e^{-2\kappa\hat{\varphi}}
 + R \, e^{\kappa \hat\varphi}} \right)\,.
 \label{eta}
\end{eqnarray}
Plots of these expressions are given in Figure \ref{NewFig}, which
show three qualitatively different regimes.

\medskip\noindent{\em Slow-Roll Regime}

\medskip\noindent
Both slow roll parameters are naturally exponentially small in the
regime $R^{1/3} \ll e^{-\hat\varphi/2} \ll 1$. In this regime it
is the term $e^{-\kappa\hat\varphi/2}$ that dominates in
(\ref{SCALA}), and so the dynamics is effectively governed by the
approximate potential
\begin{equation}
 V \simeq \frac{\mathcal{C}_2}{\left\langle \mathcal{V}
 \right\rangle^{10/3}}\left(3
 - 4 \, e^{-\kappa\hat{\varphi}/2}
 \right) \,. \label{SCALARE}
\end{equation}
This resembles a standard potential for large-field inflation,
which drives the field to evolve towards smaller
values\footnote{It would be interesting to see how our
inflationary mechanism fits in the general analysis of
supergravity conditions for inflation performed in \cite{marta}.}.
The slow-roll parameters (\ref{eps}) and (\ref{eta}) in this
regime simplify to
\begin{eqnarray}
 \varepsilon  &\simeq&
 \frac{8}{3 \left[3
 \, e^{\kappa\hat{\varphi}/2} -4\right]^{2}},
 \label{Eps} \\
 \eta &\simeq& -\frac{4}{3\left[
 3 \, e^{\kappa\hat{\varphi}/2} -4\right]} \,,
 \label{Eta}
\end{eqnarray}
and for all $\hat\varphi$ in this regime we have the interesting
relation
\begin{equation} \label{epsvsetarelation}
 \varepsilon \simeq \frac{3 \,\eta^2}{2} \,.
\end{equation}

\medskip\noindent{\em Small-$\hat\varphi$ Regime}

\medskip\noindent
The slow-roll conditions break down once $\hat\varphi$ is small
enough that the two negative exponentials are comparative in size.
to produce a zero in $\eta$. An inflection point occurs in this
regime, located where
\begin{equation}
 \left( \frac{\partial^{2} V}{\partial \hat{\varphi}^{2}}
 \right)_{\hat\varphi_{ip}} \simeq \frac{4\mathcal{C}_2}{3 \langle
 \mathcal{V}\rangle^{10/3}}
 \left( -  e^{-\kappa\hat{\varphi}_{ip}/2}
 + 4 \, e^{-2\kappa \hat{\varphi}_{ip}}
 \right) = 0,
\end{equation}
and so
\be
 \hat{\varphi}_{ip} = \frac{1}{\sqrt{3}} \ln \left(
 \frac{16 \, \mathcal{C}_2}{\mathcal{C}_1} \right)
 \simeq \frac{\ln 4 }{\sqrt{3}} \simeq 0.8004..\,.
\ee
As Figure \ref{NewFig} shows, to the left of this point
$\varepsilon$ grows quickly, while at the inflection point
$\hat{\varphi} = \hat{\varphi}_{ip}$, we have $\varepsilon_{ip} =
1.464$ and $\eta_{ip} = 0$. Just to the right of this, at $\hat
\varphi_{end} = 1$ we have $\varepsilon_{end} = 0.781$ and
$\eta_{end} = -0.256$, making this as good a point as any to end
inflation. (In what follows we verify numerically that our results
are not sensitive to precisely where we end inflation in this
regime.)

\medskip\noindent{\em Large-$\hat\varphi$ Regime}

\medskip\noindent
Once $R \, e^{\kappa \hat\varphi} \gg 3$ the positive exponential
dominates the potential, eq.~(\ref{SCALA}), which becomes
well-approximated by
\begin{equation}
 V \simeq \frac{m_{\varphi }^{2}}{4} \, R
 \, e^{\kappa\hat{\varphi}} \,,
\label{SCALAre}
\end{equation}
and so the slow-roll parameters plateau at constant values: $\eta
\simeq 2\varepsilon \simeq \kappa^2 = \frac43$ (as is seen in
Figure \ref{NewFig}). This shows that the slow-roll conditions
also break down for $\kappa\hat{\varphi} \simeq \ln(1/R)$,
providing an upper limit to the distance over which the slow roll
occurs (and so also on the number of \efold ings, $N_e$).

An interesting feature of transition to this large-$\hat\varphi$
regime is the necessity for $\eta$ to change sign. This is
interesting because, as figure \ref{NewFig} shows, $\varepsilon$
is still small where it does, and so there is a slow-roll region
for which $\eta \gg \varepsilon > 0$. This regime is unusual
because it allows $n_s > 1$ (see Figure \ref{Fig:nsComparison}),
unlike generic single-field inflationary models. In practice, in
what follows we choose horizon exit to occur for $\hat\varphi$
smaller than this, due to the current observational preference for
$n_s < 1$. A precise upper limit on $\hat\varphi$ this implies can
be defined as the inflection point where $\eta$ vanishes due to
the competition between the $e^{\kappa\hat{\varphi}}$ and
$e^{-\kappa\hat{\varphi}/2}$ terms of the potential. This occurs
when $e^{-\kappa\hat\varphi/2} \simeq R \, e^{\kappa
\hat\varphi}$, or $\hat{\varphi}(R) \simeq \hat{\varphi}_0(R) :=
-\ln(R)/\sqrt{3}$.

\begin{figure}[ht]
\begin{center}
\epsfig{file=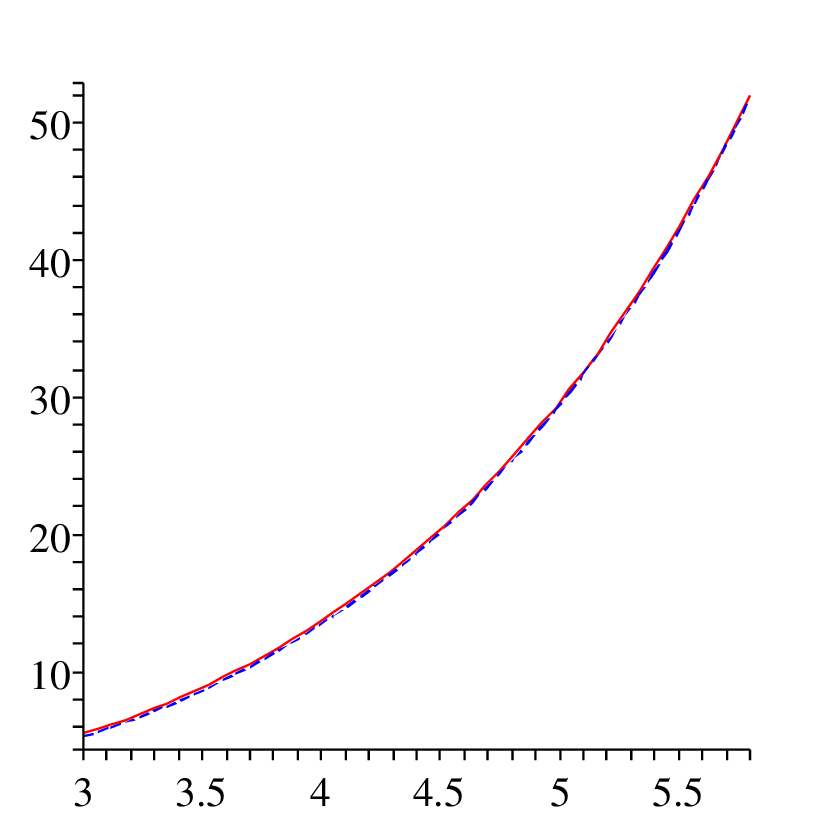, height=60mm,width=67mm}
\epsfig{file=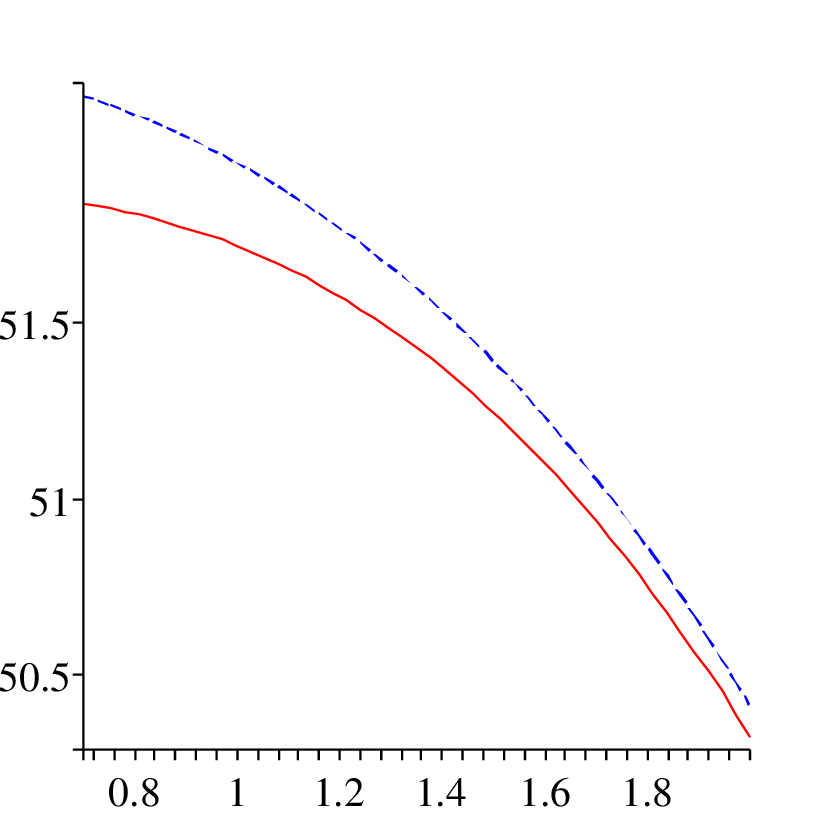, height=60mm,width=67mm} \caption{
 Plot of the number of $e$-foldings, $N_e$, vs $\hat\varphi_*$ (left)
 and $\hat\varphi_{end}$ (right) for $R=0$. The inflection point occurs at
  $\hat\varphi_{ip} \simeq 0.8$ and $\hat\varphi_{end} = 1$ in the left-hand
  plot. $\hat\varphi_* = 5.7$ in the right-hand plot. The solid (red) curves
  are computed using the full
  potential (3.33) while the dashed (blue) curves are computed
  using the approximate potential (3.38).}
   \label{Fig:efolds}
\end{center}
\end{figure}

We may now ask whether the slow-roll regime is large enough to
allow 60 \efold ings of inflation. The number of \efold ings $N_e$
occurring during the slow-roll regime can be computed using the
approximate potential, eq.~\pref{SCALARE}, which gives
\begin{equation}
 N_{e}=\int_{\hat{\varphi}_{end}}^{\hat{\varphi}_{\ast}}
 \frac{V}{V'} \; \exd \hat{\varphi}
 \simeq \frac{\sqrt{3}}{4} \int_{\hat{\varphi}_{end}}^{\hat{
 \varphi}_{\ast }} \left[3 \, e^{\kappa\hat{\varphi}/2}
 -4 \right] \, \exd\hat{\varphi}
 = \left[ \frac94 \, e^{\kappa\hat{\varphi}/2}
 - \sqrt3 \, \hat\varphi
 \right]_{\hat{\varphi}_{end}}^{\hat{\varphi}_{\ast }} \,,
 \label{Nefunc}
\end{equation}
where $e^{\kappa\hat{\varphi}_{end}} \simeq 16\, \mathcal{C}_2 /
\mathcal{C}_1 \simeq 4  \ll e^{\kappa\hat\varphi_*}$ represents
the onset of the small-$\hat\varphi$ regime, as described above,
and $\hat\varphi = \hat\varphi_*$ denotes the value of
$\hat\varphi$ at horizon exit. Figure \ref{Fig:efolds} shows how
the number of \efold ings depends on the assumed field value
during horizon exit, as well as the insensitivity of this result
to the assumed point where inflation ends. This shows that
interesting inflationary applications require $\hat\varphi$ to
roll through an interval of at least O(5).

\begin{figure}[ht]
\begin{center}
\epsfig{file=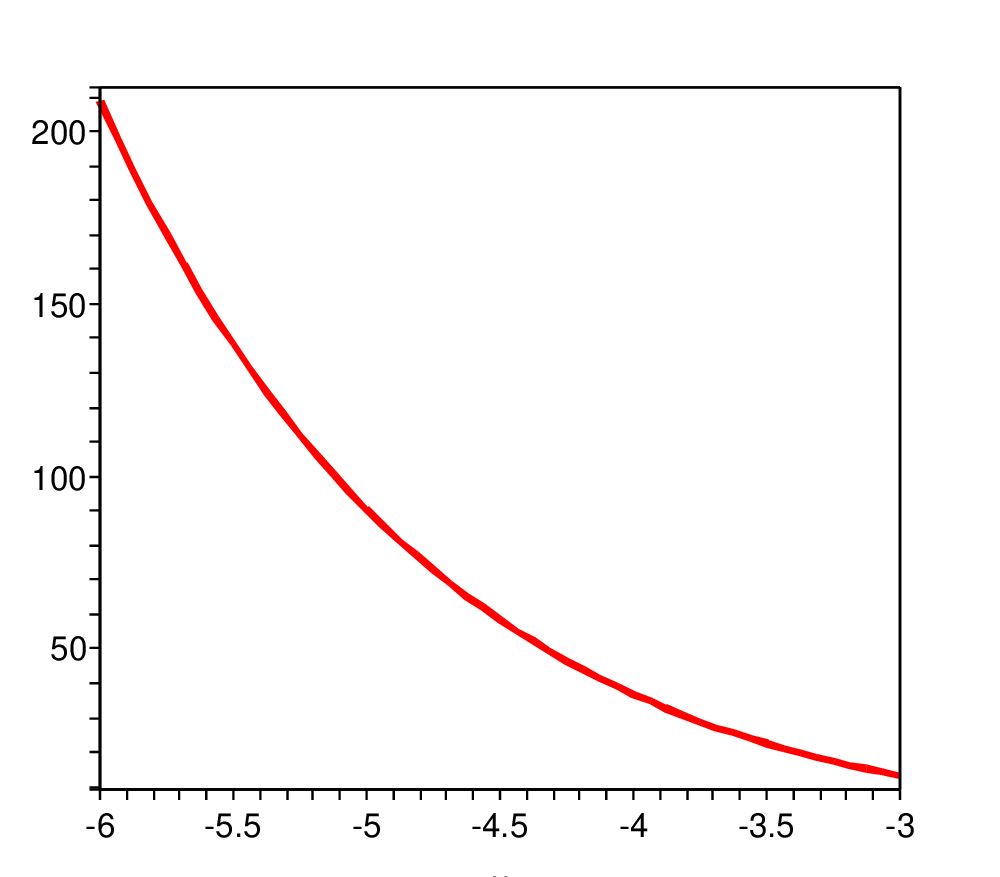, height=60mm,width=67mm} \caption{
 Plot of the maximum number of $e$-foldings, $N_e^{max}$, vs
 $x=\log_{10}R$, defined by the condition
 $\hat\varphi_* = \hat\varphi_0(R)$ as described in the text.
 The integration takes $\hat\varphi_{end} = 1$, and the
 curves are computed
  using the approximate potential (3.38).}
   \label{Fig:Nemax}
\end{center}
\end{figure}

An estimate for the upper limit to $N_e$ that can be obtained as a
function of $R$ can be found by using $\hat\varphi_* =
\hat\varphi_0(R)$ in eq.~\pref{Nefunc}. This leads to
\begin{equation}
 N_e^{max} \simeq \frac94 \left( R^{-1/3} - 2 \right)
 - \left[ \ln \left( \frac{1}{R} \right) - \ln 8 \right]
  \,,
 \label{Nemax}
\end{equation}
This result is plotted in Figure \ref{Fig:Nemax}, and shows that
more than 60 \efold ings of inflation requires $R \lsim 3 \times
10^{-5}$.

The validity of the $\alpha'$ and $g_s$ expansions also set a
limit to how large $\hat{\varphi}_*$ can be taken, since the
exponential growth of $\delta V_{(g_s)}$ for large $\hat\varphi$
would eventually allow it to become larger than the lower-order
contributions, $\delta V_{(sp)}+\delta V_{(\alpha')}$.
Microscopically this arises because $\hat{\varphi}\to\infty$
corresponds to $\tau_1\to\infty$ and $\tau_2\to 0$, leading to the
failure of the expansion of $\delta V^{KK}_{(g_s),\tau_2}$ in
inverse powers of $\tau_2$. However, as is argued in Appendix
\ref{Appendix B}, it is the slow-roll condition $\eta \ll 1$ that
breaks down first as $\hat\varphi$ increases, and so provides the
most stringent upper edge to the inflationary regime. For the two
sample sets SV1 and SV2 given in the Tables, we obtain $R \simeq
2.3\cdot 10^{-6}$, and this gives $\hat{\varphi}_{max} \simeq
12.4$ (in particular allowing more than 60 \efold ings of
inflation).

\subsubsection{Observable footprints}

We now turn to the observable predictions of the model. These
divide naturally into two types: those predictions depending only
on the slow roll parameters, which are insensitive to the
underlying potential parameters; and those which also depend on
the normalization of the inflationary potential, and so depend on
more of the details of the underlying construction.

\subsubsection*{Model-independent predictions}

The most robust predictions are for those observables whose values
depend only on the slow roll parameters, such as the spectral
index and tensor-to-scalar ratio, which are given as functions of
the slow-roll parameters (evaluated at horizon exit) by
\be \label{rnsslowroll}
    n_s = 1 + 2\eta_* - 6\varepsilon_* \qquad \hbox{and} \qquad
    r =16 \, \varepsilon_* \,.
\ee
In general, as can be seen from (\ref{eps}) and (\ref{eta}), the
two quantities $\varepsilon_*$ and $\eta_*$ are functions of two
parameters, $\hat\varphi_*$ and $R$; hence $n_s = n_s( \hat
\varphi_*, R)$ and $r=r(\hat\varphi_*,R)$. However we have also
seen that having a significant number of \efold ings requires $R
\ll 1$, and so to a good approximation $n_s = n_s(\hat\varphi_*)$
and $r = r(\hat\varphi_*)$, unless $\hat\varphi_*$ is large enough
that $R e^{\kappa\hat\varphi_*}$ cannot be neglected.

\begin{figure}[ht]
\begin{center}
\epsfig{file=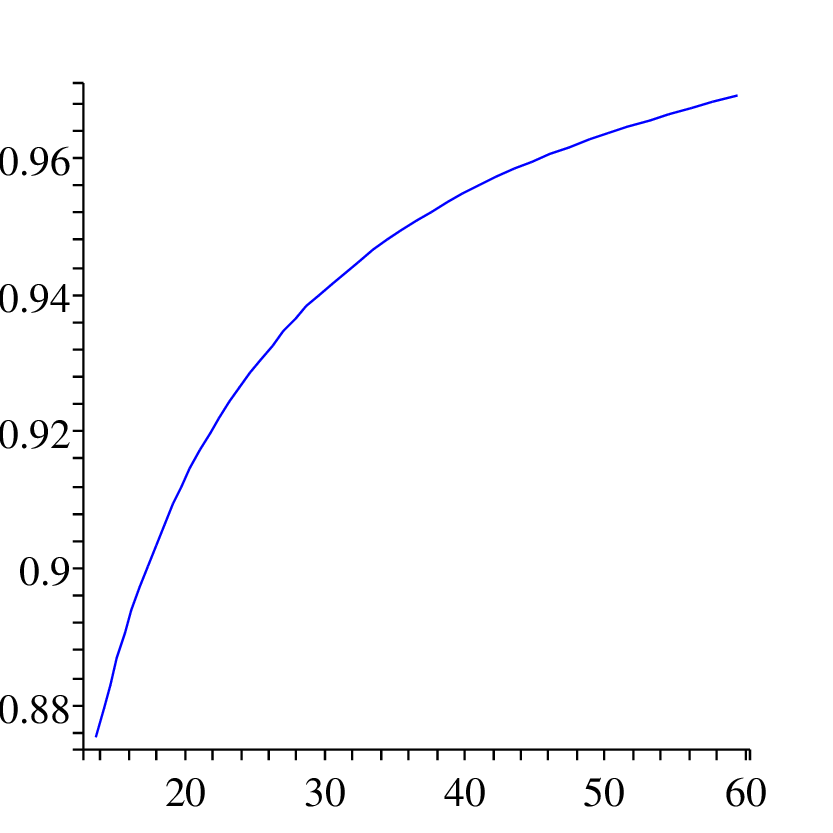, height=40mm,width=47mm}
\epsfig{file=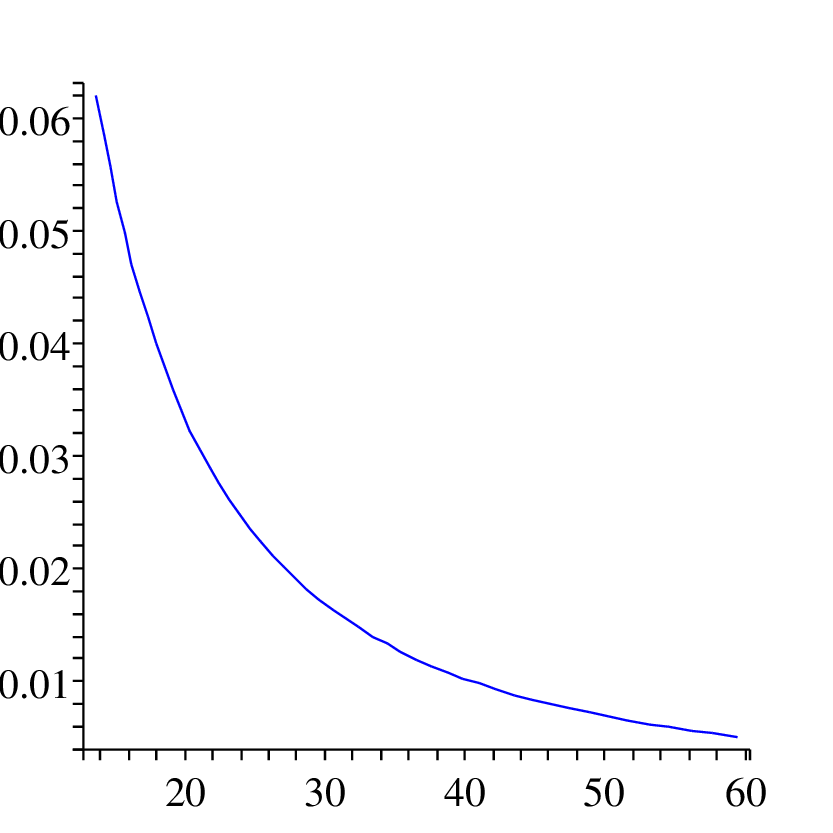, height=40mm,width=47mm}
\epsfig{file=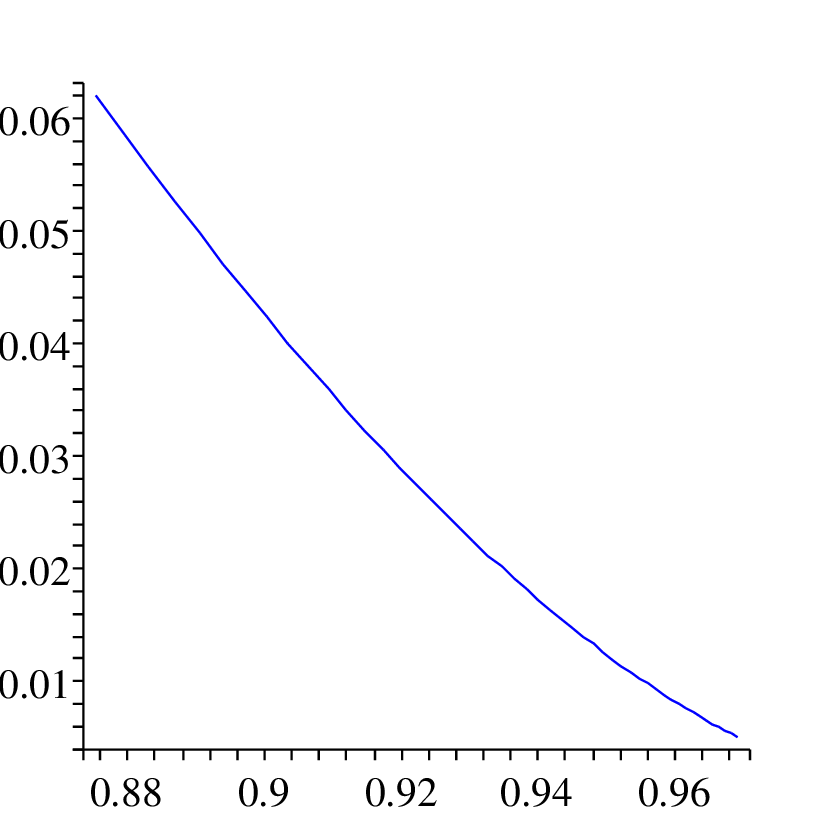, height=40mm,width=47mm}
 \caption{
 A plot of $n_s$ (left panel) and $r$ (center panel) vs the number
 of $e$-foldings, $N_e$. The right panel plots the correlation $r$
 vs $n_s$ that results when $N_e$ is eliminated, resembling simple
 single-field large-field models.}
   \label{Fig:randns}
\end{center}
\end{figure}

For small $R$ we find the robust correlation predicted amongst
$r$, $n_s$ and $N_e$, as described in the introduction. The
implied relation between $r$ and $n_s$ is most easily found by
using the relation $\varepsilon_* = \frac32 \, \eta_*^2$,
eq.~\pref{epsvsetarelation} in eq.~\pref{rnsslowroll} and dropping
$\varepsilon_*$ relative to $\eta_*$ in $n_s -1$:
\begin{equation}
 r \simeq 6 (n_s - 1)^2 \,,
\end{equation}
showing that a smaller ratio of tensor-to-scalar perturbations,
$r$, correlates with larger $n_s$. Figure \ref{Fig:randns} plots
the predictions for $r$ and $n_s$ that are obtained in this way.

\begin{figure}[ht]
\begin{center}
\epsfig{file=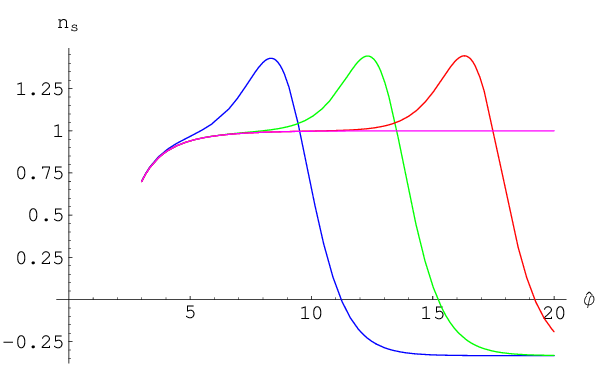, height=60mm,width=70mm} \caption{
 Plots of the spectral index $n_s$
 vs $\hat\varphi$ for $R=0$ (purple curve),
 $R=10^{-8}$ (red curve), $R=10^{-6}$ (green
 curve), and $R=10^{-4}$ (blue curve).}
   \label{Fig:nsComparison}
\end{center}
\end{figure}

Deviations from this correlation arise for large enough
$\varphi_*$, for which $N_e$ approaches the maximum number of
\efold ings possible, and this is illustrated in Figure
\ref{Fig:nsComparison}, which plots $n_s$ vs $\hat\varphi_*$ for
several choices of $R$. (Notice in particular the excursion to
values $n_s > 1$ shown in the figure for $\hat\varphi_* \simeq
\hat\varphi_0(R)$ when $R \ne 0$, as discussed above.) In the
extreme case where $\hat\varphi_* = \hat\varphi_0(R)$ we have
$\eta_* \simeq 0$ and $\varepsilon_* \simeq \frac23 \, R^{2/3}$,
leading to
\begin{equation}
 r \simeq \frac{32}{3} \, R^{2/3}
 \qquad \hbox{and} \qquad
 n_s \simeq 1-4 \, R^{2/3} \,.
\end{equation}
Recall that $N_e^{max} \gsim 60$ implies $R \lsim 3 \times
10^{-5}$, and in the extreme case $R \simeq 3 \times 10^{-5}$ the
above formulae lead to $r \simeq 0.01$ and $n_s \simeq 0.996$.
Should $r \simeq 0.01$ be observed and ascribed to this scenario,
the close proximity of horizon exit to the beginning of inflation
would likely imply other observable implications for the CMB,
along the lines of those discussed in refs.~\cite{tPsignals}.

\subsubsection*{Model-dependent predictions}

We next turn to those predictions which depend on the
normalization, $V_0$, of the inflaton potential, and so depend
more sensitively on the parameters of the underlying supergravity.

\medskip \noindent {\it Number of $e$-foldings:}
The first model-dependent prediction is the number of \efold ings
itself, since this depends on the value $\hat\varphi_*$ taken by
the scalar field at horizon exit. Indeed we have already seen that
the constraint that there be enough distance between
$\hat\varphi_*$ and $\hat\varphi_{end}$ to allow many \efold ings
of inflation imposes upper limits on parameters such as $R$. The
strongest such limit turned out to be the requirement that $n_s$
be low enough to agree with observed values (see the discussion
surrounding eq.~\pref{Nemax}). For numerical comparison of our
benchmark parameter sets we formalise this by requiring
$\hat\varphi < \hat{\varphi}_{max}$, defined as the value for
which $n_s < 0.974$, since this is the 68\% C.L observational
upper bound (for small $r$). Table 4 then lists the maximal number
of \efold ings that are possible given the constraint
$\hat\varphi_* < \hat\varphi_{max}$ for the models given in Tables
1 and 2.

\begin{figure}[ht]
\begin{center}
\begin{tabular}{c||c|c|c}
  & LV & SV1 & SV2 \\
  \hline\hline
  $\langle \varphi \rangle$ & 12.02 & 1.9 & 1.7 \\
  $\hat\varphi_{max}$ & 6.3 & 6.14 & 6.16 \\
  $N_e^{max}$ & 72 & 64 & 64 \\
  $A_{COBE}$ & $2.1 \cdot 10^{-45}$
             & $1.2 \cdot 10^{-7}$ & $2.8 \cdot 10^{-7}$ \\
  $\mathcal{R}_{cv}$ & 1201.6 & 29.7 & 12.2 \\
\end{tabular}\\
\vspace{0.3cm}{{\bf Table {4}:} Model parameters for the
inflationary potential. $N_e^{max}$ denotes the number of \efold
ings computed when rolling from $\hat\varphi_{max}$ to
$\hat\varphi = 1$. $A_{COBE}$ is calculated at $N_e\simeq 60$ and we set $K_{cs}=3\ln g_s \simeq -3.6$.}
\end{center}
\end{figure}

But how many \efold ings of inflation are required is itself a
function of both the inflationary energy scale and the
post-inflationary thermal history. For instance, suppose the
inflaton energy density, $\rho_{inf}\sim M_{inf}^4=V_{end}$,
rethermalises during a re-reheating epoch during which the
equation of state is $p=w\rho$, at the end of which the
temperature is $T_{rh}$, and after this the radiation-dominated
epoch lasts right down to the present epoch. With these
assumptions, $M_{inf}$, $T_{rh}$, $w$ and $N_{e}$ are related
by\footnote{We thank Daniel Baumann for identifying an error in
this formula in an earlier version.}
\begin{equation}
 N_e\simeq 62+\ln\left(\frac{M_{inf}}{10^{16}
 GeV}\right)-\frac{\left(1-3w\right)}{3\left(1+w\right)}
 \ln\left(\frac{M_{inf}}{T_{rh}}\right). \label{cosmology}
\end{equation}
This formula is obtained by equating the product $aH$ at horizon
exit during inflation and horizon re-entry in the cosmologically
recent past, $a_{he}H_{he}=a_0 H_0$, and using the intervening
cosmic expansion to relate these two quantities to $N_e$, $T_{rh}$
and $M_{inf}$ \cite{liddle}. In particular it shows (if $w <
\frac13$) that lower reheat temperatures (for fixed $M_{inf}$)
require smaller $N_e$. For instance, if $M_{inf} \simeq 10^{16}$
GeV and $w=0$ then an extremely low reheat temperature, $T_{rh}
\simeq 1$ GeV, allows $N_e \simeq 50$.

\begin{figure}[ht]
\begin{center}
\begin{tabular}{|c|c|c|c|c|}
$T_{rh}$ (GeV) & $N_e$ & $n_s$ & $r$ \\
\hline \hline
$10^{10}$ & 57 & 0.9702 & 0.0057 \\
\hline
$5\cdot 10^{7}$ & 55 & 0.9690 & 0.0060 \\
\hline
$10^{5}$ & 53 & 0.9676 & 0.0064 \\
\hline
$5\cdot 10^{3}$ & 52 & 0.9669 & 0.0066 \\
\end{tabular} \\\smallskip
{\bf Table {5}:} Predictions for cosmological observables as a
function of $T_{rh}\leq 10^{10} GeV$ fixing $M_{inf}=5\cdot
10^{15} GeV$ (for $w=0$ and $R=2.3\cdot 10^{-6}$).
\end{center}
\end{figure}

Given that $M_{inf}$ is constrained by the requirement that
inflation generate the observed primordial scalar fluctuations
(see below), eq.~\pref{cosmology} is most usefully read as giving
the post-inflationary reheat temperature that is required to have
modes satisfying $k = (aH)_*$ be the right size to be re-entering
the horizon at present. That is, given a measurement of $n_s$ one
can invert the prediction $n_s(N_e)$ to learn $N_e$, and so also
$r$ and the two slow roll parameters, $\varepsilon_*$ and
$\eta_*$. Then computing $M_{inf}$ from the amplitude of
primordial fluctuations allows eq.~\pref{cosmology} to give
$T_{rh}$. In particular, eq.~\pref{cosmology} represents an
obstruction to using the cosmology (without assuming more
complicated reheating) if $N_e$ is too low, since the required
$T_{rh}$ would be so low as to be ruled out.

\begin{figure}[hb]
\begin{center}
\epsfig{file=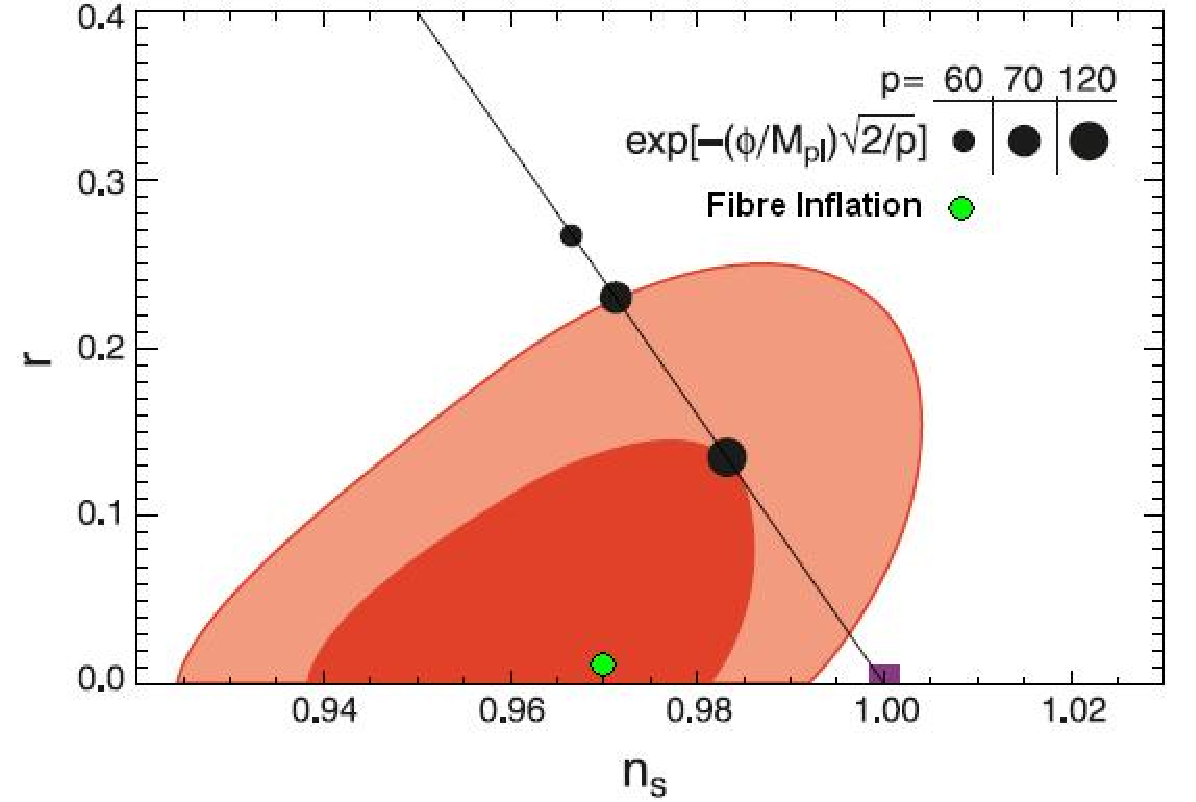, height=70mm,width=100mm}
\caption{The contours show the 68\% and 95\% CL derived from
WMAP+BAO+SN in the $(r-n_s)$ plane.} \label{Fig8}
\end{center}
\end{figure}

A few illustrative values are listed in Table 5, which assumes a
matter-dominated reheating epoch ($w=0$) and takes $M_{inf}=4
\times 10^{15}$ GeV, to compute $N_e\simeq 57$ and $T_{rh}$ as a
function of $n_s$ and $r$. These all show respectable reheat
temperatures, with $10^3 \, \hbox{GeV} < T_{rh} < 10^{10}$ GeV,
with the upper bound motivated by the requirement that gravitini
not be overproduced during reheating \cite{sarkar}. Furthermore,
as shown in Figure \ref{Fig8}, these values for $n_s$ and $r$ that
are predicted lie well within the observably allowed range.
Furthermore $r$ is large enough to allow detection by forthcoming
experiments such as EPIC, BPol or {\it CMBPol}
\cite{Verde,rBounds}.

\medskip\noindent{\it Amplitude of Scalar Perturbations:}
It is not impressive to have relatively large values for the
tensor-to-scalar ratio, $r$, unless the amplitude of primordial
scalar perturbations are themselves observably large. Since this
depends on the size of Hubble scale at horizon exit, it is
sensitive to the constant $V_0 = m_\varphi^2/4 = \mathcal{C}_2/
\V^{10/3}$ that pre-multiplies the inflationary potential. The
condition that we reproduce the COBE normalisation for primordial
scalar density fluctuations, $\delta_{H} = 1.92 \cdot 10^{-5}$,
can be expressed as:
\begin{equation}
 A_{COBE}\equiv\left(\frac{g_s\, e^{K_{cs}}}{8\pi}\right)
 \left(\frac{V^{3/2}}{V'}\right)^{2} \simeq 2.7\cdot
 10^{-7}, \label{cobe}
\end{equation}
where the prefactor $\left(g_s\,e^{K_{cs}}/8\pi\right)$ is the correct
overall normalisation of the scalar potential obtained from
dimensional reduction \cite{LVS}.

As Table 4 shows, it is possible to obtain models with many \efold
ings and which satisfy the COBE normalisation condition, but this
clearly prefers relatively large values for $g_s$ and $1/\V$, and
so tends to prefer models whose volumes are not inordinately
large. It is then possible to evaluate the inflationary scale as (setting $K_{cs}=3\ln g_s\simeq -3.6$):
\begin{equation}
 M_{inf}=V^{1/4}_{end}\simeq
 V_0^{1/4}M_P=\left(\frac{\mathcal{C}_{2}}{8\pi}\right)^{1/4}\frac{g_s}
 {\langle\mathcal{V}\rangle^{5/6}}M_P\sim 5\cdot 10^{15} GeV,
\end{equation}
as can be deduced from Table 6 which summarises the different
inflationary scales obtained for the models SV1 and SV2 with
smaller values for the overall volume. These results were used
above in Table 5 to determine the correlation between observables
and reheat temperature.

\begin{figure}[ht]
\begin{center}
\begin{tabular}{c||c|c}
  & SV1 & SV2 \\
  \hline\hline
  $\mathcal{C}_2$ & 5157.35 & 9946.73 \\
  $\langle\mathcal{V}\rangle$ & 1709.55 & 1626.12 \\
  $N_e^{max}$ & 64 & 64 \\
  $M_{inf}$ & $5.5 \cdot 10^{15}$ & $6.8 \cdot 10^{15}$ \\
\end{tabular}\\
\vspace{0.3cm}{{\bf Table {6}:} Inflationary scales for models
with large $r$ (setting $K_{cs}=3\ln g_s\simeq -3.6$).}
\end{center}
\end{figure}

We have seen that although the Fibre Inflation mechanism can
naturally produce inflation with detectable tensor modes if the
moduli start at large enough values for $\hat\varphi$ ({\em i.e.}
high-fibre models), the generic such model ({\em e.g.} the LV
model of the Tables) predicts too small a Hubble scale during
inflation to have observable fluctuations. Such models may
nonetheless ultimately prove to be of interest, either by using
alternative mechanisms \cite{DiffMechs} to generate perturbations,
or as a way to generate a second, shorter and relatively late
epoch of inflation \cite{2ndInf} (as might be needed to eliminate
relics in the later universe).

\subsection{Two-field cosmological evolution}

Given that the resulting volumes, $\mathcal{V}\gtrsim 10^3$, are
not extremely large, one could wonder whether the approximations
made above are fully justified or not. We pause now to re-examine
in particular the assumption that $\V$ and $\tau_3$ remain fixed
at constant values while $\tau_1$ rolls during inflation. We first
identify the combination of parameters that controls this
approximation, and then re-analyse the slow roll with these fields
left free to move. This more careful treatment justifies our use
of the single-field approximation elsewhere.

\subsubsection{Inflaton back-reaction onto $\V$ and $\tau_3$}

Recall that the approximation that $\V$ and $\tau_3$ not move is
justified to the extent that the $\tau_1$-independent stabilising
forces of the potential $\delta V_{(\alpha')}$ remain much
stronger than the forces in $\delta V_{(g_s)}$ that try to make
$\V$ and $\tau_3$ also move. And this hierarchy of forces seems
guaranteed to hold because of the small factors of $g_s$ and
$1/\V^{1/3}$ that suppress the string-loop contribution relative
to the $\alpha'$ corrections. However we also see, from
\pref{cobe} and Table 4, that observably large primordial
fluctuations preclude taking $g_s\,e^{K_{cs}}/\V^{10/3}$ to be too small --
at least when they are generated by the standard mechanism. This
implies a tension between the COBE normalization and the validity
of our analysis at fixed $\V$, whose severity we now try to
estimate.

Since the crucial issue is the relative size of the forces on $\V$
due to $\delta V_{(\alpha')}$, $\delta V_{(sp)}$ and $\delta
V_{(g_s)}$, we first compare the derivatives of these potentials.
Keeping in mind that it is the variable $\vartheta_v \sim \ln \V$
that satisfies the slow-roll condition, we see that the relevant
derivative to be compared is $\V \partial /\partial \V$.
Furthermore, since it is competition between derivatives of
$\delta V_{(sp)}$ and $\delta V_{(\alpha')}$ in eq.~(\ref{ygfdo})
that determines $\V$ in the leading approximation, it suffices to
compare the string-loop potential with only the $\alpha'$
corrections, say. We therefore ask when
\be
 \left| \V \frac{\partial \delta V_{(g_s)}}{\partial \V}
 \right| \ll \left| \V \frac{\partial
 \delta V_{(\alpha')}}{\partial \V}
 \right| \,,
\ee
or when
\begin{equation}
 \frac{10 \, \mathcal{C}_2}{\V^{10/3}}
 \ll \frac{9 \, \xi}{4g_{s}^{3/2}}
 \frac{W_{0}^{2}}{\mathcal{V}^{3}} \,,
\label{Valfa}
\end{equation}
where we take $3 \gg 4 \, e^{-\kappa\hat\varphi/2}$ during
inflation when simplifying the left-hand side. Grouping terms we
find the condition
\begin{equation}
 \mathcal{R}_{cv} :=
 \left(\frac{9 \xi W_{0}^{2}}{40 g_{s}^{3/2}} \right)
 \frac{\mathcal{V}^{1/3}}{\mathcal{C}_2}
 \simeq \left( \frac{9 \xi \zeta^{4/3}}{40 g_{s}^{3/2}} \right)
 \frac{\mathcal{V}^{1/3}}{A}
 \gg 1 \,,
 \label{stab}
\end{equation}
which is clearly satisfied if we can choose $g_s$ and $1/\V^{1/3}$
to be sufficiently small. The value for $\mathcal{R}_{cv}$
predicted by the benchmark models of Tables 1 and 2 is given in
Table 4. This Table shows that large $\mathcal{R}_{cv}$ is much
larger in large-$\V$ models, as expected, with $\mathcal{R}_{cv} >
10^3$ in the LV model. By contrast, $\mathcal{R}_{cv} \gsim 10$
for inflationary parameter choices (SV1 and SV2) that satisfy the
COBE normalisation. Although these are large, the incredible
finickiness of inflationary constructions leads us, in the next
section, to study the multi-field problem where the volume modulus
is free to roll in addition to the inflaton. Be doing so we hope
to widen the parameter space of acceptable inflationary models.

\subsubsection{Relaxing the Single-Field Approximation}

\begin{figure}[ht]
\begin{center}
\epsfig{file=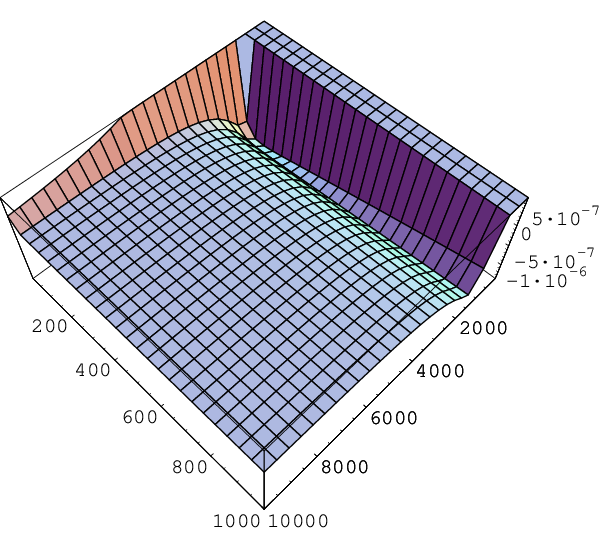, height=60mm,width=60mm}
\epsfig{file=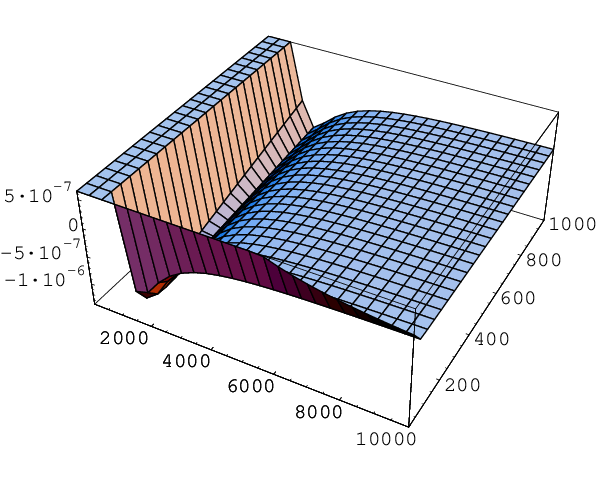, height=60mm,width=60mm} \caption{Two
  views of the
  inflationary trough representing the potential as a function of the
  volume and $\tau_1$ for $R=0$. The rolling is mostly in the $\tau_1$ direction
  (`north-west' direction in the left-hand figure and `south-west'
  direction in the
  right-hand figure).} \label{Fig:trough}
\end{center}
\end{figure}

In this section we redo the inflationary analysis without making
the single-field approximation. We start from the very general
scalar potential, whose form is displayed in Figure
\ref{Fig:trough},
\begin{equation}
 V=\mu_{1}\frac{\sqrt{\tau _{3}}}{\mathcal{V}}e^{-2a_{3}\tau
 _{3}}-\mu_{2}W_{0}\frac{\tau _{3}e^{-a_{3}\tau
 _{3}}}{\mathcal{V}^{2}}
 +\mu_{3}\frac{W_{0}^{2}}{\mathcal{V}^{3}}+\frac{\delta
 _{up}}{\mathcal{V}^{4/3}}+\frac{D}{\mathcal{V}^{3}\sqrt{\tau
 _{3}}}+\left( \frac{A}{\tau
 _{1}^{2}}-\frac{B}{\mathcal{V}\sqrt{\tau _{1}}}+\frac{C\tau
 _{1}}{\mathcal{V}^{2}}\right)
 \frac{W_{0}^{2}}{\mathcal{V}^{2}}\,.
\label{ilpot}
\end{equation}
Here
\begin{equation}
 \mu_{1}\equiv\frac{8a_{3}^{2}A_{3}^{2}}{3\alpha\gamma},\text{ \ \
 \ }\mu_{2}\equiv 4 a_{3}A_{3},\text{ \ \ \
 }\mu_{3}\equiv\frac{3\xi}{4 g_{s}^{3/2}}.
\end{equation}
Recall that the correction proportional to $D$ does not depend on
$\tau_1$ which is mostly the inflaton, but it can change the
numerical value obtained for $\tau_3$ and $\V$ at the minimum.
However, for $D=g_s^2\left(\mathcal{C}_3^{KK}\right)^2\ll 1$ this
modification is negligible. Thus we set $D=0$ from now on.

The result for $\V$ obtained by solving ${\partial V}/{\partial
\tau _{3}}=0$, in the limit $a_{3}\tau _{3}\gg 1$, reads
\begin{equation}
 \mathcal{V} =\frac{2\mu_{2}W_{0}}{\mu_{1}} \sqrt{\tau _{3}}
 \left( \frac{1- a_3 \tau_3}{1 - 4 a_3 \tau_3} \right)
 e^{a_{3}\tau_{3}}
 \simeq \frac{\mu_{2}W_{0}}{2\mu_{1}}\sqrt{\tau _{3}}
 \, e^{a_{3}\tau_{3}}.  \label{tautre}
\end{equation}
Solving eq.~\pref{tautre} for $\tau_3$, we obtain the result
\be
\label{app:tau3vsV}
    a_3 \tau_3 \simeq a_3 \tau_3 + \ln \left(\frac{\sqrt\tau_3}{2}
    \right)  := \ln \left(
    c \V \right) \,,
\ee
where $c = 2 a_3 A_3/(3\alpha \gamma \, W_0)$. Here the first
approximate equality neglects the slowly-varying logarithmic
factor, bearing in mind that in most of our applications we find
$\sqrt{\tau_3} \simeq 2$. Using this to eliminate $\tau_3$ in
(\ref{ilpot}) then gives the following approximate potential for
$\V$ and $\tau_1$
\begin{equation}
 V=\left[-\mu_{4}(\ln
 \left(c\mathcal{V}\right))^{3/2}+\mu_{3}\right]
 \frac{W_{0}^{2}}{\mathcal{V}^{3}}+\frac{\delta
 _{up}}{\mathcal{V}^{4/3}}+\left( \frac{A}{\tau
 _{1}^{2}}-\frac{B}{\mathcal{V}\sqrt{\tau _{1}}}+\frac{C\tau
 _{1}}{\mathcal{V}^{2}}\right) \frac{W_{0}^{2}}{\mathcal{V}^{2}}, \label{PPpot1}
\end{equation}
where $\mu_{4} = \frac32 \alpha\gamma a_3^{-3/2}$.

Given that we set $\tau_3$ at its minimum, $\partial_{\mu}
\tau_3=0$, and so the non canonical kinetic terms look like
(\ref{Lkin2}). In order now to study inflation, we let the two
fields $\mathcal{V}$ and $\tau_1$ evolve according to the
cosmological evolution equations for non-canonically normalised
scalar fields:
\begin{equation}
 \left\{
 \begin{array}{c}
 \ddot{\varphi}^{i}+3H\dot{\varphi}^{i}
 +\Gamma_{jk}^{i}\dot{\varphi}^{j}\dot{
 \varphi}^{k}+g^{ij}\frac{\partial V}{\partial \varphi ^{j}}=0, \\
 H^{2}=\left( \frac{\dot{a}}{a}\right) ^{2}=\frac{1}{3}\left(
 \frac{1}{2} g_{ij}\dot{\varphi}^{i}\dot{\varphi}^{j}+V\right),
 \end{array}
 \right.
\end{equation}
where $\varphi_i$ represents the scalar fields ($\mathcal{V}$ and
$\tau_1$ in our case), $a$ is the scalar factor, and $\Gamma
_{jk}^{i}$ are the target space Christoffel symbols using the
metric $g_{ij}$ for the set of real scalar fields $\varphi^i$ such
that $\frac{\partial^2 K}{\partial \Phi^{I}
\partial_{\mu} \Phi^{*J}} \partial^{\mu}\Phi^I\partial\Phi^{*J}
=\frac{1}{2}g_{ij}\partial_{\mu}\varphi^i\partial^{\mu}\varphi^j$.

For numerical purposes it is more convenient to write down the
evolution of the fields as a function of the number $N_e$ of
\efold ings rather than time. Using
\begin{equation}
 a(t)=e^{N_e},\textit{ \ \ \ \ \ \ \ \ \ \
 }\frac{d}{dt}=H\frac{d}{dN_e} ,
\end{equation}
we avoid having to solve for the scale factor, instead directly
obtaining $\mathcal{V}(N_e)$ and $\tau_1(N_e)$. The equations of
motion are (with $'$ denoting a derivative with respect to $N_e$):
\begin{eqnarray}
 \tau _{1}^{\prime \prime } &=&-\left( \mathcal{L}_{kin}+3\right)
 \left( \tau _{1}^{\prime }+2\tau _{1}^{2}\frac{V_{,
 \tau_{1}}}{V}+\tau _{1}\mathcal{V} \frac{V_{,\mathcal{V}}}{V}\right)
 +\frac{\tau _{1}^{\prime 2}}{\tau _{1}},
 \notag \\
 \mathcal{V}^{\prime \prime } &=&-\left( \mathcal{L}_{kin}+3\right)
 \left( \mathcal{V}^{\prime }+\tau _{1}\mathcal{V}\frac{V_{,
 \tau_{1}}}{V}+\frac{3
 \mathcal{V}^{2}}{2}\frac{V_{,\mathcal{V}}}{V}\right)
 +\frac{\mathcal{V} ^{\prime 2}}{\mathcal{V}},
\end{eqnarray}
We shall focus on the parameter case SV2, for which a numerical
analysis of the full potential gives:
\begin{equation}
 \left\langle \mathcal{V}\right\rangle =1413.26,\text{ \ \ \
 }\left\langle \tau _{1}\right\rangle =6.77325,\text{ \ \ \ }
 \delta_{up}=0.082.
\end{equation}
To evaluate the initial conditions, we fix $\tau_1\gg\left\langle
\tau _{1}\right\rangle$ and then we work out numerically the
minimum in the volume direction $\langle \V \rangle = \langle \V
\rangle( \tau_1)$.

Notice that, in general, in the case of unwarped up-lifting
$\frac{\delta_{up}}{\V^2}$, the volume direction develops a
run-away for large $\tau_1$, whereas the potential is well behaved
for the case with warped up-lifting $\frac{ \delta_{up}
}{\V^{4/3}}$. Thus we set the following initial conditions:
\begin{equation}
 \tau_1(0)=5000\text{ \ \ }\Rightarrow\text{ \ \
 }\V(0)\equiv\langle\mathcal{V}\rangle(\tau_1=5000)=1841.25,\text{
 \ \ \ }\tau_1'(0)=0,\text{ \ \ \ }\V'(0)=0.
\end{equation}
We need to check now that for this initial point we both get
enough \efold ings and the spectral index is within the allowed
range. In order to do this, we start by recalling the
generalisation of the slow-roll parameter $\varepsilon$ in the
two-field case:
\begin{equation}
 \varepsilon=-\frac{\left(V_{,\tau_1}\dot{\tau_1}+V_{,\mathcal{V}}\dot{\mathcal{V}}\right)^2}{4\mathcal{L}_{kin}V^2},
\end{equation}
and so it becomes a function of the number of \efold ings. In the
case SV2, $\varepsilon\ll 1$ for the first 65 \efold ings as it is
shown in figure \ref{Fig1} below.

\begin{figure}[ht]
\begin{center}
\epsfig{file=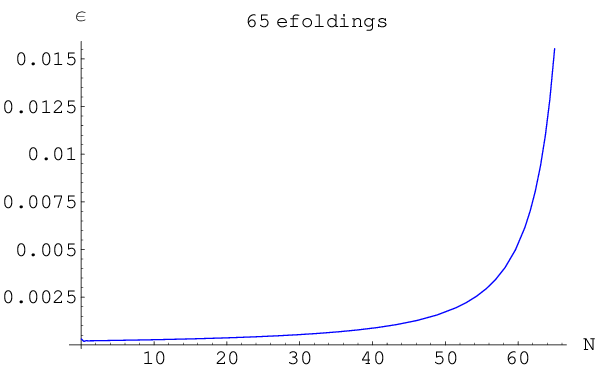, height=50mm,width=60mm}
\epsfig{file=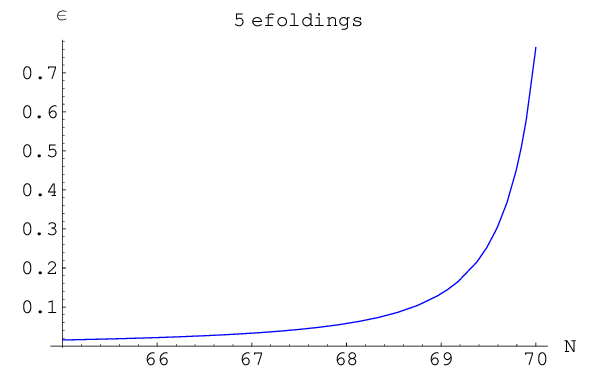, height=50mm,width=60mm}
\caption{$\varepsilon$ versus $N$ for (left) the first 65 \efold
ings of inflation and (right) the last 5 \efold ings.}
\label{Fig1}
\end{center}
\end{figure}

However $\varepsilon$ grows faster during the last 5 \efold ings
until it reaches the value $\varepsilon\simeq 0.765$ at $N=70$ at
which point the slow-roll approximation ceases to be valid and
inflation ends. This can be seen in figure \ref{Fig1}. (From here
on we save $N_e$ to refer to the physical number of \efold ings,
and denote by $N$ the variable that parameterises the cosmological
evolution of the fields).

\begin{figure}[ht]
\begin{center}
\epsfig{file=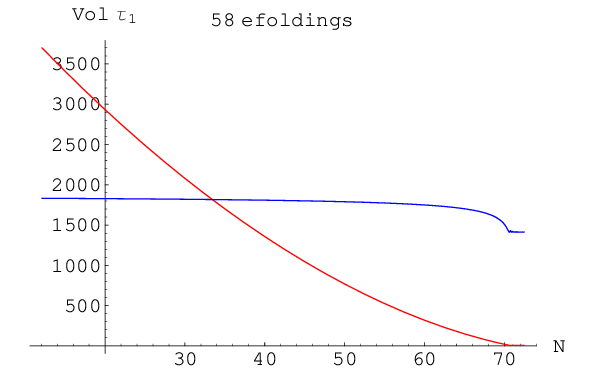, height=70mm,width=100mm}
\caption{$\tau_1$ (red curve) and $\mathcal{V}$ (blue curve)
versus $N$ for the last 58 \efold ings of inflation.} \label{Fig3}
\end{center}
\end{figure}

Therefore focusing on horizon exit at 58 \efold ings before the
end of inflation, we need to start at $N=12$. We also find
numerically that at horizon exit $\varepsilon(N=12)=0.0002844$
which corresponds to a tensor-to-scalar ration $r=4.6\cdot
10^{-3}$. Figure \ref{Fig3} shows the cosmological evolution of
the two fields during the last 58 \efold ings of inflation before
the fields start oscillating around the minimum. It it clear how
the motion is mostly along the $\tau_1$ direction, as expected.

\begin{figure}[ht]
\begin{center}
\epsfig{file=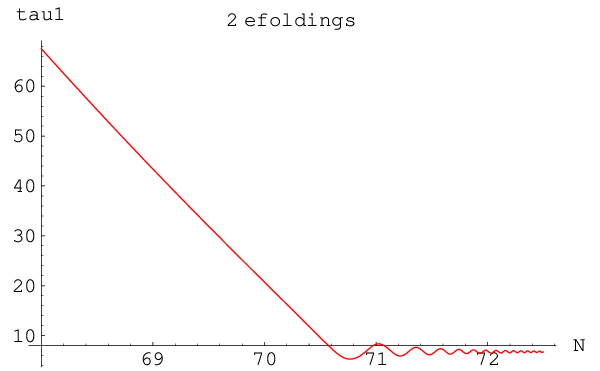, height=60mm,width=67mm}
\epsfig{file=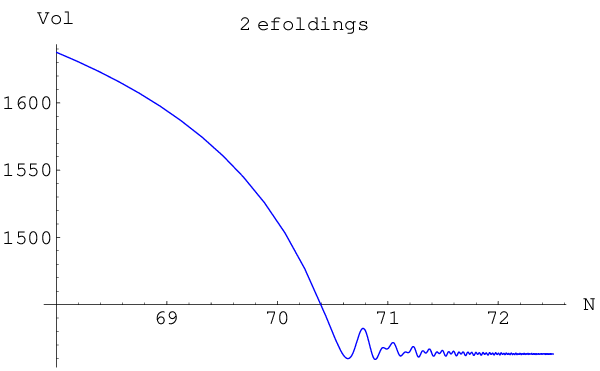, height=60mm,width=67mm} \caption{
 Plot of the $\tau_1$ (red curve on the left)
 and $\V$ (blue curve on the right) vs $N$ for the last 2 \efold ings of inflation.}
   \label{Fig4}
\end{center}
\end{figure}

Figure \ref{Fig4} gives a blow-up of the $\tau_1$ and $\V$
trajectory close to the minimum for the last 2 \efold ings of
inflation, where it is evident how the fields oscillate before
sitting at the minimum.

\begin{figure}[ht]
\begin{center}
\epsfig{file=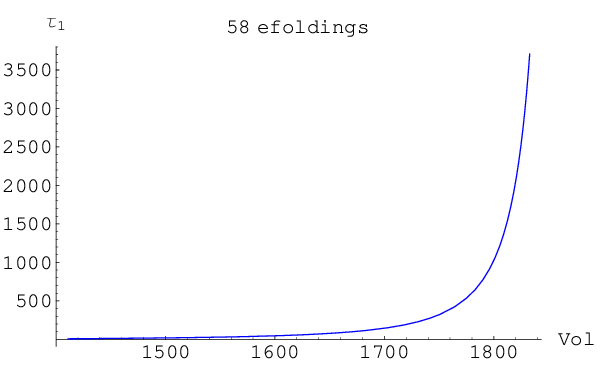, height=60mm,width=67mm}
\epsfig{file=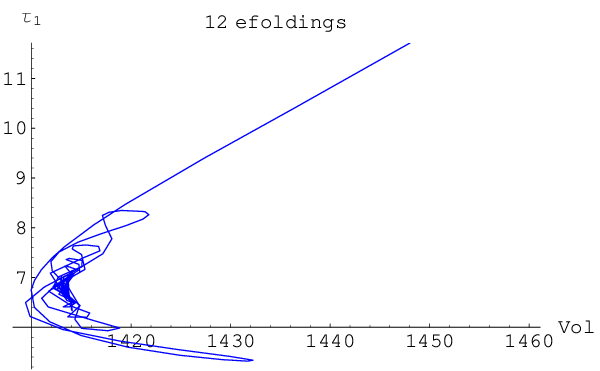, height=60mm,width=67mm}
\caption{Path of the inflation trajectory in the $\tau_1$-$\V$
space for the last 58 (left) and 12 (right) \efold ings of
inflation.}
   \label{Fig5}
\end{center}
\end{figure}

Finally, figure \ref{Fig5} illustrates the path of the inflation
trajectory in the $\tau_1$-$\V$ space.

\begin{figure}[ht]
\begin{center}
\epsfig{file=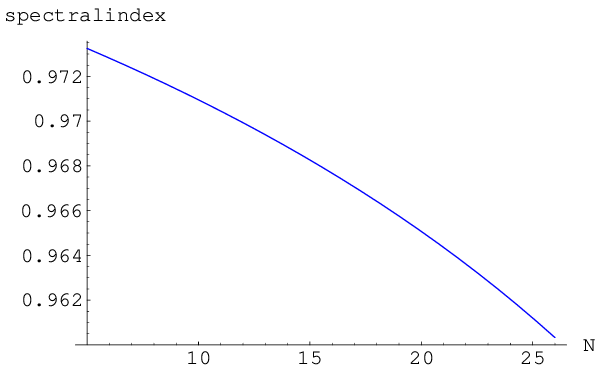, height=50mm,width=60mm}
\epsfig{file=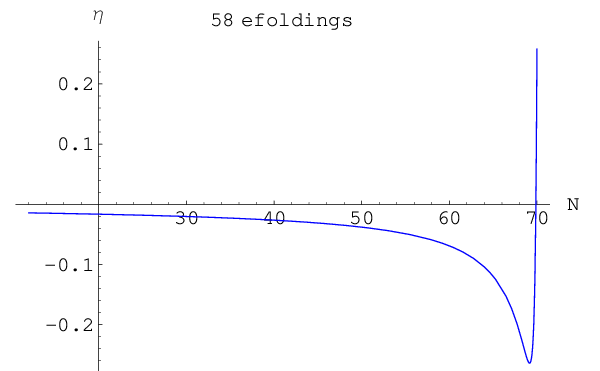, height=50mm,width=60mm} \caption{Left: $n_s$
versus $N$ between 65 and 44 \efold ings before the end of
inflation. Right: $\eta$ versus $N$ during the last 58 \efold ings
of inflation.} \label{Fig6}
\end{center}
\end{figure}

To consider the experimental predictions of Fibre Inflation we
need to make sure that the inflaton is able to generate the
correct amplitude of density fluctuations. After multiplying the
scalar potential (\ref{PPpot1}) by the proper normalisation factor
$g_s\,e^{K_{cs}}/(8\pi)$, the COBE normalisation on the power spectrum of
scalar density perturbations is given by
\begin{equation}
 \sqrt{P}\equiv\frac{\sqrt{g_s}\,e^{K_{cs}/2}}{20\sqrt{3}\,\pi^{3/2}}\sqrt{\frac{V}{\varepsilon}}=2\cdot
 10^{-5},
\end{equation}
where both $V$ and $\varepsilon$ have to be evaluated at horizon
exit for $N=12$ corresponding to $N_e=58$. We find numerically
that the COBE normalisation is perfectly matched:
\begin{equation}
 \text{at \ }N=12\text{: \ }\tau_1=3710.5,\textit{ \ \
 }\V=1832.74,\textit{ \ }\Rightarrow\textit{ \ }V=6.1\cdot
 10^{-7}\textit{ \ }\Rightarrow\textit{ \ }\sqrt{P}=2.15\cdot
 10^{-5}. \notag
\end{equation}
We need also to evaluate the spectral index which is defined as
\begin{equation}
 n_s=1+\frac{d \ln{P(k)}}{d \ln{k}}\simeq 1+\frac{d \ln{P(N)}}{d
N},
\end{equation}
where the latter approximation follows from the fact that
$k=aH\simeq He^{H}$ at horizon exit, so $d\ln{k}\simeq dN$. In
figure \ref{Fig6} we plot the spectral index versus $N$ around
horizon exit, namely between 65 and 44 \efold ings before the end
of inflation. It turns out that $n_s(N=12)=0.96993$, and so our
starting point is within the experimentally allowed region for the
spectral index.

We also checked that the second slow-roll parameter $\eta$,
obtainable from $\eta=(n_s+6\varepsilon-1)/2$, is always less than
unity during the last 58 \efold ings as shown in figure \ref{Fig6}
below. It is interesting to notice that $\eta$ vanishes very close
to the end of inflation for $N=69.88$. This is perfect agreement
with the presence of the inflection point previously found in the
fixed-volume approximation.

The inflationary scale evaluated at the end of inflation turns out
to be
\begin{equation}
 M_{inf}=V^{1/4}_{end}M_P=V(N=70)^{1/4}M_P=5.2\cdot 10^{16} GeV,
\end{equation}
and so, using (\ref{cosmology}) for $w=0$, we deduce that we can
obtain $N_e=58$ if $T_{rh}=2.27\cdot 10^{9} GeV$ which is
correctly below $10^{10} GeV$ to solve the gravitino problem.
Finally we conclude that we end up with the following experimental
predictions:
\begin{equation}
 n_s\simeq 0.970,\textit{ \ \ \ }r\simeq4.6\cdot 10^{-3} \,,
\end{equation}
in agreement with our earlier single-field results.

\subsection{Naturalness}

Finally, we return to the issue of the stability of the
inflationary scenario presented here to various kinds of
perturbations, and argue that it is much more robust than are
generic inflationary mechanisms because of the control afforded by
the LARGE volume approximation.\footnote{We thank Liam McAllister
for several helpful conversations on this point.}

There are several reasons why inflationary models are generically
sensitive to perturbations of various kinds, of which we list
several.

\medskip\noindent{\it Dimension-six Operators and the
$\eta$ problem:}

\medskip\noindent A generic objection to the stability of an
inflationary scenario rests on the absence of symmetries
protecting scalar masses. This line of argument
\cite{Copeland:1994vg} grants that it is possible to arrange a
regime where the scalar potential is to a good approximation
constant, $V = V_0$, chosen to give the desired inflationary
Hubble scale, $3 M_p^2 H^2 = V_0$. It then asks whether there are
dangerous higher-dimension interactions in the effective theory
that are small enough to allow an effective field theory
description, but large enough to compete with the extraordinarily
flat inflationary potential.

In particular, since $V_0$ is known (by assumption) not to be
precluded by symmetries of the problem, and since scalar masses
are notoriously difficult to rule out by symmetries, one worries
about the possibility of the following dimension-six combination
of the two:
\be
    {\cal L}_{\rm eff} = \frac{1}{M^2} \; V_0 \varphi^2 \,,
\ee
where $\varphi$ is the canonically normalised inflaton and $M$ is
a suitable heavy scale appropriate to any heavy modes that have
been integrated out. Such a term is dangerous, even with $M \simeq
M_p$, because it contributes an amount $V_0/M_p^2 \simeq H^2$ to
the inflaton mass, corresponding to $\delta \eta \simeq {\cal
O}(1)$. Supersymmetric versions of this argument use the specific
form $V_F = e^K \, U$, where $U$ is constructed from the
superpotential, and argue that inflation built on regions of
approximately constant $U$ get destabilised by generic $\delta K
\simeq \varphi^*\varphi$ corrections to the K\"ahler potential.

A related question that is specific to the large-field models
required for large tensor fluctuations asks what controls the
expansion of the effective theory in powers of $\varphi$ if fields
run over a range as large as $M_p$.

We believe that neither of these problems arise in the Fibre
Inflation models considered here. First, both arguments rely on
generic properties of an expansion in powers of $\varphi$, which
is strictly a valid approximation only for small excursions about
a fixed point in field space. As the previous paragraph points
out, such an expansion cannot be used for large field excursions
and one must instead identify a different small parameter with
which to control calculations. In the present instance this small
parameter is given both by powers of $1/\V$ and by powers of
$g_s$, since these control the underlying string perturbation
theory and low-energy approximations. In particular, as we find
explicitly in Appendix \ref{Appendix A}, in the supersymmetric
context these ensure that perturbations to $K$ have the form
$\delta e^K \simeq \delta (1/\V^2) \simeq -2\delta \V/\V^3$, and
it is the suppression by additional powers of the LARGE volume
that makes such corrections less dangerous than they would
generically be.

Normally when dangerous corrections are suppressed by a small
expansion parameter, the suppression can be traced to additional
symmetries that emerge in the limit that the parameter vanishes.
But for large-volume expansions, the corrections vanish strictly
in the de-compactification limit, $\V \to \infty$, which {\em
does} enjoy many symmetries (like higher-dimensional general
covariance) that are not evident in the effective
lower-dimensional theory. It would be worth understanding in more
detail whether the natural properties of the large-volume
expansion can be traced to these additional symmetries of higher
dimensions.

\medskip\noindent{\it Integrating out sub-Planckian modes:}

\medskip\noindent There is a more specific objection, related to the
above. Given the potential sensitivity of inflation to
higher-dimension operators, this objection asks why the inflaton
potential is not destabilised by integrating out the many heavy
particles that are likely to live above the inflationary scale,
$M_I \simeq V_0^{1/4}$ and below the Planck scale? (See, for
instance, ref.~\cite{tPsignals} for more specific variants of this
question.)

In particular, for string inflation models one worries about the
potential influence of virtual KK modes, since these must be
lighter than the string and 4D Planck scales, and generically
couple to any inflaton field. For Fibre Inflation this question
can be addressed fairly precisely, since virtual KK modes are
included in the string loop corrections that generate the
inflationary potential in the first place.

There are generically two ways through which loops can introduce
the KK scale into the low-energy theory. First, the lightest KK
masses enter as a cutoff for the virtual contribution of the very
light states that can be studied purely within the 4D effective
theory. These states contribute following generic contributions to
the inflaton potential
\be \label{4Dloop}
    \delta V_{inf}^{4D} \simeq c_1 \hbox{STr} \, M^4 + c_2 m_{3/2}^2
    \hbox{STr} \, M^2 + \cdots \,,
\ee
where $c_1$ and $c_2$ are dimensionless constants, the gravitino
mass, $m_{3/2}$, measures the strength of supersymmetry breaking
in the low-energy theory, and the super-traces are over powers of
the generic 4D mass matrix, $M$, whose largest elements are of
order the KK scale, $M_{KK}$. In general low-energy supersymmetry
ensures $c_1 = 0$, making the second term the leading
contribution.

Now comes the important point. In the LARGE volume models of
interest, we know that $m_{3/2} \sim \V^{-1}$, and we know that
$M_{KK}$ is suppressed relative to the string scale by
$\V^{-1/6}$, and so in Planck units $M_{KK} \sim \V^{-2/3}$. These
together imply that $\delta V_{inf}^{4D} \sim \V^{-10/3}$, in
agreement with the volume-dependence of the loop-generated
inflationary potential discussed above.

But $\delta V_{inf}$ also potentially receives contributions from
scales larger than $M_{KK}$ and these cannot be described by the
4D loop formula, eq.~\pref{4Dloop}. These must instead be computed
using the full higher-dimensional (string) theory, potentially
leading to the dangerous effective interactions in the low-energy
theory. This calculation is the one that is explicitly performed
for torii in \cite{bhk} and whose properties were estimated more
generally in \cite{07040737,cicq}. Their conclusion is that such
effective contributions {\em do} arise in the effective 4D theory,
appearing there as contributions to the low-energy K\"ahler
potential. The contributions from open-string loops wrapped on a
cycle whose volume is $\tau$ have the generic form
\be
    \delta K \simeq \frac{1}{\V} \left[a_1 \sqrt\tau+ \frac{a_2}{\sqrt{\tau}}
    + \cdots \right] \,,
\ee
where it is $1/\tau$ that counts the loop expansion.

These two terms can potentially give contributions to the scalar
potential, and if so these would scale with $\V$ in the following
way,
\be
    \delta V^{\rm he}_{inf} \sim \frac{a_1}{\V^{8/3}} +
    \frac{a_2}{\V^{10/3}} + \cdots \,.
\ee
Notice that the first term is therefore potentially dangerous,
scaling as it does like $M_{KK}^4$. However a simple calculation
shows that the contribution of a term $\delta K \propto
\tau^\omega/\V$ gives a contribution to $V_F$ of the form $\delta
V_F \propto \left(\omega - \frac12 \right) \V^{-8/3}$ \cite{cicq},
implying that the leading correction to $K$ happens to drop out of
the scalar potential (although it does contribute elsewhere in the
action).

These calculations show how LARGE volume and 4D supersymmetry can
combine to keep the potentially dangerous loop contributions of KK
and string modes from destabilising the inflaton potential. We
regard the study of how broadly this mechanism might apply
elsewhere in string theory as being well worthwhile.

\section{Conclusions}

This year the Planck satellite is expected to start a new era of
CMB observations, and to be joined over the next few years by
other experiments aiming to measure the polarization of the cosmic
microwave background and to search for gravitational waves. We
have presented a new class of explicit string models, with moduli
stabilisation, that both agrees with current observations and can
predict observable gravitational waves, most probable not at
Planck but at future experiments. Many of the models' inflationary
predictions are also very robust against changes to the underlying
string/supergravity parameters, and in particular predict a
definite correlation between the scalar spectral index, $n_s$, and
tensor-to-scalar ratio, $r$. It is also encouraging that these
models realise inflation in a comparatively natural way, inasmuch
as a slow roll does not rely on fine-tuning parameters of the
potential against one another.

Other important features of the model are:
\begin{itemize}
\item The comparative flatness of the inflaton direction,
$\OmegA$, is guaranteed by general features of the modulus
potential that underly the LARGE volume constructions. These
ultimately rely on the no-scale structure of the lowest-order
K\"{a}hler potential and the fact that the leading $\alpha'$
corrections depend on the K\"{a}hler moduli only through the
Calabi-Yau volume.
\item The usual $\eta$ problem of generic supergravity theories is
also avoided because of the special features of the no-scale
LARGE-volume structure. In particular, the expansion of the
generic $e^K = \V^{-2}$ factor of the $F$-term potential are
always punished by the additional powers of $1/\V$, which they
bring along: $\delta e^K = -2 \V^{-3} \delta \V$. This result is
explicitly derived in Appendix \ref{Appendix A}.
\item The exponential form of the inflationary potential is a
consequence of two things. First, the loop corrections to $K$ and
$V$ dependend generically on powers of $\OmegA$ and the volume.
And second, the leading-order K\"{a}hler potential gives a kinetic
term for $\OmegA$ of the form $(\partial \ln \OmegA)^2$, leading
to the canonically normalised quantity $\varphi$, with $\OmegA=
e^{\kappa\hat\varphi}$, with $\kappa = 2/\sqrt3$. So we know the
potential can have a typical large-field inflationary form, $V =
K_1 - K_2 e^{-\kappa_1\hat\varphi} + K_3 e^{-\kappa_2 \hat\varphi}
+ \cdots $, without knowing any details about the loop
corrections.
\item The robustness of some of the predictions then follows
because the coefficients $K_{i}$ turn out to be proportional to
one another. They are proportional because of our freedom to shift
$\varphi$ so that $\hat\varphi = 0$ is the minimum of $V$, and our
choice to uplift this potential so that it vanishes at this
minimum. The two conditions $V(0) = V'(0) = 0$ impose two
conditions amongst the three coefficients $K_1$, $K_2$ and $K_3$
(where three terms in the potential are needed to have a minimum).
The remaining normalisation of the potential can then be expressed
without loss of generality in terms of the squared mass,
$m^2_\varphi = V''(0)$.
\item The exact range of the field $\hat\varphi$ depends only on
the ratio of two parameters ($B/A$) of the underlying
supergravity. This quantity is typically much greater than one due
to the string coupling dependence of this ratio, leading to
`high-fiber' models for which $\hat\varphi$ can naturally run
through trans-Planckian values. $B/A \gg 1$ also suffices to
ensure that the minimum $\langle \varphi \rangle$ lies inside the
K\"ahler cone. But the range of $\hat\varphi$ also cannot be too
large, since it depends only logarithmically on $B/A$. This
implies that $\hat\varphi$ at most rolls through a few Planck
scales, which can allow $50-60$ \efold ings, or even a bit more.
This makes the models potentially sensitive to details of the
modulus dynamics at horizon exit, along the lines of
\cite{tPsignals}, since this need not be deep in an inflationary
regime.
\item The COBE normalisation is the most constraining restriction
to the underlying string/supergravity parameters. In particular,
as usual, it forbids the volume from being very large because it
restricts the string scale to be of the order of the GUT scale.
This leads to the well known tension between the scale of
inflation and low-energy supersymmetry \cite{cklq}. Of course,
this conclusion assumes the standard production mechanism for
primordial density fluctuations, and it remains an interesting
open question whether alternative mechanisms might allow a broader
selection of inflationary models in this class. In particular,
this makes the development of a reheating mechanism particularly
pressing for this scenario.
\item The model is extremely predictive since the requirement of
generating the correct amplitude of scalar perturbations fixes the
inflationary scale of the order the GUT scale, which, in turn,
fixes the numbers of \efold ings. Lastly the number of \efold ings
is correlated with the cosmological observables and we end up with
the general prediction: $n_s\simeq 0.970$ and $r\simeq 0.005$. We
find examples with $r\simeq 0.01$ and $n_s \simeq 1$ also to be
possible, but only if horizon exit occurs very soon after the
onset of inflation.
\end{itemize}
For these reasons, even though the string-loop corrections to the
K\"{a}hler potential are not fully known for general Calabi-Yau
manifolds, because they come as inverse powers of K\"{a}hler
moduli and the dilaton we believe the results we find here are
likely to be quite generic. Of course, it would in any case be
very interesting to have more explicit calculations of the loop
corrections to K\"{a}hler potentials in order to better understand
this scenario. Furthermore even though blow-up modes are very
common for Calabi-Yau manifolds, it would be useful to have
explicit examples of K3 fibration Calabi-Yau manifolds with the
required intersection numbers.

During Fibre Inflation an initially large K3 fiber modulus
$\tau_1$ shrinks, with the volume $\V=t_1\tau_1$ approximately
constant. Consequently, the value of the 2-cycle modulus $t_1$,
corresponding to the base of the fibration, must grow during
inflation. This forces us to check that $t_1$ is not too small at
the start of inflation, in particular not being too close to the
singular limit $t_1\to 0$ where perturbation theory breaks down.
We show in appendix \ref{Appendix B} that the inflationary region
can start sufficiently far away from this singular limit. The more
restrictive limit on the range of the inflationary regime is the
breakdown of the slow-roll conditions as $t_1$ gets smaller,
arising due to the growth of a positive exponentials in the
potential when expressed using canonical variables. One can
nonetheless show that natural choices of the underlying parameters
can guarantee that enough \efold ings of inflation are achieved
before reaching this region of field space.

It is worth emphasising that, independent of inflation and as
mentioned in section 3.2,  we have also shown that our scenario
allows for the LARGE volume to be realised in such a way that
there is a hierarchy of scales in the K\"ahler moduli, allowing
the interesting possibility of having two dimensions much larger
than the rest and  making contact with the potential
phenomenological and cosmological implications of two large extra
dimensions scenarios\cite{add,cc}.

We do not address the issues of initial conditions, which in our
case ask why the other fields start initially near their minimum,
and why inflationary modulus should start out high up a fiber. As
for K\"{a}hler modulus inflation, one argument is that {\em any}
initial modulus configuration must evolve towards its stabilised
value, and so if the last modulus to reach is minimum happens to
be a fibre modulus we expect this inflationary mechanism to be
naturally at work.

\vspace{4.8mm}
\newpage

\section*{Acknowledgements}

We are  indebted to Joe Conlon for multiple insightful
discussions, critical remarks and suggestions. We also thank
Daniel Baumann, Markus Berg, Jim Cline, Keshav das Gupta, Marta
Gomez-Reino, Michele Liguori, Liam Macallister, Anshuman Maharana,
Juan Maldacena, Marieke Postma, Toni Riotto, Gary Shiu, Eva
Silverstein, Licia Verde and Alex Westphal for useful
conversations. MC is partially funded by St John's College, EPSRC
and CET. FQ is partially funded by STFC and a Royal Society
Wolfson merit award. CB receives partial research support from
NSERC of Canada, CERN and McMaster University, and wishes to thank
the Center for Theoretical Cosmology (CTC) at Cambridge University
for its generous hospitality while this work was in progress.
Research at the Perimeter Institute is supported in part by the
Government of Canada through NSERC and by the Province of Ontario
through MRI.

\appendix
\section{Higher order corrections to the inflationary potential}
\label{Appendix A}

In this appendix we derive explicitly the leading corrections to
the fixed-volume approximation, which give rise to higher order
operators. We show that these operators do not introduce an $\eta$
problem since they are suppressed by inverse powers of the overall
volume.

\subsection{Derivation of the $\tau_1$ dependent shift of
  $\langle\mathcal{V}\rangle$}

\bigskip We start from the very general scalar potential
(\ref{PPpot1}):
\begin{equation}
 V=\left[-\mu_{4}(\ln
 \left(c\mathcal{V}\right))^{3/2}+\mu_{3}\right]
 \frac{W_{0}^{2}}{\mathcal{V}^{3}}+\frac{\delta
 _{up}}{\mathcal{V}^{2}}+\left( \frac{A}{\tau
 _{1}^{2}}-\frac{B}{\mathcal{V}\sqrt{\tau _{1}}}\right)
 \frac{W_{0}^{2}}{\mathcal{V}^{2}}, \label{app:pot1}
\end{equation}
where we have set $C=0$ since the loop corrections proportional to
$C$ turn out to be numerically small in the cases of interest,
both to finding the minimum in $\tau_{1}$ and to the inflationary
region. We now minimise this potential to obtain $\langle \V
\rangle$, first turning off the loop potential to investigate how
the uplifting term changes the minimum for $\V$. We follow this by
a perturbative study of the additional $\tau_1$-dependence
generated by the loop corrections: $\left\langle
\mathcal{V}\right\rangle =\mathcal{V}_{0} + \delta
\mathcal{V}(\tau _{1})$.

\subsubsection*{Uplifting only}

In the absence of loop corrections the potential reads
\begin{equation}
 V=\left[ -\mu_{4}(\ln
 \left(c\mathcal{V}\right))^{3/2}+\mu_{3}\right]
 \frac{W_{0}^{2}}{\mathcal{V}^{3}}+\frac{\delta
 _{up}}{\mathcal{V}^{2}}, \label{app:Pot1}
\end{equation}
where the up-lifting term is chosen to ensure
\begin{equation}
 \langle V \rangle=\left[-\mu_{4}(\ln
 \left(c\mathcal{V}_0\right))^{3/2}+\mu_{3}\right]
 \frac{W_{0}^{2}}{\mathcal{V}_0^{3}}+\frac{\delta
 _{up}}{\mathcal{V}_0^{2}}=0, \label{app:POt1}
\end{equation}
and so
\begin{equation}
 \delta_{up}=\left[\mu_{4}(\ln
 \left(c\mathcal{V}_0\right))^{3/2}-\mu_{3}\right]
 \frac{W_{0}^{2}}{\mathcal{V}_0}\,. \label{app:up-lifting}
\end{equation}
Here $\V_0$ satisfies $\left.\partial V/\partial\V\right|_{\V_0} =
0$, and so must solve
\begin{equation}
 \frac{4\delta_{up}\mathcal{V}_0}{\mu_{4}W_0^2}+\frac{6\mu_{3}}{\mu_{4}}+3
 (\ln\left(c\mathcal{V}_0\right))^{1/2}-6(\ln\left(c\mathcal{V}_0\right))^{3/2}=0\,.
 \label{app:citala}
\end{equation}

This is most simply analysed once it is rewritten as
\begin{equation}
 \psi+ p\, (\ln\left(c\mathcal{V}_0\right))^{1/2}
 -(\ln\left(c\mathcal{V}_0\right))^{3/2}=0,
\label{app:lunga}
\end{equation}
with
\begin{equation}
 \psi :=\frac{\mu_3}{\mu_4}=\frac{\xi}{2\,\alpha\gamma}
 \left(\frac{a_3}{g_s}\right)^{3/2} \,,
\end{equation}
and the parameter $p$ takes the value $p = \frac32$ if we evaluate
$\delta_{up}$ using (\ref{app:up-lifting}), or $p = \frac12$ if we
take $\delta_{up} = 0$. Tracking the dependence on $p$ therefore
allows us to understand the sensitivity of the result to the
presence of the uplifting term.

The exact solution of (\ref{app:lunga}) is
\begin{equation}
 \ln \left( c\mathcal{V}_{0}\right) = \frac{\left[12p +\left( 108
 \psi + 12 \sqrt{81 \psi^2 - 12 p^{3}} \right)^{2/3} \right]^2} {36
 \left(108 \psi + 12 \sqrt{81 \psi^{2}-12 p^{3}} \right)^{2/3}} \,,
 \label{app:prima}
\end{equation}
which approaches the $p$-independent result
\begin{equation}
 \ln \left( c \mathcal{V}_0 \right) \simeq \psi^{2/3} =a_3
 \left( \frac{\hat\xi}{2\, \alpha \gamma} \right)^{2/3} \,,
\end{equation}
when $\psi \gg 1$, in agreement with eq.~\pref{app:tau3vsV}
together with expression \pref{x} for $\tau_3$. This shows that we
may expect the uplifting corrections to $\V_0$ to be small when
$\psi \gg 1$.

\subsubsection*{Including loop corrections}

The potential now is given by (\ref{app:pot1}) and so the presence
of the loops will generate a $\tau_1$ dependent shift of
$\mathcal{V}$ such that
\begin{equation}
 \mathcal{V}=\mathcal{V}_0+\delta \mathcal{V}(\tau_{1}),\text{ \
 with \ }\delta \mathcal{V}(\tau_{1})\ll \mathcal{V}_0\text{ \
 }\forall \tau_1.
\end{equation}
In order to calculate $\delta\mathcal{V}(\tau_1)$ at leading
order, let us solve the minimisation equation for the volume
taking into account that now that we have turned on the string
loops, we need to replace $\delta_{up}$ by $\delta_{up}' =
\delta_{up} + \mu_{up}$, where $\mu_{up}$ is the constant needed
to cancel the contribution of the loops to the cosmological
constant.
\begin{equation}
 \frac{\partial V}{\partial \mathcal{V}}=0\Longleftrightarrow
 \frac{4A\mathcal{V}}{\mu_4\tau_1^2}-\frac{6B}{\mu_4\sqrt{\tau_1}}
 +4\frac{\delta_{up}'\mathcal{V}}{\mu_{4}W_0^2}+6\frac{\mu_{3}}{\mu_{4}}+3
 (\ln\left(c\mathcal{V}\right))^{1/2}-6(\ln\left(c\mathcal{V}\right))^{3/2}=0.
\end{equation}
We notice that the logarithm in the previous expression can be
expanded as follows
\begin{equation}
 \ln \left( c\mathcal{V}\right)=\ln \left(
 c\mathcal{V}_{0}\right)+\frac{\delta\mathcal{V}(\tau_1)}{\mathcal{V}_0},
\end{equation}
and by means of another Taylor series and the result
(\ref{app:citala}), we are left with
\begin{gather}
 \left( 4\frac{\delta
 _{up}\mathcal{V}_{0}}{\mu_{4}W_{0}^{2}}+4\frac{\mu
 _{up}\mathcal{V}_{0}}{\mu_{4}W_{0}^{2}}+\frac{4A\mathcal{V}_{0}}{%
 \mu_{4}\tau _{1}^{2}}+\frac{3}{2}\left( \ln \left( c\mathcal{V}%
 _{0}\right) \right) ^{-1/2}-9\left( \ln \left(
 c\mathcal{V}_{0}\right) \right) ^{1/2}\right) \frac{\delta
 \mathcal{V}(\tau _{1})}{\mathcal{V}_{0}}=
 \notag \\
 -\frac{4A\mathcal{V}_{0}}{\mu_{4}\tau _{1}^{2}}+\frac{6B}{\mu_{4}%
 \sqrt{\tau _{1}}}-4\frac{\mu
 _{up}\mathcal{V}_{0}}{\mu_{4}W_{0}^{2}}.
\end{gather}
Now recalling the expression (\ref{app:up-lifting}) for
$\delta_{up}$ combined with (\ref{app:lunga}), we obtain
\begin{equation}
 \frac{\delta \mathcal{V}(\tau _{1})}{\mathcal{V}_{0}}=\frac{\left(
 \frac{4A \mathcal{V}_0}{\mu_{4}\tau
 _{1}^{2}}-\frac{6B}{\mu_{4}\sqrt{\tau _{1} }}+4\frac{\mu
 _{up}\mathcal{V}_{0}}{\mu_{4}W_{0}^{2}}\right) }{\left(3\left( \ln
 \left( c\mathcal{V}_{0}\right) \right) ^{1/2}-\frac{3}{2}\left(
 \ln \left( c\mathcal{V}_{0}\right) \right) ^{-1/2}-\frac{4A
 \mathcal{V}_0}{\mu_{4}\tau _{1}^{2}}-4\frac{\mu _{up}\mathcal{V
 }_{0}}{\mu_{4}W_{0}^{2}}\right)}. \label{th}
\end{equation}
We can still expand the denominator in (\ref{th}) and working at
leading order we end up with
\begin{equation}
 \frac{\delta \mathcal{V}(\tau _{1})}{\mathcal{V}_{0}}=\frac{\left(
 \frac{4A \mathcal{V}_0}{\mu_{4}\tau
 _{1}^{2}}-\frac{6B}{\mu_{4}\sqrt{\tau _{1} }}+4\frac{\mu
 _{up}\mathcal{V}_{0}}{\mu_{4}W_{0}^{2}}\right) }{\left(3\left( \ln
 \left( c\mathcal{V}_{0}\right) \right) ^{1/2}-\frac{3}{2}\left(
 \ln \left( c\mathcal{V}_{0}\right) \right) ^{-1/2}\right)}.
\label{tho}
\end{equation}
We have now all the ingredients to work out the canonical
normalisation.

\subsection{Canonical normalisation}

As we have seen in the previous subsection of this appendix,
$\mathcal{V}$ and $\tau_{3}$ will both have a $\tau_1$ dependent
shift of the form
\begin{eqnarray}
\mathcal{V} &=&\mathcal{V}_{0}+\delta\mathcal{V}(\tau_{1}), \\
\tau_{3}
&=&\frac{\ln\left(c\mathcal{V}_0\right)}{a_3}+\frac{\delta\mathcal{V}(\tau_1)}{a_3\mathcal{V}_0},
\end{eqnarray}
which will cause $\partial_{\mu}\mathcal{V}$ and
$\partial_{\mu}\tau_3$ not to vanish when we study the canonical
normalisation of the inflaton field $\tau_1$ setting both
$\mathcal{V}$ and $\tau_{3}$ at its $\tau_1$ dependent minimum.
Thus we have
\begin{eqnarray}
\partial_{\mu }\mathcal{V} &=&\frac{\partial\left(\delta\mathcal{V}(\tau_1)\right)}{\partial\tau_1} \partial _{\mu }\tau _{1},
\label{1}
\\
\partial _{\mu }\tau _{3} &=&\frac{1}{a_3\mathcal{V}_0}
\frac{\partial\left(\delta\mathcal{V}(\tau_1)\right)}{\partial\tau_1}
\partial _{\mu }\tau _{1}. \label{2}
\end{eqnarray}
The non canonical kinetic terms look like
\begin{align}
-\mathcal{L}_{kin}& =\frac{1}{4}\frac{\partial ^{2}K}{\partial
\tau _{i}\partial \tau _{j}}\partial _{\mu }\tau _{i}\partial
^{\mu }\tau _{j}
\notag \\
& =\frac{3}{8\tau _{1}^{2}}\left( 1-\frac{2\alpha \gamma
}{3}\frac{\tau _{3}^{3/2}}{\mathcal{V}}\right) \partial _{\mu
}\tau _{1}\partial ^{\mu }\tau _{1}-\frac{1}{2\mathcal{V}\tau
_{1}}\left( 1-\alpha \gamma \frac{\tau
_{3}^{3/2}}{\mathcal{V}}\right) \partial _{\mu }\tau _{1}\partial
^{\mu }
\mathcal{V}  \notag \\
& +\frac{1}{2\mathcal{V}^{2}}\partial _{\mu }\mathcal{V}\partial
^{\mu } \mathcal{V}-\frac{3\alpha \gamma }{2}\frac{\sqrt{\tau
_{3}}}{\mathcal{V}^{2}}
\partial _{\mu }\tau _{3}\partial ^{\mu }\mathcal{V}+\frac{3\alpha \gamma }{8
}\frac{1}{\mathcal{V}\sqrt{\tau _{3}}}\partial _{\mu }\tau
_{3}\partial
^{\mu }\tau _{3}  \notag \\
& \simeq \frac{3}{8\tau _{1}^{2}}\partial _{\mu }\tau _{1}\partial
^{\mu }\tau _{1}-\frac{1}{2\mathcal{V}\tau _{1}}\partial _{\mu
}\tau _{1}\partial ^{\mu
}\mathcal{V}+\frac{1}{2\mathcal{V}^{2}}\partial _{\mu }\mathcal{V}
\partial ^{\mu }\mathcal{V}  \notag \\
& \qquad \qquad -\frac{3\alpha \gamma }{2}\frac{\sqrt{\tau
_{3}}}{\mathcal{V} ^{2}}\partial _{\mu }\tau _{3}\partial ^{\mu
}\mathcal{V}+\frac{3\alpha \gamma
}{8}\frac{1}{\mathcal{V}\sqrt{\tau _{3}}}\partial _{\mu }\tau
_{3}\partial ^{\mu }\tau _{3}.  \label{LKin}
\end{align}
Now using (\ref{1}) and (\ref{2}), we can derive the leading order
correction to the canonical normalisation in the constant volume
approximation:
\begin{equation}
-\mathcal{L}_{kin}=\frac{3}{8\tau _{1}^{2}}\left[
1-\frac{4\tau_1}{3}\frac{\partial}{\partial\tau_1}\left(\frac{\delta\mathcal{V}(\tau_1)}{\mathcal{V}_0}\right)
\right] \partial _{\mu }\tau _{1}\partial ^{\mu }\tau
_{1}=\frac{1}{2}\partial _{\mu }\varphi
\partial ^{\mu }\varphi,
\end{equation}
where $\varphi$ is the canonically normalised inflaton. Writing
$\varphi =g(\tau _{1})$ we deduce the following differential
equation
\begin{equation}
\frac{\partial g(\tau_1)}{\partial
\tau_1}=\frac{\sqrt{3}}{2\tau_1}\sqrt{1-\frac{4\tau_1}{3}\frac{\partial}
{\partial\tau_1}\left(\frac{\delta\mathcal{V}(\tau_1)}{\mathcal{V}_0}\right)},
\end{equation}
which, after expanding the square root, admits the straightforward
solution
\begin{equation}
\varphi =\frac{\sqrt{3}}{2}\ln \tau _{1}-\frac{1}{\sqrt{3}}\left(
\frac{ \delta \mathcal{V}(\tau _{1})}{\mathcal{V}_{0}}\right)
=\frac{\sqrt{3}}{2} \ln \tau _{1}\left[ 1-\frac{2}{3\ln \tau
_{1}}\left( \frac{\delta \mathcal{V} (\tau
_{1})}{\mathcal{V}_{0}}\right) \right],  \label{4}
\end{equation}
where the leading order term reproduces what we had in the main
text. We still need to invert this relation to get $\tau_1$ as a
function of $\varphi$ and then plug this result back in the
potential. We can write this function as
\begin{equation}
\tau _{1}=e^{2\varphi/\sqrt{3}}\left(1+h(\varphi)\right),
\label{44}
\end{equation}
where $h(\varphi)\ll 1$.

At this point we can substitute (\ref{44}) in (\ref{4}) and by
means of a Taylor expansion, derive an equation for $h(\varphi)$:
\begin{gather}
\varphi =\varphi -\varphi \left[ \frac{2}{3\ln \tau _{1}}\left(
\frac{\delta \mathcal{V}(\tau _{1})}{\mathcal{V}_{0}}\right)
\right] _{\tau
_{1}=e^{2\varphi /\sqrt{3}}}+\frac{\sqrt{3}}{2}h(\varphi )+...  \notag \\
\Longrightarrow \text{ \ }h(\varphi )\text{ }=\frac{2}{3}\text{\
}\left. \left( \frac{\delta \mathcal{V}(\tau
_{1})}{\mathcal{V}_{0}}\right) \right\vert _{\tau _{1}=e^{2\varphi
/\sqrt{3}}},
\end{gather}
where we have imposed that the two first order corrections cancel
in order to get the correct inverse function. Therefore the final
canonical normalisation of $\tau_1$ which goes beyond the constant
volume approximation reads
\begin{equation}
\tau _{1}=e^{2\varphi /\sqrt{3}}\left[ 1+\frac{2}{3}\left. \left(
\frac{ \delta \mathcal{V}(\tau _{1})}{\mathcal{V}_{0}}\right)
\right\vert _{\tau _{1}=e^{2\varphi /\sqrt{3}}}\right].
\label{444}
\end{equation}

\subsection{Leading correction to the inflationary slow roll}

In order to derive the full final inflationary potential at
leading order, we have now to substitute
$\mathcal{V}=\mathcal{V}_0+\delta\mathcal{V}(\tau_1)$ in
(\ref{app:pot1}) to obtain a function of just $\tau_{1}$. After
two subsequent Taylor expansions, the potential reads
\begin{equation}
V=\left[-\mu_{4}(\ln \left(
c\mathcal{V}_{0}\right))^{3/2}\left(1+\frac{3\delta\mathcal{V}(\tau_1)}{2\mathcal{V}_0\ln
\left( c\mathcal{V}_{0}\right)}\right)+\mu_{3}+\frac{\delta
_{up}'\left(\mathcal{V}_0+\delta
\mathcal{V}(\tau_{1})\right)}{W_0^{2}}+ \frac{A\mathcal{V}}{\tau
_{1}^{2}}-\frac{B}{\sqrt{\tau _{1}}}\right]
\frac{W_{0}^{2}}{\mathcal{V}^{3}}. \notag
\end{equation}
Recalling the expression (\ref{app:up-lifting}) for $\delta_{up}$,
the leading contribution of the non-perturbative and $\alpha'$ bit
of the scalar potential cancels against the up-lifting term and we
are left with the expansion of $V_{(np)}+V_{(\alpha')}+V_{(up)}$
plus the loops:
\begin{equation}
V=\left[ -\frac{3\mu_{4}}{2}(\ln \left( c\mathcal{V}_{0}\right)
)^{1/2}\frac{\delta \mathcal{V}(\tau
_{1})}{\mathcal{V}_{0}}+\frac{\delta _{up}\delta \mathcal{V}(\tau
_{1})}{W_{0}^{2}}+\frac{\mu _{up}\mathcal{V}}{W_{0}^{2}}
+\frac{A\mathcal{V} }{\tau _{1}^{2}}-\frac{B}{\sqrt{\tau
_{1}}}\right] \frac{W_{0}^{2}}{\mathcal{V}^{3}}.
\label{interessante}
\end{equation}

It is now very interesting to notice in the previous expression
that the leading order expansion of the non-perturbative and
$\alpha'$ bit of the potential cancels against the expansion of
the up-lifting term. In fact from (\ref{interessante}), we have
that
\begin{equation}
\delta V_{(np)}+\delta V_{(\alpha')}=-\frac{3\mu_{4}}{2}(\ln
\left( c\mathcal{V}_{0}\right) )^{1/2}\frac{\delta
\mathcal{V}(\tau _{1})}{\mathcal{V}_{0}}
\frac{W_{0}^{2}}{\mathcal{V}^{3}},
\end{equation}
along with
\begin{equation}
\delta V_{(up)}=\frac{\delta _{up}\delta \mathcal{V}(\tau
_{1})}{\mathcal{V}^{3}}=\frac{3\mu_{4}}{2}(\ln \left(
c\mathcal{V}_{0}\right) )^{1/2}\frac{\delta \mathcal{V}(\tau
_{1})}{\mathcal{V}_{0}} \frac{W_{0}^{2}}{\mathcal{V}^{3}},
\end{equation}
where the last equality follows from (\ref{app:up-lifting}) and
(\ref{app:lunga}). This result was expected since we fine tuned
$V_{(up)}$ to cancel $V_{(np)}+V_{(\alpha')}$ at
$\mathcal{V}=\mathcal{V}_0$ and then we have applied the same
shift $\mathcal{V}=\mathcal{V}_0+\delta\mathcal{V}(\tau_{1})$ to
both of them, so clearly still obtaining a cancellation.

Thus we get the following \textit{exact} result for the
inflationary potential
\begin{equation}
V_{inf}=\left[\frac{\mu _{up}\mathcal{V}}{W_{0}^{2}}
+\frac{A\mathcal{V} }{\tau _{1}^{2}}-\frac{B}{\sqrt{\tau
_{1}}}\right] \frac{W_{0}^{2}}{\mathcal{V}^{3}}.
\end{equation}
It is now possible to work out the form of $\mu_{up}$. The minimum
for $\tau_1$ lies at
\begin{equation}
\langle\tau_1\rangle=\left(\frac{4A}{B}\mathcal{V}\right)^{2/3},
\end{equation}
and so by imposing $\langle V_{inf}\rangle=0$ we find
\begin{equation}
\mu_{up}=\frac{3}{A^{1/3}}\left(\frac{B}{4}\right)^{4/3}\frac{W_0^2}{\mathcal{V}^{4/3}}.
\end{equation}
We can now expand again $\mathcal{V}$ around $\mathcal{V}_0$ and
obtain:
\begin{equation}
\mu_{up}=\frac{3}{A^{1/3}}\left(\frac{B}{4}\right)^{4/3}\frac{W_0^2}{\mathcal{V}_0^{4/3}}
\left(1-\frac{4}{3}\frac{\delta\mathcal{V}(\tau_1)}{\mathcal{V}_0}\right),
\label{mu}
\end{equation}
along with
\begin{equation}
V_{inf}=V^{(0)}+\delta V,
\end{equation}
where
\begin{equation}
V^{(0)}=\left[\frac{3}{A^{1/3}}\left(\frac{B}{4}\right)^{4/3}\frac{1}{\mathcal{V}_0^{1/3}}
+\frac{A\mathcal{V}_0 }{\tau _{1}^{2}}-\frac{B}{\sqrt{\tau
_{1}}}\right] \frac{W_{0}^{2}}{\mathcal{V}_0^{3}},
\end{equation}
is the inflationary potential derived in the main text in the
approximation that the volume is $\tau_1$-independent during the
inflationary slow roll, and
\begin{equation}
\delta
V=\left(\frac{\delta\mathcal{V}(\tau_1)}{\mathcal{V}_0}\right)
\left[-\frac{10}{A^{1/3}}\left(\frac{B}{4}\right)^{4/3}\frac{1}{\mathcal{V}_0^{1/3}}
-2\frac{A\mathcal{V}_0 }{\tau _{1}^{2}}+3\frac{B}{\sqrt{\tau
_{1}}}\right] \frac{W_{0}^{2}}{\mathcal{V}_0^{3}},
\end{equation}
is the leading order correction to that approximation.

Now that we have an expression for the up-lifting term $\mu_{up}$
given by (\ref{mu}), we are able to write down explicitly the form
of the shift of $\mathcal{V}$ due to $\tau_1$ (\ref{tho}) at
leading order:
\begin{equation}
\frac{\delta \mathcal{V}(\tau _{1})}{\mathcal{V}_{0}}=\frac{\left(
\frac{4A \mathcal{V}_0}{\mu_{4}\tau
_{1}^{2}}-\frac{6B}{\mu_{4}\sqrt{\tau _{1}
}}+\frac{3B^{4/3}}{\mu_{4}\left(4A\right)^{1/3}}\frac{1}{\mathcal{V}_0^{1/3}}\right)
}{\left(3\left( \ln \left( c\mathcal{V}_{0}\right) \right)
^{1/2}-\frac{3}{2}\left( \ln \left( c\mathcal{V}_{0}\right)
\right) ^{-1/2}\right)}. \label{thor}
\end{equation}
Notice that the other possible source of correction to $V^{(0)}$
is the modification of the canonical normalisation of $\tau_1$ due
to $\delta\mathcal{V}(\tau_1)$ given by (\ref{444}). Let us
therefore evaluate now the contribution coming from this further
correction. Working just at leading order, we have to substitute
(\ref{444}) in $V^{(0)}$ and then expand obtaining
\begin{equation}
V^{(0)}=V_{inf}^{(0)}+\delta V^{(0)},
\end{equation}
whereas we can just substitute $\tau_1=e^{2\varphi/\sqrt{3}}$ in
$\delta V$ since an expansion of this term would be subdominant.
At the end, we find that
\begin{equation}
V_{inf}=V_{inf}^{(0)}+\delta V_{inf},
\end{equation}
where
\begin{equation}
V_{inf}^{(0)}=\left[ \frac{3}{A^{1/3}}\left( \frac{B}{4}\right)
^{4/3}
\frac{1}{\mathcal{V}_{0}^{1/3}}+A\mathcal{V}_{0}e^{-4\varphi
/\sqrt{3} }-Be^{-\varphi /\sqrt{3}}\right]
\frac{W_{0}^{2}}{\mathcal{V}_{0}^{3}}, \label{55}
\end{equation}
is the canonically normalised inflationary potential used in the
main text in the constant volume approximation. Moreover, $\delta
V^{(0)}$ and $\delta V$ turn out to have the same volume scaling
and so their sum will give the full final leading order correction
to $V_{inf}^{(0)}$:
\begin{gather}
\delta V_{inf}=\delta V^{(0)}+\delta V,  \notag \\
\delta V_{inf}=-\frac{10}{3}\left. \left( \frac{\delta
\mathcal{V}(\tau _{1}) }{\mathcal{V}_{0}}\right) \right\vert
_{\tau _{1}=e^{2\varphi /\sqrt{3}}} \left[ \frac{3}{A^{1/3}}\left(
\frac{B}{4}\right) ^{4/3}\frac{1}{\mathcal{V}
_{0}^{1/3}}+A\mathcal{V}_{0}e^{-4\varphi /\sqrt{3}}-Be^{-\varphi
/\sqrt{3}} \right] \frac{W_{0}^{2}}{\mathcal{V}_{0}^{3}}.
\label{45}
\end{gather}

Comparing (\ref{55}) with (\ref{45}), we notice the interesting
relation
\begin{equation}
\delta V_{inf}=-\frac{10}{3}\left. \left( \frac{\delta
\mathcal{V}(\tau _{1}) }{\mathcal{V}_{0}}\right) \right\vert
_{\tau _{1}=e^{2\varphi /\sqrt{3}}}V_{inf}^{(0)},
\end{equation}
which implies
\begin{equation}
V_{inf}=V_{inf}^{(0)}\left[1-\frac{10}{3}\left. \left(
\frac{\delta \mathcal{V}(\tau _{1}) }{\mathcal{V}_{0}}\right)
\right\vert _{\tau _{1}=e^{2\varphi /\sqrt{3}}}\right].
\label{VeryGood}
\end{equation}
This last relation shows a special instance of the general
mechanism discussed in the main text of how this model avoids the
$\eta$-problems that normally plague inflationary potentials. In
particular, the corrections from the one loop potential is seen to
enter in the volume-suppressed combination
$\delta\mathcal{V}/\mathcal{V}_0\ll 1$, ensuring that their
contribution to the inflationary parameters $\varepsilon$ and
$\eta$ is negligible.

\section{Loop Effects at High Fibre}
\label{Appendix B}

In this section we investigate in more detail what happens at the
string loop corrections when the K3 fibre gets larger and larger
and simultaneously the $\mathbb{C}P^1$ base approaches the
singular limit $t_1\to 0$. One's physical intuition is that loop
corrections should signal the approach to this singular point. In
fact, we show here that the Kaluza-Klein loop correction in
$\tau_2$ is an expansion in inverse powers of $\tau_2$ which goes
to zero when $\tau_1\to\infty\Leftrightarrow t_1\to 0$, as can be
deduced from (\ref{ts}). Therefore the presence of the singularity
is signaled by the blowing-up of these corrections. We then
estimate the value $\tau_1^*$ below which perturbation theory
still makes sense and so we can trust our approximation in which
we consider only the first term in the 1-loop expansion of $\delta
V^{KK}_{\tau_2,1-loop}$ and we neglect all the other terms of the
expansion along with higher loop effects. However it will turn out
that, still in a region where $\tau_1<\tau_1^*$, $\delta
V^{KK}_{\tau_2}$, corresponding to the positive exponential in
$V$, already starts to dominate the potential and stops inflation.

Let us now explain the previous claims more in detail. Looking at
the expressions (\ref{LOOP}) for all the possible 1-loop
corrections to $V$, we immediately realise that both $\delta
V^{KK}_{(g_s),\tau_1}$ and $\delta V^{W}_{(g_s),\tau_1\tau_2}$
goes to zero when the K3 fibre diverges since
$t^*=\sqrt{\lambda_1\tau_1}$. Therefore these terms are not
dangerous at all. Notice that there is no correction at 1-loop of
the form $1/\left(t_1\mathcal{V}^3\right)$ because, just looking
at the scaling behaviour of that term, we realise that it should
be a correction due to the exchange of winding strings at the
intersection of two stacks of $D7$ branes given by $t_1$, but the
topology of the K3 fibration is such that there are no 4-cycles
which intersect in $t_1$, and so these corrections are absent.

However the sign at 1-loop that there is a singularity when
$\tau_1\to\infty\Leftrightarrow t_1\to 0$, is that $\delta
V^{KK}_{(g_s),\tau_2}$ blows-up. In fact, following our previous
analysis \cite{cicq}, the contribution of $\delta
K^{KK}_{\tau_2,1-loop}$ at the level of the scalar potential is
given by the following expansion:
\begin{eqnarray}
\delta V_{\tau_2,1-loop}^{KK} &=&\sum\limits_{p=1}^{\infty }\left(
\alpha _{p}g_s^{p}\left( \mathcal{C}_{2}^{KK}\right)
^{p}\frac{\partial^{p}\left(K_{0}\right) }{\partial \tau
_{2}^{p}}\right) \frac{W_{0}^{2}}{\mathcal{V}
^{2}}\text{ }  \notag \\
\text{\ \ with }\alpha _{p} &=&0\text{ }\Longleftrightarrow p=1.
\label{App:V at 1-loop}
\end{eqnarray}
The vanishing coefficients of the first contribution is the
`extended no-scale structure.' Hence we obtain an expansion in
inverse powers of $\tau_2$:
\begin{eqnarray}
\delta V_{\tau _{2},1-loop}^{KK} &=&\left[ \alpha _{2}\left(
\frac{\rho }{ \tau _{2}}\right) ^{2}+\alpha _{3}\left( \frac{\rho
}{\tau _{2}}\right)
^{3}+...\right] \frac{W_{0}^{2}}{\mathcal{V}^{2}}  \notag \\
\text{with }\rho  &\equiv &g_{s}C_{2}^{KK}\ll 1\text{ and }\alpha
_{i}\sim \mathcal{O}(1)\text{ }\forall i.  \label{App:ewe}
\end{eqnarray}
We can then see that, since from (\ref{ts}) when
$\tau_1\to\infty\Leftrightarrow t_1\to 0$, $\tau_2\to 0$, all the
terms in the expansion (\ref{App:ewe}) diverge and perturbation
theory breaks down. Thus the region where the expansion
(\ref{App:ewe}) is under control is given by
\begin{equation}
\frac{\rho}{\tau_2}\leq 2\cdot 10^{-2}\textit{ \
}\Leftrightarrow\textit{ \
}\frac{\mathcal{V}}{\alpha\sqrt{\tau_1}}\geq 50 g_s
C^{KK}_2\textit{ \ }\Leftrightarrow\textit{ \
}\tau_1\leq\sigma_1\mathcal{V}^2\text{ \ with \
}\sigma_1\equiv\left(50 \alpha g_s C^{KK}_2\right)^{-2}.
\label{constr1}
\end{equation}
We need still to evaluate what happens at two and higher loop
level. The behaviour of the 1-loop corrections was under rather
good control since it was conjectured from a generalisation of an
exact toroidal calculation \cite{07040737} and it was tested by a
low energy interpretation in \cite{cicq}. However there is no
exact 2-loop calculation for the toroidal case which we could try
to generalise to an arbitrary Calabi-Yau. Thus the best we can do,
is to constrain the scaling behaviour of the 2-loop corrections
from a low energy interpretation. A naive scaling analysis
following the lines of \cite{cicq}, suggests that
\begin{equation}
\frac{\partial^{2}\left( \delta K_{\tau_2,2-loops}^{KK}\right)
}{\partial \tau_2^{2}}\sim \frac{g_s}{16\pi ^{2}}\frac{1}{\tau_2
}\frac{\partial ^{2}\left( \delta K_{\tau_2,1-loop}^{KK}\right)
}{\partial \tau_2^{2}}, \label{App:2 loops}
\end{equation}
and so $\delta K_{\tau_2,2-loops}^{KK}$ is an homogeneous function
of degree $n=-4$ in the 2-cycle moduli, exactly as $\delta
K_{\tau_1\tau_2,1-loop}^{W}$. Given that
\begin{equation}
\frac{\partial\left(\delta
K_{\tau_2,1-loop}^{KK}\right)}{\partial\tau_2}=-g_sC_{2}^{KK}\frac{\partial^2
\left(K_{tree}\right)}{\partial\tau_2^2},
\end{equation}
equation (\ref{App:2 loops}) takes the form
\begin{equation}
\frac{\partial ^{2}\left( \delta K_{\tau_2,2-loops}^{KK}\right)
}{\partial \tau_2^{2}}\sim -\frac{g_s^2C_{2}^{KK}}{16\pi
^{2}}\frac{1}{\tau_2}\frac{\partial^3
\left(K_{tree}\right)}{\partial\tau_2^3}. \label{App:2loops}
\end{equation}
The previous relation and the homogeneity of the K\"{a}hler
metric, produce then the following guess for the Kaluza-Klein
corrections at 2 loops
\begin{equation}
\delta K_{\tau_2,2-loops}^{KK}\sim -\frac{g_s^2C_{2}^{KK}}{16\pi
^{2}}\frac{\partial^2 \left(K_{tree}\right)}{\partial\tau_2^2},
\end{equation}
that at the level of the scalar potential would translate into
\begin{equation}
\delta
V_{\tau_2,2-loops}^{KK}=\frac{g_s^{2}C_2^{KK}}{8\pi^2}\left[\frac{1}{\tau_{2}^{2}}
+\mathcal{O}\left(\frac{1}{\tau_2^3}\right)
\right]\frac{W_0^2}{\mathcal{V}^2}. \label{ewef}
\end{equation}
We notice that (\ref{ewef}) has the same behaviour of
(\ref{App:ewe}) apart from the suppression factor $\left( 8\pi
^{2}\mathcal{C}^{KK}_2\right) ^{-1}\sim \mathcal{O}(10^{-2})$.
This is not surprising since the leading contribution of $\delta
K_{\tau_2,1-loop}^{KK}$ in $V$ is zero due to the extended
no-scale but the leading contribution of $\delta
K_{\tau_2,2-loops}^{KK}$ in $V$ is non-vanishing. Thus we conclude
that in the region $\tau_1\ll\sigma_1\mathcal{V}^2$ both higher
terms in the 1-loop expansion (\ref{App:ewe}) and higher loop
corrections (\ref{ewef}) are subleading with respect to the first
term in (\ref{App:ewe}) which we considered in the study of the
inflationary potential.

However, writing everything in terms of the canonically normalised
inflaton field $\hat{\varphi}$ expanded around the minimum, the
first term in (\ref{App:ewe}) turns into the positive exponential
which, as we have seen in section \ref{SlowRoll}, destroys the
slow-roll conditions when it starts to dominate the potential at
$\hat{\varphi}_{max}=12.4$ for $R=2.3\cdot 10^{-6}$. In this point
$\delta V^{W}_{(g_s),\tau_1\tau_2}$ is not yet completely
subleading with respect to $\delta V^{KK}_{(g_s),\tau_2}$ and so
the slow-roll conditions are still satisfied. The form of this
bound in terms of $\tau_1$ can be estimated as follows:
\begin{equation}
\tau _{1}=\left\langle \tau _{1}\right\rangle e^{2\hat{\varphi}
_{\max }/\sqrt{3} }=\left( \frac{4A}{B}\right) ^{2/3}y_{max
}^{2}\mathcal{V}^{2/3}\mathit{\ } \Leftrightarrow \mathit{\ }\tau
_{1}\leq \sigma _{2}\mathcal{V}^{2/3}\text{ \ with \ }\sigma
_{2}\equiv 4.2\cdot 10^{6}\left( \frac{A}{B}\right) ^{2/3}.
\label{constr2}
\end{equation}

\begin{figure}[ht]
\begin{center}
\epsfig{file=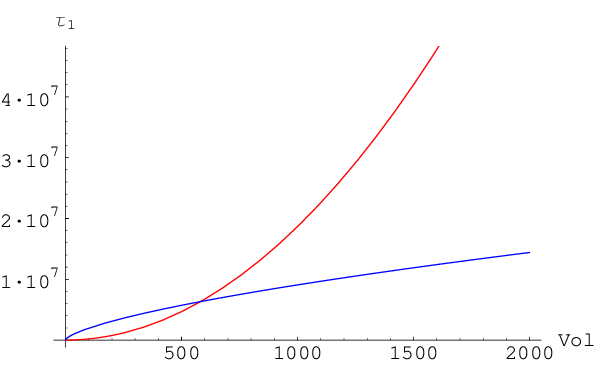, height=80mm,width=90mm} \caption{Plots of
the constraints $\tau_1^{max}=\sigma_1\mathcal{V}^2$ (red curve)
and $\tau_1^{max}=\sigma_2\mathcal{V}^{2/3}$ (blue curve) in the
($\tau_1$, $\mathcal{V}$) plane for the case SV2.}
   \label{Fig:walls}
\end{center}
\end{figure}

Let us now compare the bound (\ref{constr1}) with (\ref{constr2})
to check which is the most stringent one that constrains the field
region available for inflation. The value of the volume at which
the two bounds are equal is
\begin{equation}
\mathcal{V}_*=\left(\frac{\sigma_2}{\sigma_1}\right)^{3/4}=1.65\cdot
10^{7}\alpha g_s^{5/2}\frac{C^{KK}_1
(C^{KK}_2)^{3/2}}{(C^W_{12})^{1/2}}.
\end{equation}
Using the natural choice of parameter values made in the main text
(for SV2 for example), $\mathcal{V}_*=582$, and so, since we
always deal with much larger values of the overall volume, we
conclude that the most stringent constraint is (\ref{constr2}) as
can be seen from figure \ref{Fig:walls}.

Therefore the final situation is that, when the K3 fibre gets
larger, $\delta V^{KK}_{(g_s),\tau_2}$ starts dominating the
potential and ruining inflation well before one approaches the
singular limit in which the perturbative expansion breaks down and
these corrections blow-up to infinity.

\end{document}